%% file: main.tex
\documentclass[aps,prb,reprint,superscriptaddress,longbibliography]{revtex4-2}

\usepackage[T1]{fontenc}
\usepackage{amsmath,amssymb,amsthm,bm,graphicx,physics,mathtools}
\usepackage{xcolor}
\usepackage{hyperref}
\usepackage[capitalize]{cleveref}
\usepackage{multirow}
\usepackage{booktabs}
\usepackage{dsfont}
\hypersetup{colorlinks=true,linkcolor=blue,citecolor=blue,urlcolor=blue}
\usepackage{comment}

\newcommand{\kn}{\mathbf{k}}

\newcommand\isum{ \int_0^{2\pi} \frac{d\phi}{2\pi} \int_0^\pi \frac{\sin \theta \, d\theta}{2\pi} \int_0^\infty \frac{k^2 \, dk}{2\pi}}
\newcommand{\intsh}{\int_0^{2\pi} \frac{d\phi}{2\pi} \int_0^\pi \frac{\sin \theta \, d\theta}{2\pi}}
\newcommand{\intshone}{\int_0^{2\pi} \frac{d\phi_1}{2\pi} \int_0^\pi \frac{\sin \theta_1 \, d\theta_1}{2\pi}}
\newcommand{\gzero}{ \sum_{r,s,p,q} \Gamma^0_\mathrm{rspq} e^{i(r\theta_1 + s\phi_1 + p\theta_2 + q\phi_2)}}

\newcommand{\gone}{ \sum_{r,s,p,q} \Gamma_\mathrm{rspq} e^{i(r\theta_1 + s\phi_1 + p\theta_2 + q\phi_2)}}
\newcommand{\kp}{\mathbf{k'}}
\newcommand{\q}{\mathbf{q}}
\usepackage{minitoc}
\begin{document}

\title{\textcolor{black}{Localization and universality of three-dimensional pseudospin-$s$ fermions}}

\author{Arpan Gupta}
\affiliation{Department of Physics, Indian Institute of Technology Delhi, Hauz Khas, New Delhi 110016, India.}
\author{Gargee Sharma}
\affiliation{Department of Physics, Indian Institute of Technology Delhi, Hauz Khas, New Delhi 110016, India.}

\begin{abstract}
Quantum interference of electrons in disordered conductors is a sensitive probe of the internal structure of quasiparticles, revealing universal signatures of symmetry through weak localization (WL) and weak antilocalization (WAL). While these phenomena are well understood for the conventional Schr\"odinger and Dirac-Weyl fermions, their fate in the broader class of multifold chiral fermions remains largely unexplored. 
We develop a unified framework for semiclassical transport and quantum interference in three-dimensional disordered fermions with an arbitrary pseudospin $s$. 
Starting from a general short-range matrix disorder $\mathcal{M}$, we derive compact expressions for elastic lifetimes and ladder vertex corrections for arbitrary pseudospin with multiband effects, and then show that in the scalar-disorder limit while the Drude conductivity is strongly pseudospin and helicity dependent, in contrast, the leading quantum interference correction exhibits a striking universality: its magnitude remains identical to that of conventional diffusive metals and Weyl fermions, while its sign is determined solely by the parity of $2s$, placing half-integer pseudospins in the symplectic class (WAL) and integer pseudospins in the orthogonal class (WL). We also analyze the role of interband and intervalley scattering for $s=3/2$. By solving the resulting coupled Bethe-Salpeter equations, we demonstrate that channel mixing suppresses WAL and drives a crossover toward localization. Our results establish a general theory of localization across the full pseudospin hierarchy, revealing an interplay between internal geometry, symmetry class, and transport universality.
\end{abstract}

\maketitle
%%%%%%%%%%%%%%%%%%%%%%%%%%%%%%%%%%%%%%%%%%%%%%%%%%%%%%%%%%%%%%
\section{Introduction}
An exact quantum-mechanical treatment of electrons undergoing multiple scattering from a random distribution of impurities is, in general, an intractable many-body problem. The first key conceptual breakthrough came with Anderson~\cite{anderson1958absence,abrahams1979scaling}, who showed localization in a disordered tight-binding model using non-perturbative, self-consistent arguments based on the suppression of wavefunction amplitudes.
Since this seminal work, localization theory has become a cornerstone of condensed matter physics. It is now well established that sufficiently strong disorder can localize electronic wave functions, leading to a disorder-driven metal–insulator transition known as Anderson localization (AL). In the weak-disorder regime, and at sufficiently low temperatures where inelastic scattering is suppressed, a systematic perturbative expansion around the classical Drude conductivity reveals that even the leading quantum correction encodes remarkably rich physics. Specifically, constructive interference between time-reversed scattering paths enhances the return probability, giving rise to weak localization (WL), widely regarded as the precursor to strong localization~\cite{bergmann1984weak, lee1985disordered,akkermans2007mesoscopic,altshuler1980magnetoresistance,chakravarty1986weak}. Remarkably, this interference can instead become destructive when the internal geometric structure of Bloch-states imprints a nontrivial phase on time-reversed paths. This possibility was first elucidated by Hikami, Larkin, and Nagaoka in the context of spin-orbit coupled metals~\cite{hikami1980spin}, and has since been predicted and observed in systems ranging from graphene to topological insulators and Weyl semimetals~\cite{lu2015weak, suzuura2002crossover,khveshchenko2006electron,mccann2006weak,gorbachev2007weak,wu2007weak,gorbachev2007weak,tikhonenko2008weak,tikhonenko2008weak,tkachov2011weak,lu2011competition,lu2013intervalley,lu2014finite,fu2019quantum}. The resulting destructive interference manifests as weak antilocalization (WAL). These twin phenomena are now understood as universal infrared signatures of symmetry, encoded in the structure of the \textit{Cooperon}, which is the propagator that describes the coherent return probability via time-reversed paths~\cite{hikami1980spin}.

In topological Weyl and Dirac semimetals, the low-energy quasiparticles are no longer conventional Schr\"odinger electrons with a quadratic dispersion, but are instead massless Dirac fermions possessing a well-defined chirality and a nontrivial topological structure~\cite{castro2009electronic,armitage2018weyl,vafek2014dirac}. A defining feature of these systems is spin–momentum locking, which perfectly suppresses exact backscattering in the presence of smooth, long-range disorder that preserves chirality. More precisely, intravalley scattering processes acquire a geometric phase of $\pi$ between time-reversed paths, leading to destructive interference and hence WAL. In contrast, short-range disorder induces intervalley scattering between nodes of opposite chirality, restoring channels where this phase cancellation is absent and thereby driving the system towards WL-like behavior~\cite{lu2015weak,mccann2006weak}. Closely related physics arises in three-dimensional topological insulators, where surface states described by a single massless Dirac cone exhibit robust WAL due to spin–momentum locking, but in sufficiently thin films, hybridization between the top and bottom surfaces generates a finite Dirac mass, opening a gap and modifying the Berry phase, leading to a crossover from WAL to WL~\cite{lu2011competition,rosen2019absence}. 

In recent years, a remarkable zoo of gapless fermionic excitations has been discovered in condensed matter systems, which transcends the conventional spin-$1/2$ Dirac and Weyl varieties familiar from high-energy physics~\cite{bradlyn2016beyond}. In crystalline solids, certain symmetries can stabilize multifold band degeneracies (three-, four-, six-fold and so on), whose low-energy descriptions are generalized chiral Hamiltonians characterized by an effective pseudospin-$s$~\cite{bradley1972c,wieder2016double,ezawa2016pseudospin,bradlyn2016beyond,chang2018topological}. These higher-fold chiral fermions, which have no direct counterpart in relativistic vacuum field theory, exhibit qualitatively new features emerging from their enlarged internal Hilbert space. In particular, they can carry higher topological charge, display unconventional optical selection rules, and host anomalous transport responses that go beyond those of simple Weyl fermions~\cite{ahmad2025longitudinal}. Such quasiparticles have been experimentally realized in CoSi~\cite{takane2019observation}, RhSi~\cite{sanchez2019topological}, and AlPt~\cite{schroter2019chiral}, where spectroscopic probes have investigated the corresponding topological surface states. Complementary platforms have also emerged in engineered systems: for instance, pseudospin-1 excitations, often termed Maxwell fermions, have been proposed and realized in photonic and cold-atom optical lattice settings~\cite{zhu2017emergent}. 

This development raises a natural and largely open question: how does quantum localization evolve across the full pseudospin-$s$ hierarchy? While pseudospin-$1/2$ Weyl fermions and pseudospin-$1$~\cite{singh2023quantum,miao2023weak} fermions have been studied in isolation, a unified analytical framework capable of treating an arbitrary pseudospin-$s$ that extracts the relevant semiclassical transport coefficients, identifies the surviving diffusive Cooperon modes, and clarifies the role of interband and intervalley mixing has been missing. This question is compelling not only because multifold fermions continue to proliferate in topological materials, but also because the pseudospin-$s$ Hamiltonian provides a rare setting in which one can attempt a genuine all-$s$ transport theory rather than a collection of isolated case studies.

In this work, we develop such a framework for three-dimensional disordered pseudospin-$s$ fermions. Starting from a rotationally invariant low-energy Hamiltonian of the form $H\sim \mathbf{k}\cdot\mathbf{S}$, we formulate the problem in the helicity basis, and treat short-range disorder in its most general form rather than helicity preserving. We show that the disorder matrix elements acquire a universal angular structure that can be expressed in terms of Wigner rotation functions~\cite{sakurai2020modern,varshalovich1988quantum,edmonds1996angular}, enabling a unified treatment of scattering processes across all helicity sectors with multiband effects. This formulation allows us to obtain compact general expressions for the elastic lifetime, the transport lifetime, and the ladder vertex renormalization for arbitrary pseudospin and helicity bands in the presence of general disorder. In the scalar disorder limit, we find that the semiclassical (Drude) response is strongly pseudospin dependent-- increasing $s$ progressively suppresses backscattering and enhances conductivity. In contrast, remarkably, the leading interference correction retains the same magnitude as in the conventional Weyl problem and in standard diffusive metals, while its sign is determined solely by the parity of $2s$, placing half-integer pseudospins in the symplectic class and integer pseudospins in the orthogonal class. Furthermore, we go beyond the purely diagonal problem and address the first nontrivial multichannel problem, i.e., intervalley and interband scattering, specifically in the case of $s=3/2$. In this case the Cooperon becomes a matrix object in the combined band and valley space, and the standard single-mode Cooperon ansatz is no longer sufficient. By solving the resulting coupled Bethe–Salpeter equations using an iterative ansatz, we demonstrate that interband and intervalley mixing suppress the WAL channel and drive a crossover toward negative magnetoresistance, with the critical scattering strength correlated with the corresponding backscattering probability. Our work thus provides a unified analytical framework for localization and quantum interference across the full pseudospin-$s$ hierarchy.

The remainder of this paper is organized as follows. In Sec.~\ref{sec2}, we introduce the model for pseudospin-$s$ fermions and construct the helicity eigenstates. In Sec.~\ref{sec3}, we develop the disorder formalism for a general short-range matrix potential $\mathcal{M}$ and derive compact expressions for the elastic scattering rates. Sec.~\ref{sec4} is devoted to the calculation of vertex corrections, where we obtain the renormalized current operator and evaluate the Drude conductivity.
In Sec.~\ref{sec5}, we present the quantum-interference correction, highlighting the emergence of universal behavior in the scalar-disorder limit and its dependence on the parity of $2s$. In Sec.~\ref{sec6}, we go beyond the single-channel approximation and study the multichannel $s=\tfrac{3}{2}$ case, including the effects of interband and intervalley scattering. Finally, we summarize our results and discuss their implications in Sec.~\ref{sec7}. Technical details and auxiliary derivations are provided in the Appendices.
\section{Model and symmetry}\label{sec2}
%%%%%%%%%%%%%%%%%%%%%%%%%%%%%%%%%%%%%%%%%%%%%%%%%%%%%%%%%%%%%%
We consider a system hosting low-energy fermionic excitations of arbitrary pseudospin$-s$, in which the spin degrees of freedom are linearly coupled to the particle momentum. The corresponding low-energy effective Hamiltonian is given by
\begin{equation}
H_s(\mathbf{k}) = \hbar v\,\mathbf{k}\cdot\mathbf{S},
\label{eq:H}
\end{equation}
where $v$ is a characteristic velocity, and $\mathbf{S}=(S_x,S_y,S_z)$ denotes the spin-$s$ representation of the SU(2) generators that satisfy the algebra $[S_i,S_j]=i\epsilon_
{ijk} S_k$, while spin conservation implies $S^2 = s(s+1)\mathbb{I}$.  
Owing to its rotational invariance, the Hamiltonian
depends only on the relative orientation between the spin and momentum degrees of freedom.
One can re-express the Hamiltonian as 
\begin{equation}
H_s(\mathbf{k}) = \hbar v\,k\,(\hat{\mathbf{k}}\cdot\mathbf{S});
\end{equation}
the energy spectrum is determined by the
eigenvalues of the spin projection operator and takes the form
\begin{equation}
E_{s'}(\mathbf{k}) = \hbar v\, s' k, \qquad s' = -s,-s{+}1,\ldots,s-1,s.
\label{eq:Ebands}
\end{equation}
The eigenstates of the Hamiltonian are thus essentially helicity eigenstates. The spectrum consists of $2s{+}1$ linearly dispersing bands that intersect at a single
degenerate point $\mathbf{k}=0$. For integer values of $s$, a dispersionless (flat) band
appears at zero energy corresponding to $s'=0$, whereas for half-integer $s$, such a band is absent.

The helicity eigenstates, defined up to a momentum-dependent phase, are constructed as
\begin{equation}
\ket{\mathbf k;s,s'}=R(\hat{\mathbf k})\ket{s,s'}
   =\sum_{m=-s}^{s} D^{s}_{ms'}(\phi,\theta)\ket{s,m},
\label{eq:wf}
\end{equation}
where the operator $R(\hat{\mathbf k})=e^{-i\phi S_z}e^{-i\theta S_y}$, aligns the $z$ axis with the direction $\hat{\mathbf k}=(\sin\theta\cos\phi,\sin\theta\sin\phi,\cos\theta)$; \(D^{s}_{m s'}(\phi,\theta)=e^{-im\phi}d^{s}_{m s'}(\theta)\), and the reduced Wigner functions \(d^{s}_{m s'}(\theta)\) obey the following orthogonality relations~\cite{sakurai2020modern,varshalovich1988quantum}:
\begin{align*}
    \sum_{m=-s}^{s}d^{s}_{m s'}(\theta)d^{s}_{m s''}(\theta)=\delta_{s's''},\\ 
\sum_{s'=-s}^{s}d^{s}_{m s'}(\theta)d^{s}_{m' s'}(\theta)=\delta_{m m'}.
\label{eq:dorth}
\end{align*}

For the topmost band \(s'=s\),
\begin{equation}
d^{s}_{m s}(\theta)
  =\sqrt{\binom{2s}{s+m}}
     \!\left(\cos\frac{\theta}{2}\right)^{s+m}
     \!\left(\sin\frac{\theta}{2}\right)^{s-m},
\label{eq:pascal}
\end{equation}
whose coefficients are Pascal's triangle numbers. The general relation for arbitrary $s$ and $s'$ is given in Eq.~\ref{eq:A1_A}. 
Note that each eigenstate is a coherent superposition of \(2s{+}1\) spin projections whose relative weights follow Eq.~\ref{eq:wf}. With increasing \(s\), the spinor texture becomes progressively smoother, and the pseudospin winding on the Fermi surface becomes more densely packed. Therefore, near-backscattering ($|\theta-\theta'|\sim \pi-\epsilon$ where $\epsilon$ is a small number) diminishes systematically with increasing $s$, specifically, $\left|\langle \mathbf{k};s,s' \mid {\mathbf{k}';s,s''} \rangle\right|^2
\sim \left({\epsilon}/{2}\right)^{2|s'+s''|}$. This trend that is also explicitly reflected in the transport and vertex results we derive later. 

It is also fruitful to comment on the behavior of pseudospins by evaluating their fluctuations about the mean value. While the pseudospin has a well-defined projection along its mean direction, the transverse components exhibit fluctuations that grow as $\sim\sqrt{s}$.
Since the magnitude of the pseudospin itself scales as 
$s$, the relative fluctuations decrease as $\sim 1/\sqrt{s}$ As a result, in the large-$s$ limit the spinor becomes increasingly sharply localized around its mean direction, recovering the behavior of a spinless metal.

%In the large-\(s\) limit, the spinor becomes narrowly peaked and approaches a classical vector, recovering the behaviour of a spinless metal, and the increasing dimension of the internal space introduces richer angular structure in both the Berry connection and in scattering, and consequently new interference channels in quantum transport. The topology of the pseudospin texture controls whether interference enhances or suppresses backscattering, giving weak localisation (WL) or antilocalisation (WAL). The special cases of the Hamiltonian are the spinless Schrödinger particle \(s=0\), and the massless Dirac or Weyl fermion (\(s=\tfrac12\)).
%This rotation-group representation provides all overlap factors required for disorder averaging and interference calculations. It also makes clear that the universality of WL/WAL originates purely from the SU(2) representation structure rather than from model-specific details. The algebra below will be developed identically for both two and three dimensions. In two dimensions one simply sets \(\theta=\pi/2\), so the momentum is parameterized only by the azimuthal angle~\(\phi\); in three dimensions the full angular dependence enters the integrals.%

%%%%%%%%%%%%%%%%%%%%%%%%%%%%%%%%%%%%%%%%%%%%%%%%%%%%%%%%%%%%%%
\section{Elastic disorder and scattering rates}\label{sec3}
We introduce a disorder described by
\begin{align}
U(\mathbf r)=\sum_i u_0\,\delta(\mathbf r-\mathbf R_i) \mathcal{M}, \nonumber \\
\qquad
\overline{U(\mathbf r)U(\mathbf r')}=
n_i u_0^2\,\delta(\mathbf r-\mathbf r')\mathcal{M}^2,
\label{eq:disorder}
\end{align}
where $n_i$ is the impurity density, $u_0$ is the scattering strength, and $\mathcal{M}$ is a general square matrix of dimension $(2s+1)$ and couples the different bands due to its off-diagonal structure.
We expand $\mathcal{M}$ in terms of irreducible tensor operators~\cite{edmonds1996angular,varshalovich1988quantum,sakurai2020modern}:
\begin{equation}
\mathcal M=\sum_{L=0}^{2s}\sum_{Q=-L}^{L} M_{LQ}\,T_Q^{(L)},
\label{eq:M_tensor_expansion}
\end{equation}
where the tensor operators satisfy
\begin{equation}
\langle s,s''|T^{(L)}_{Q}|s,s'\rangle
=
\sqrt{2L+1}\,
(-1)^{s+s''-L}
\begin{pmatrix}
s & L & s\\
s' & Q & -s''
\end{pmatrix},
\label{eq:T_matrix_element1}
\end{equation}
$(\cdots)$ being the Wigner's 3-j symbol, which is related to the Clebsch-Gordon coefficients via
\begin{align}
C^{\,s\, m'}_{\,s\, m,\; L\, Q}
=
(-1)^{s - L + m'}\,
\sqrt{2s+1}\,
\begin{pmatrix}
s & L & s \\
m & Q & -m'
\end{pmatrix}.
\end{align}
The disorder-averaged Green’s functions for helicity $s'$ are given by:
\begin{equation}
G^{R/A}_{s'}(\mathbf k,E)
 =\frac{1}{E-\hbar v s'k\pm i\hbar/(2\tau_{s'})},
\label{eq:GRGA}
\end{equation}
where the scattering rate $1/\tau_{s'}$ is, includes both intra- and interband processes. The self-energy in the first Born approximation is
\begin{equation}
\Sigma^R_{s'}(\mathbf{k})
=n_i u_0^2
\sum_{s''}\sum_{\mathbf k'}
\left|\langle \mathbf{k}', s'' | \mathcal{M} |\mathbf{k}, s' \rangle\right|^2
G^0_{s''}(\mathbf k' ),
\label{eq:born}
\end{equation}
where $G^0_{s''}(\mathbf{k}')$ is the bare Green's function, and the corresponding scattering rate is given by
\begin{equation}
\frac{1}{\tau_{s'}}(\mathbf{k},\omega)
= -\frac{2}{\hbar} \mathrm{Im},\Sigma^R_{s'}(\mathbf{k},\omega).
\end{equation}
The overlap can be written in terms of the Wigner $D$-matrix as~\cite{sakurai2020modern,varshalovich1988quantum,edmonds1996angular}
\begin{equation}
\langle \mathbf{k}', s''| \mathcal{M} |\mathbf{k}, s' \rangle
=
\sum_{m_1,m_2}
D^{s *}_{m_1 s''}(\hat{\mathbf k}')
\, {M}_{m_1 m_2} \,
D^{s}_{m_2 s'}(\hat{\mathbf k})
\label{eq:general_overlap_main}
\end{equation}
where $M_{m' m} = \langle s, m' |\mathcal{M} | s, m \rangle$. Employing the properties of Wigner-D functions, and the Tensor operators, we evaluate the following closed-form expression for the scattering time (see Appendix~\ref{AppA} for derivation): 
\begin{equation}
\frac{1}{\tau_{s'}}
=
\frac{2\pi}{\hbar} n_i u_0^2
\frac{1}{(2s+1)^2}
\sum_{s''} N_F^{(s'')}
\sum_Q |\mathcal A_{s'}(Q)|^2,
\end{equation}
where 
\begin{equation}
\mathcal A_{s'}(Q)
=
\sum_L \sqrt{2L+1}\,M_{LQ}\,
C^{\,s,\;s'+Q}_{\,s,\;s',\;L,\;Q},
\label{eq:ASQ1}
\end{equation}
and $N_F^{(s')}$ is the density of states in the helicity band, given by
\begin{align}
N_F^{(s')}= \frac{E_F^2}{2\pi^2(\hbar v)^3|s'|^3}.
\end{align}
Interestingly, after angular averaging, the scattering process no longer retains information about the specific final helicity state, and the scattering rate is governed only by the phase space available through the density of states. The helicity dependence is instead carried by the symmetry-resolved angular momentum transfer channels $Q$ via the Clebsch-Gordon selection rules. 

For scalar disorder ($\mathcal{M}=\mathcal{I}$), $L=Q=0$ are the only nonzero components, and therefore $\mathcal A_{s'}(Q)
=
\delta_{Q,0}\,M_{00}\,
C^{\,s,\,s'}_{\,s,\,s',\,0,\,0}$, and 
\begin{equation}
\frac{1}{\tau_{s'}}
=
\frac{2\pi}{\hbar} n_i u_0^2
\frac{1}{(2s+1)}
\sum_{s''} N_F^{(s'')}.
\end{equation}
\begin{comment}
Here we consider a scalar disorder i.e. $M = \mathds{1}$, and  the overlap satisfies
$
\langle \mathbf{k'}, s,s''\mid\mathbf{k},s,s'\rangle 
= \langle \mathbf{k'}, s,s' \mid \mathbf{k},s,s''\rangle^{\,*},
$
we obtain
\begin{equation}
\big|\braket{\mathbf k',s,s''}{\mathbf k,s,s'}\big|^2 =
\bigl| d^{\,s}_{s''s'}(\beta) \bigr|^2.
\end{equation}
The overlap depends only on the relative rotation angle $\beta$, which is the angle between the two momentum directions:
\begin{equation}
\cos\beta=\hat{{\mathbf k}}\cdot\hat{{\mathbf k}'}.
\end{equation}
Thus, for a general spin-$s$ system, the spinor overlap is fully determined by the Wigner-$D$ matrix elements $d^{\,s}_{s''s'}(\beta)$.

By evaluating the self-energy at the Fermi level $E_F$ within the Born approximation Eq.~\eqref{eq:born}, and carrying out the radial integration, we obtain
\begin{equation}
\frac{1}{\tau_{s''\to s'}}
 =\frac{2\pi n_i u_0^2 N_F^{(s')}}{4\pi\hbar}
 \int d\Omega\,
  \big|d^{s}_{s''s'}(\theta)\big|^2,
\label{eq:tau_s_1}
\end{equation}
where $N_F^{(s')}$ is the density of states in the helicity band, given by
\begin{align}
N_F^{(s')}= \frac{E_F^2}{2\pi^2(\hbar v)^3|s'|^3}.
\end{align}
Using the identity 
\(\int d\Omega\,|d^{s}_{s''s'}(\theta)|^2 = {4\pi}/{2s+1}\)~\cite{varshalovich1988quantum},
we obtain
\begin{equation}
\tau_{s''\to s'}=\frac{(2s+1) \hbar}{2\pi n_i u_0^2 N_F^{(s')}}.
\label{eq:tau_simple_1}
\end{equation}
Thus, the scattering rate is directly proportional to the density
of states, which is in agreement with the Eq.~\ref{eq:final_scatt_time1}
\end{comment}
The transport scattering time involves an additional angular weighting factor $(1-\cos\theta)$ that suppresses forward scattering~\cite{bruus2004many}. For scalar disorder, the valley resolved time is
\begin{equation}
\frac{1}{\tau^{\mathrm{tr}}_{s''\to s'}}
 =2\pi n_i u_0^2 N_F^{(s')}
  \int d\Omega\,
  \big|d^{s}_{s''s'}(\theta)\big|^2(1-\cos\theta).
\label{eq:tautr}
\end{equation}
The angular integrals can be evaluated using the algebra of Wigner
functions~\cite{varshalovich1988quantum}, yielding
\begin{equation}
\frac{1}{\tau^{\mathrm{tr}}_{s'}}
 =\frac{1}{\tau_{s'}}
   \Big(1- \frac{s'^2}{s(s+1)}\Big).
\label{eq:tau_result}
\end{equation}
This reproduces familiar limits:
$\tau^{\mathrm{tr}}=(3/2)\tau$ for $s=\tfrac12$, and
$\tau^{\mathrm{tr}}=2\tau$ ($\tau^{\mathrm{tr}}=\tau$) for the $s=1$, respectively.  
As $s$ increases, the prefactor approaches unity, indicating that forward
scattering dominates for higher pseudospin and backscattering
is geometrically suppressed.

%The same result follows from the Boltzmann equation. Writing
%\(f_{s'}(\hat{\mathbf k})
%=f_0-\partial_\epsilon f_0\,A_{s'}\,\hat{\mathbf %k}\!\cdot\!\mathbf E\)
%and inserting the collision kernel
%\(W(\theta)\propto |d^{s}_{s's'}(\theta)|^2\),
%one obtains the angular average
%\(\langle\cos\theta\rangle
% = s'^2/[s(s+1)]\),
%yielding Eq.~(\ref{eq:tau_result}) directly.
%This semiclassical picture highlights that the pseudospin texture acts
%as a geometric filter suppressing backscattering, consistent with the
%quantum derivation.

\begin{figure}
\centering
\includegraphics[width=\linewidth]{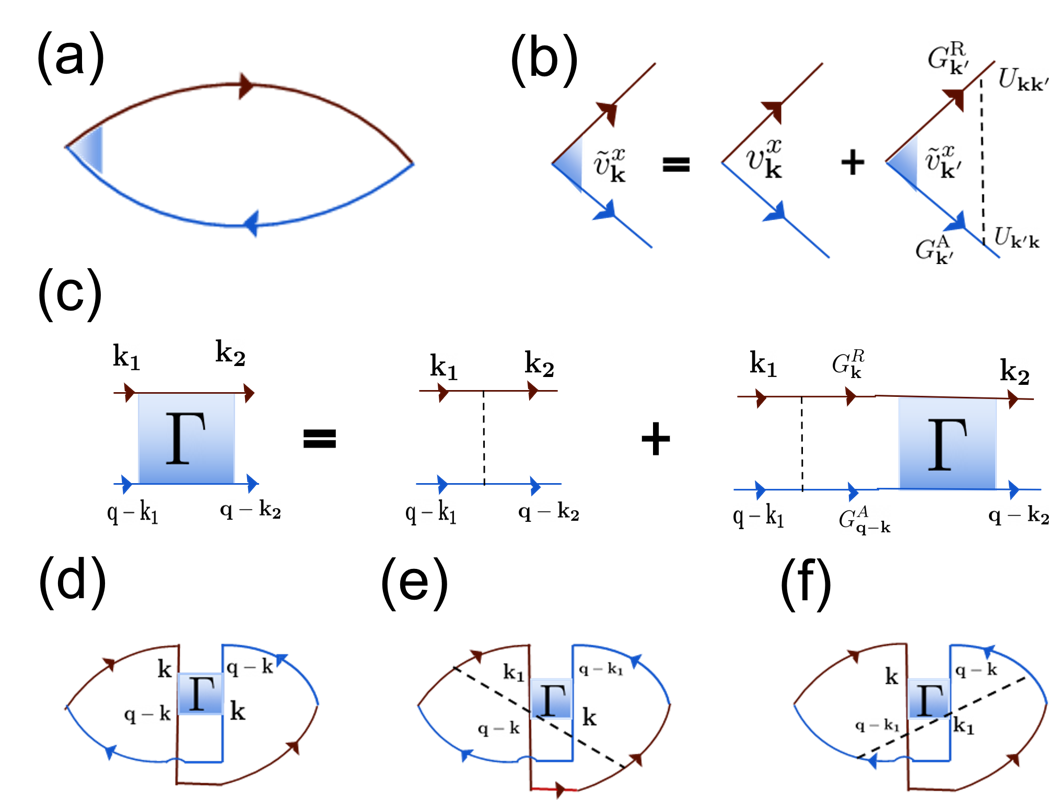}
\caption{Feynman diagrams for the conductivity of a 3D Weyl semimetal: (a) Drude conductivity, (b) vertex correction, (c) Bethe–Salpeter equation for the Cooperon propagator, and (d) bare Hikami box relating the conductivity correction to the Cooperon propagator, with (e) and (f) the corresponding dressed Hikami boxes. Solid lines represent the disorder-averaged Green’s functions $G^{R/A}_{\mathbf{k}}$, while dashed lines denote disorder scattering.}
\label{fig:Feyman_diagram}
\end{figure}
%%%%%%%%%%%%%%%%%%%%%%%%%%%%%%%%%%%%%%%%%%%%%%%%%%%%%%%%%%%%%%
\section{Vertex corrections and Ward identity}\label{sec4}
%%%%%%%%%%%%%%%%%%%%%%%%%%%%%%%%%%%%%%%%%%%%%%%%%%%%%%%%%%%%%%
The impurity potential not only broadens the quasiparticle spectrum but also renormalizes the current operator through multiple scattering. 
This renormalization is required by charge conservation and is formally encoded in the Ward identity, which relates the self-energy and the current vertex~\cite{velicky1969theory,rammer1986quantum}. 
Diagrammatically, this corresponds to summing the ladder series of impurity lines in the particle–hole channel, leading to a Bethe–Salpeter equation for the dressed current vertex as shown in Fig.~\ref{fig:Feyman_diagram}(b).

The Ward identity for a disordered system may be written as
\begin{align}
&\Omega\,\Gamma^{0}_{s'}(\mathbf{k},\omega;\mathbf{q},\Omega)
-\mathbf{q}\cdot\bm{\Gamma}_{s'}(\mathbf{k},\omega;\mathbf{q},\Omega)
=\nonumber\\&
G^{-1}_{s'}(\mathbf{k}+\mathbf{q},\omega+\Omega)
-
G^{-1}_{s'}(\mathbf{k},\omega),
\label{eq:Ward_general}
\end{align}
which expresses charge conservation by relating the density and current components of the full vertex to the single-particle Green's function. In the static, long-wavelength limit, this reduces to
\begin{equation}
\bm{\Gamma}_{s'}(\mathbf{k})
\propto
\frac{\partial G^{-1}_{s'}(\mathbf{k},\omega)}{\partial \mathbf{k}},
\label{eq:Ward_static}
\end{equation}
ensuring that the vertex corrections consistently incorporate the same scattering processes that determine the quasiparticle lifetime.

The disorder kernel that enters the self-energy also enters the Bethe-Salpeter equation for the vertex. Defining
\begin{equation}
W_{s' s''}(\hat{\mathbf k},\hat{\mathbf k}')
\equiv
\big|
\langle \mathbf{k},s'|\mathcal{M}|\mathbf{k'},s'' \rangle
\big|^2,
\label{eq:Wkernel_general_1}
\end{equation}
the ladder equation for the $\beta$-component of the vertex reads
\begin{align}
\Gamma_{s',\beta}(\hat{\mathbf k})
&=
v_{s',\beta}(\hat{\mathbf k}) \nonumber\\
&\quad
+
n_i u_0^2
\sum_{s''}
\mathcal J_{s''}
\int\frac{d\Omega'}{4\pi}\,
W_{s's''}(\hat{\mathbf k},\hat{\mathbf k}')
\,
\Gamma_{s'',\beta}(\hat{\mathbf k}'),
\label{eq:BS_general_1}
\end{align}
with
\begin{equation}
\mathcal J_{s'}
=
\int \frac{k^{d-1}dk}{(2\pi)^d}\,
G^R_{s'}(\mathbf k,E_F)G^A_{s'}(\mathbf k,E_F)
=
\frac{2\pi}{\hbar}N_F^{(s')}\tau_s'.
\label{eq:Jlambda_general_1}
\end{equation}
Within the same Born-plus-ladder approximation, the self-energy and vertex are therefore built from the same impurity correlator, so the construction is conserving and compatible with the Ward identity~\cite{rammer1991quantum, mahan2013many}.
For a generic matrix disorder $\mathcal{M}$, the kernel $W_{s's''}(\hat{\mathbf k},\hat{\mathbf k}')$ is not purely a function of the relative angle, and the dressed vertex is not characterized by a single scalar renormalization factor. The natural generalization is therefore to expand the vertex in spherical harmonics:
\begin{equation}
\Gamma_{s',\beta}(\hat{\mathbf k})
=
\sum_{\ell m}\eta^{(\beta)}_{s';\ell m}\,
Y_{\ell m}(\hat{\mathbf k}),
\label{eq:Gamma_harmonic_expansion_1}
\end{equation}
while the bare velocity contains only the $\ell=1$ sector. Equation~\eqref{eq:BS_general_1} then becomes a linear system in the combined band-harmonic space,
\begin{equation}
\eta^{(\beta)}_{s';\ell m}
=
\eta^{(\beta,0)}_{s';\ell m}
+
\sum_{s''\ell'm'}
\mathcal K^{\ell m,\ell' m'}_{s's''}\,
\eta^{(\beta)}_{s'';\ell'm'},
\label{eq:eta_matrix_general_1}
\end{equation}
with 
\begin{align}
\mathcal K^{\ell m,\ell' m'}_{s's''}
&=
n_i u_0^2\,\mathcal J_{s''}
\int\!\frac{d\Omega}{4\pi}\frac{d\Omega'}{4\pi}\,
Y_{\ell m}^{*}(\hat{\mathbf k})\,
W_{s's''}^{\hat{\mathbf k},\hat{\mathbf k}'}
\,Y_{\ell' m'}(\hat{\mathbf k}'),
\nonumber\\
\mathrm{and\quad}\eta^{(\beta,0)}_{s';\ell m}
&=
\int \frac{d\Omega}{4\pi}\, v_{s',\beta}(\hat{\mathbf k})
Y^*_{\ell m}(\hat{\mathbf k}).
\label{eq:Kmatrix_general_!}
\end{align}
Thus, the general vertex renormalization is a matrix
\begin{equation}
\eta^{(\beta)}
=
\left(\mathds{1}-\hat{\mathcal K}\right)^{-1}
\eta^{(\beta,0)},
\label{eq:eta_general_solution_1}
\end{equation}
rather than a single number.
Since our system is isotropic, the dressed vertex remains parallel to the bare velocity and only the $\ell=1$ component survives. Then one may write
\begin{equation}
\Gamma_{s',\beta}(\hat{\mathbf k})
=
\eta_{s'}\,v_{s',\beta}(\hat{\mathbf k}),
\label{eq:eta_scalar_ansatz1}
\end{equation}
and Eq.~\eqref{eq:BS_general_1} reduces to
\begin{equation}
\eta_{s'}
=
1+
n_i u_0^2 \mathcal J_{s'}
\
 \int \frac{d\Omega'}{4\pi}\,
W_{s's'}(\hat{\mathbf k}',\hat{\mathbf k})\,\cos\theta'
\,\eta_{s'}.
\label{eq:eta_isotropic_start1}
\end{equation}
and the velocity correction factor can be written compactly as
\begin{equation}
\eta_{s'}
=
\frac{1}{1-\mu_{s'}},
\qquad
\mu_{s'}
\equiv
\frac{
\displaystyle \int \frac{d\Omega'}{4\pi}\,
W_{s's'}(\hat{\mathbf k}',\hat{\mathbf k})\,\cos\theta'
}{
\displaystyle \int \frac{d\Omega'}{4\pi}\,
W_{s's'}(\hat{\mathbf k}',\hat{\mathbf k})
}.
\label{eq:eta_mu_general_main}
\end{equation}
We find the following closed-form of a $\mu_{s'}$ (see Appendix.~\ref{AppA_vel_correct} for complete derivation)
\begin{equation}
\mu_{s'}
=
\frac{s'}{s(s+1)}
\,
\frac{
\sum_Q (s'+Q)\,|\mathcal A_{s'}(Q)|^2
}{
\sum_Q |\mathcal  A_{s'}(Q)|^2
}
\end{equation}
where $\mathcal A_{s'}(Q)$ is define in Eq.~\ref{eq:ASQ1}. 

For scalar disorder ($\mathcal{M}=\mathcal{I}$), we obtain
\begin{equation}
\mu_{s'}=\frac{s'^2}{s(s+1)},
\qquad
\eta_{s'}=
\frac{1}{1-\dfrac{s'^2}{s(s+1)}}.
\label{eq:eta_scalar_general_1}
\end{equation}
For the outermost helicity band $s'=s$,
\begin{equation}
\eta_s=s+1,
\qquad
\tau_s^{\mathrm{tr}}=(s+1)\tau_s.
\label{eq:eta_scalar_outerband_1}
\end{equation}
Focusing on the intraband cases, for $s=\tfrac12$ this gives $\eta=3/2$ consistent with earlier results~\cite{lu2015weak}. For $s=1$, $\eta=2$; for $s=\tfrac32$, $\eta=5/2$. The increase of $\eta$ with $s$ reflects the suppression of backscattering and the dominance of forward scattering for higher pseudospins.
Substituting the dressed vertex into the Kubo formula,
\begin{equation}
\sigma^{(s')}_{\alpha\beta}
 = \frac{e^2}{2\pi}
   \int\!\frac{d^d k}{(2\pi)^d}\,
   v_\alpha(\hat{\mathbf k})\,
   G^R_{s'}\,\Gamma_{s',\beta}(\hat{\mathbf k})\,G^A_{s'},
\end{equation}
and for a system characterized by an arbitrary pseudospin $s$ with a specific helicity state $s'$,
\begin{align}
\sigma_{s'}^{s} &=
\frac{e^2}{d}\,v^2\,N_F^{(s')}\,
\tau_{ s'}\,\eta_{ s'} \nonumber\\
&= e^2 N_F^{(s')}\,\frac{v^2\tau^{\mathrm{tr}}_{s'}}{d}
\equiv e^2 N_F^{(s')} D_{s'},\nonumber\\
&=\frac{e^2\hbar v}{6\pi n_i u_0^2} (2s+1)\eta_{s'}
 \label{eq:druderesult}
\end{align}
where the transport time is
\begin{equation}
\tau^{\mathrm{tr}}_{s'}=\eta_{s'}\,\tau_{s'},
\label{eq:tau_tr_eta_1}
\end{equation}
This reproduces the Boltzmann result with the diffusion constant
$D_{s'}=v^2\tau^{\mathrm{tr}}_{s'}/d$.
This consistency with the Ward identity reflects the conserving nature of the ladder approximation, ensuring that the same scattering processes that determine the quasiparticle lifetime also control the current response.

\section{Diffusion modes and quantum interference}\label{sec5}
%%%%%%%%%%%%%%%%%%%%%%%%%%%%%%%%%%%%%%%%%%%%%%%%%%%%%%%%%%%%%%

Quantum corrections to the Drude conductivity arise from coherent
backscattering of time-reversed trajectories.  The corresponding
diagrams Fig.~\ref{fig:Feyman_diagram}(c) involve the maximally crossed impurity lines and are described
by the Cooperon propagator, which captures the diffusive dynamics of two
particles moving along time-reversed paths. Unless otherwise specified, from now on, we will restrict our discussion to the presence of a scalar disorder and consider only intraband scattering. In the next section, we discuss the effect of interband as well as intervalley scattering. 
%For pseudospin–$s$ fermions, the internal spin structure of the Cooperon is determined by the SU(2) rotation matrices $d^{s}_{m m'}(\theta)$, leading to alternating weak-localization (WL) and weak-antilocalization (WAL) behavior as the parity of $2s$ changes.

The bare Cooperon, which is the overlap of the trajectory with its the time counterpart without scattering, defined as \( \Gamma^{0}_{{\kn_1}{\kn_2}} = \langle U_{\kn_{1},\kn_{2}}U_{-\kn_{1},-\kn_{2}}\rangle \), has an angular structure (see Appendix~\ref{appen.sect-2}):
\begin{align}
\Gamma^{0}_{{\kn_1}{\kn_2}}= {n_0 u_0^2}\sum_{p}\sum_{m,n} C^0_\mathrm{mnp}e^{i(m\theta_1 + n\theta_2)}   e^{p(i\phi_1 + i\phi_2)}, 
\end{align}
where $p \in ({0,1,....,4s+1})$. There are several channels for backscattering. The time evolution of the Cooperon obeys a Bethe–Salpeter equation analogous to that of the vertex function, but in the particle–particle channel as seen in Fig.~\ref{fig:Feyman_diagram}(c):
\begin{equation}
    \Gamma_{{\kn_1}{\kn_2}} =  \Gamma^{0}_{{\kn_1}{\kn_2}} + \sum_{{\kn}} \Gamma^{0}_{{\kn_1}{\kn}}G^R_{{\kn}}G^A_{\mathbf{q}-{\kn}} \Gamma_{{\kn}{\kn_2}}\label{eq:Cooperon_BS}
\end{equation}
{where \( \Gamma_{{\kn_1}{\kn_2}} = n_0 u_0^2\sum\limits_{m,n,p} C_\mathrm{mnp}e^{i(m\theta_1 + n\theta_2)}   e^{p(i\phi_1 + i\phi_2)}\), is the dressed Cooperon. The coefficients $C_\mathrm{mnp}$ are functions of $q$, and diverge for weak localization channels i.e $C_{mnp} \to \infty$ as $ q \to 0$}.

After solving the Bethe-Salpeter equation (for details see the Appendix~\ref {appen.sect-2}), we find that $C_\mathrm{mnp}$ for each channel takes the form, 
\begin{equation}
    C_\mathrm{mnp} = \frac{1}{q^2 + Q^2_{mnp}}
\end{equation}
where \(Q_{mnp}^2\) is non-zero except  when \(p= 2s, m=n=0\). Therefore, precisely \textit{one} \(C_\mathrm{m=n=0,p = 2s}\) channel is divergent in the diffusive limit \({q}\rightarrow0\), and the full Cooperon acquires the form for spin-\(s\) with helicity \(s'\),  
\begin{equation}
\Gamma(\mathbf q) =
 \frac{e^{p(i\phi_1 - i\phi_2)}}{D_{s'} q^2 }
 ,
\label{eq:ACooperon_diff1}
\end{equation}
where $D_{s'}$ is diffusion constant.
In the limit $\mathbf{q}\to 0$, the Cooperon propagator becomes divergent and scales as $q^{-2}$. As a result, the quantum-interference correction to the Boltzmann conductivity in three dimensions takes the form
\begin{equation}
\sigma^{\mathrm{qi}}
= -\frac{e^2}{h}\frac{1}{\pi^2}
\left(\frac{1}{\ell}-\frac{1}{\ell_\phi}\right)
e^{i\pi\alpha},
\end{equation}
where $e^2/h$ is the conductance quantum, $\alpha = 2s$ with $s$ denoting the pseudospin, $\ell$ is the elastic mean free path, and $\ell_\phi$ is the phase-coherence length.

The above expression shows that, in three dimensions, the magnitude of the quantum-interference correction to the conductivity for higher-pseudospin Weyl fermions is identical to that of conventional Weyl fermions \cite{lu2015weak} and to that of a standard 3D electron gas \cite{lee1985disordered}. However, the nature of localization depends sensitively on the pseudospin: systems with half-integer pseudospin exhibit weak antilocalization, whereas those with integer pseudospin display weak localization. We therefore conclude that while the \textit{magnitude} of the quantum-interference correction is \textit{independent} of the pseudospin, the \textit{localization} behavior is determined by the \textit{pseudospin} structure.
\subsection*{Magnetic field dependence}
The magnetoconductivity for a single valley and a single band is defined as $\delta \sigma^{qi}(B) \equiv \sigma^{qi}(B) - \sigma^{qi}(0),$ where
 
\begin{align}
    \sigma^{qi}(B) &= \frac{2e^2}{h} \int_0^{1/\ell} \frac{dx}{(2\pi)^2} \bigg[ 
    \psi \left( \frac{\ell_B^2}{\ell^2} \right. \nonumber  \left. + \ell_B^2 x^2 + \frac{1}{2} \right)
    \\
   & - \psi \left( \frac{\ell_B^2}{\ell_\phi^2} \right. \left. + \ell_B^2  x^2 + \frac{1}{2} \right)
    \bigg],
    \label{Eq_sigma_B}
\end{align}
where $\ell_B = \sqrt{\hbar/4eB}$ is the magnetic length, $\psi$ is the digamma function. At low temperatures, the magnetoconductivity for a pseudospin-$s$ is proportional to the square root of the magnetic field, similar to that observed in  Weyl semimetal, spinless Schrodinger particle, and pseudospin-$1$ chiral fermions~\cite{lu2015weak,lee1985disordered, singh2023quantum}.
This is the quantum correction in conductivity due to the intravalley Cooperons. 
Consequently, integer pseudospin systems exhibit a negative
magnetoconductivity cusp, while half-integer systems show a positive
one.  The magnitude is determined by the diffusion constant $D_{s'}$
and the dephasing time, both of which are material dependent, but the
sign-dependence on $s$ is universal.

\subsection*{Summary of universality}
Time-reversal symmetry plays a central role in determining the structure of
quantum-interference corrections to transport. For generalized spin-$s$ fermions
with pseudospin $s$, time reversal is implemented as \(\mathcal{T} = e^{-i\pi S_y}\,\kappa ,\) where $\kappa$ denotes complex conjugation. This operator reverses the
pseudospin according to $\mathcal{T}\,\mathbf{S}\,\mathcal{T}^{-1} = -\mathbf{S}$,
and satisfies $ \mathcal{T}^2 = (-1)^{2s}$.
The sign of the quantum-interference correction is controlled by the factor
\begin{equation}
\zeta_s = \mathrm{sgn}\!\left(\mathcal{T}^2\right),
\end{equation}
which distinguishes constructive from destructive interference between
time-reversed scattering paths in the Cooperon channel. Explicitly,
\begin{equation}
\zeta_s =
\begin{cases}
-1, & s \ \text{is integer (weak localization)}, \\[4pt]
+1, & s \ \text{is half-integer (weak antilocalization)}.
\end{cases}
\label{eq:zeta_s}
\end{equation}

For integer pseudospin $s$, one has $\mathcal{T}^2=+1$, and the system belongs
to the orthogonal symmetry class. In this case, time-reversed paths interfere
constructively, leading to an enhanced return probability and hence weak
localization correction. In contrast, for half-integer pseudospin $s$, time-reversal
symmetry satisfies $\mathcal{T}^2=-1$, placing the system in the symplectic
class. The resulting $\pi$ Berry phase accumulated along closed
time-reversed trajectories leads to destructive interference, suppressing
backscattering and giving rise to a weak antilocalization correction.
Importantly, the classification depends only on the parity of $2s$ and is
independent of spatial dimensionality, disorder strength, or microscopic
details of the band structure. As a result, the symmetry class and the
corresponding localization behavior are universal features dictated solely by
the algebraic properties of the time-reversal operator. Consequently, the
sequence of localization behavior alternates with increasing pseudospin,
exhibiting WL for $s=1,2,\ldots$ and WAL for $s=\tfrac12,\tfrac32,\ldots$. Furthermore, \textit{the magnitude of the correction is independent of the pseudospin $s$ or the band index $s'$}. 

\begin{figure}
    \centering
    \includegraphics[width=\linewidth]{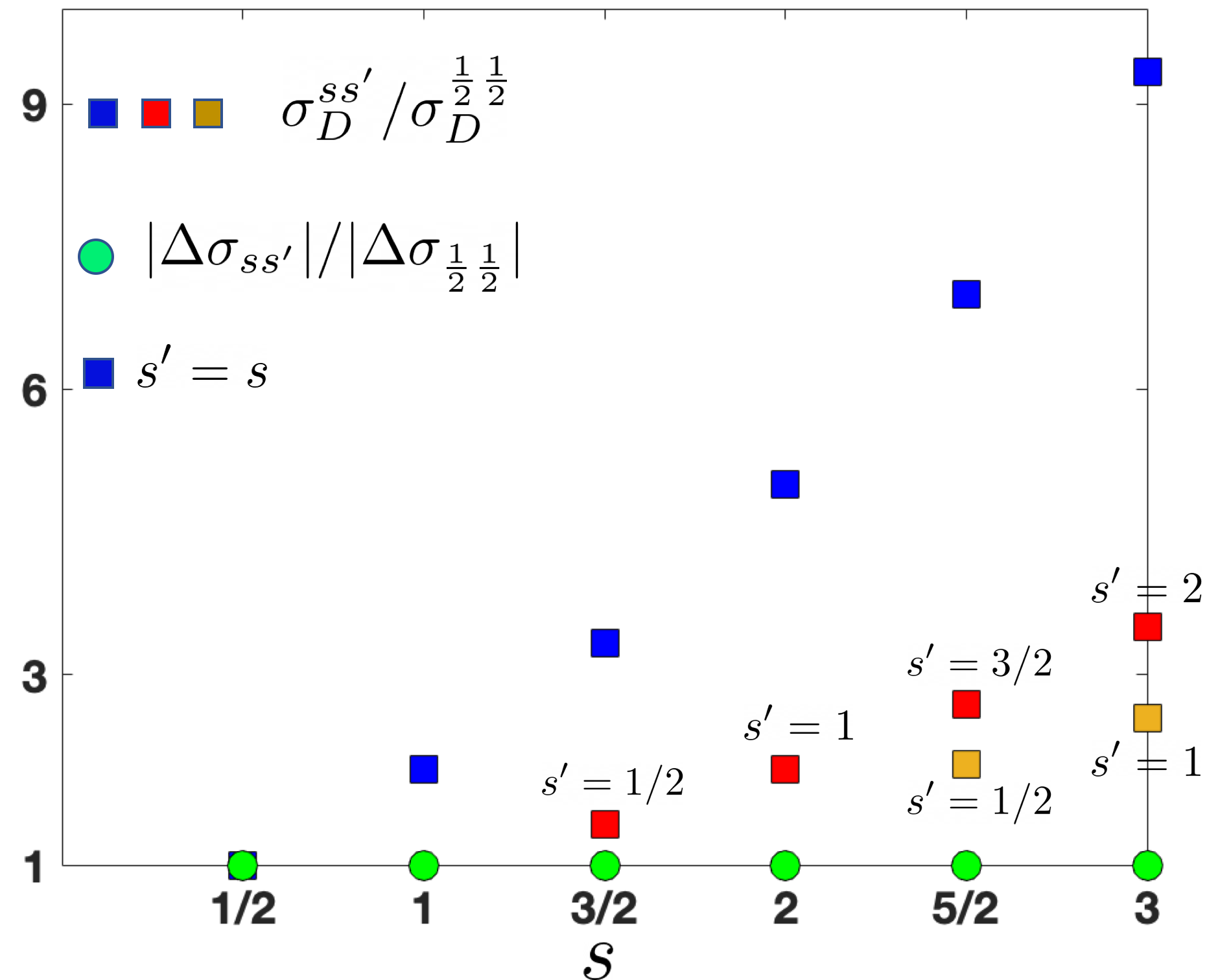}
    \caption{The relative increase of localization-induced magnetoconductivity (green dots) and the Drude conductivity (blue, red, and yellow boxes) for pseudospin-$s$ fermions.}
    \label{ratioplot}
\end{figure}

Eq.~(\ref{eq:druderesult}) demonstrates that the Kubo conductivity is explicitly sensitive to both the pseudospin magnitude and the band index, unlike the magnitude of the localization correction, which is insensitive to both these details. Specifically, $s=s'$, yields a scaling relation of $\sigma_{D}^{ss'=s} \propto (2s+1)(s+1)$.
Our results indicate that as the pseudospin increases, the Drude conductivity is significantly enhanced relative to conventional $s=1/2$ Weyl fermions.  This suggests a robust, universal nature of quantum transport corrections in higher-pseudospin systems. We also illustrate this in Fig~\ref{ratioplot}. 

An experimental test of the universality may be realized in multifold chiral semimetals such as CoSi, RhSi, and related compounds that host multiple symmetry-protected multifold fermions. By engineering chemical-potential-tuned series (via substitution, gating, or stoichiometric control), such that transport is dominated by different Fermi-surface pockets with distinct symmetry and angular-momentum structure, as the dominant transport channel is tuned across different pseudospin sectors, the Drude conductivity should vary significantly, but in contrast, the magnitude of the localization correction, extracted from the low-field magnetoconductivity cusps, should remain robust (up to the symmetry-controlled sign and dephasing scale). This separation would manifest as a strongly varying baseline conductivity accompanied by an approximately invariant low-field interference cusp. 
\begin{table*}[t]
    \centering
    \small
    \renewcommand{\arraystretch}{1.3} % Adjust row height for better readability
    \setlength{\tabcolsep}{6pt} % Adjust column spacing
    
    \begin{tabular}{l c c c c}
        \toprule
       Pseudospin($s$) & Band($m$) & Scattering time($\tau$)  & Velocity Correction($\eta$)   & Conductivity Correction  \\
       \midrule
        \multirow{2}{*}{1/2}  & $1/2$  & 2 & 3/2 & \multirow{2}{*}{WAL}  \\
            & $-1/2$ & 2 & 3/2 \\
        \midrule
        \multirow{3}{*}{1}  & $1$  & 3 & 2 & \multirow{3}{*}{WL} \\
            & $0$ & -- &  -- & \\
            & $-1$ & 3 &  2  &\\
        \midrule
        \multirow{4}{*}{3/2}  & $3/2$  & 4 & 5/2 & \multirow{4}{*}{WAL}  \\
            & $1/2$ & 4 &  15/14 & \\
            & $-1/2$ & 4  & 15/14 &\\
            & $-3/2$ & 4  & 5/2 &\\
        \bottomrule
    \end{tabular}
    \caption{\label{tab:fermions}Quantum interference contributions to conductivity for systems with pseudospin $s=1/2,\,1,$ and $3/2$. 
The table lists the band index $m$, elastic scattering time $\tau$, velocity renormalization factor $\eta$, and the resulting transport behavior, indicating weak localization or weak antilocalization in each case. The scattering time in the unit of ${ \hbar}/{2\pi n_i u_0^2 N_F^{(s')}}$.}
\end{table*}

\section{Study of four-fold degenrate pseudospin-$\tfrac32$ fermions}\label{sec6}
For pseudospin $s=\tfrac{3}{2}$, the low-energy Hamiltonian 
supports four bands labeled by $s'=\pm\tfrac{3}{2}, \pm\tfrac{1}{2}$, all degenerate at $\mathbf{k}=0$. 
Using Eq.~\eqref{eq:tau_result}, the ratio of transport to elastic scattering times is obtained as
\begin{equation}
\frac{\tau^{\mathrm{tr}}}{\tau} =
\begin{cases}
\tfrac{2}{5}, & s'=\tfrac{3}{2}, \\
\tfrac{14}{15}, & s'=\tfrac{1}{2},
\end{cases}
\qquad
\eta_{s'} =
\begin{cases}
\tfrac{5}{2}, & s'=\tfrac{3}{2}, \\
\tfrac{15}{14}, & s'=\tfrac{1}{2}.
\end{cases}
\end{equation}
The corresponding diffusion constant is given by $
D_{s'} = \frac{v^2 \tau_{s'} \eta_{s'}}{d}$, 
and the Drude conductivity then follows as $
\sigma_0 = e^2 N_F^{s'} D_{s'}$.
Since the time-reversal operator satisfies $\mathcal{T}^2 = -1$, the system belongs to the symplectic universality class and exhibits weak antilocalization (WAL). 
For a single valley of Weyl fermions, the quantum interference correction to the conductivity takes the form
\begin{equation}
\sigma^{\mathrm{qi}} = \frac{e^2}{h}\,\frac{1}{\pi^2}
\left(\frac{1}{\ell} - \frac{1}{\ell_\phi}\right),
\end{equation}
\textit{for both bands:} $s'=\tfrac{1}{2}, \tfrac{3}{2}$. 
In the presence of a perpendicular magnetic field, the magnetoconductivity is described by Eq.~\ref{Eq_sigma_B}. 

%%%%%%%%%%%%%%%%%%%%%%%%%%%%%%%%%%%%%%%%%%%%%%%%%%%%%%%%%%%%%%
\subsection{Effect of interband and intervalley scattering}
%%%%%%%%%%%%%%%%%%%%%%%%%%%%%%%%%%%%%%%%%%%%%%%%%%%%%%%%%%%%%%

Until now, we have considered scattering channels that are diagonal, i.e., the band and/or the valley indices do not change. This is not true in general, as disorder can, in principle, couple various bands and valley indices, which changes the helicity and/or chirality states of the pseudospin. For example, the Weyl fermion with pseudospin-$s$, with multiple chirality flavors, may exist at different valleys as expected by the Nielsen-Ninomiya theorem~\cite{nielsen1981no}, whereas the Weyl fermion with pseudospin-$3/2$ has multiple bands crossing the Fermi level at an arbitrary value of chemical potential. This immediately complicates the structure of quantum interference. In the presence of intervalley and interband scattering, the Cooperon is no longer a single scalar object associated with a fixed channel, but instead becomes a matrix in the combined band/valley space. Different scattering processes mix these channels, and the interference correction is governed by the coupled dynamics of all such modes. Consequently, the problem acquires an internal structure analogous to a multi-component diffusion process, where each component corresponds to a distinct sector.

In such a situation, the standard ansatz employed for diagonal scattering, where one assumes a single diffusive pole structure for the Cooperon, ceases to be valid~\cite{akkermans2007mesoscopic}. The key obstruction is that impurity scattering now generates off-diagonal correlations between different channels, so that a single-mode description cannot capture the full interference physics. Instead, the Cooperon must be determined self-consistently as a matrix object whose components are dynamically coupled through repeated scattering processes.

This naturally leads to a Bethe–Salpeter formulation in which the Cooperon is built up iteratively from the bare disorder correlators. Physically, each iteration corresponds to adding an impurity line, thereby dressing the interference propagator and progressively incorporating the mixing between different band and valley sectors.
In the present notation, the relevant Cooperon propagators $\Gamma^{m,n}_{n,m}$ and $\Gamma^{n,n}_{m,m}$ (with valley/band indices $m,n$) satisfy  
\begin{align}
\Gamma^{m,n}_{n,m}(\mathbf{k}_1, \mathbf{k}_2) &= \gamma^{m,n}_{n,m}(\mathbf{k}_1, \mathbf{k}_2) + \sum_{\mathbf{k}}  \sum_{\nu = m,n} \gamma^{m,\nu}_{n,\bar{\nu}} (\mathbf{k}_1, \mathbf{k}) \nonumber \\
&\quad \times  G_{\mathbf{k}}^{i\epsilon_n}  G_{\mathbf{q}-\mathbf{k}}^{i\epsilon_n - i\omega_m} \Gamma^{\nu ,n}_{\bar{\nu} ,m}(\mathbf{k}, \mathbf{k}_2), \nonumber \\
\Gamma^{n,n}_{m,m}(\mathbf{k}_1, \mathbf{k}_2) &= \gamma^{n,n}_{m,m}(\mathbf{k}_1, \mathbf{k}_2) +  \sum_{\mathbf{k}}  \sum_{\nu = m,n}  \gamma^{n,\nu}_{m,\bar{\nu}}(\mathbf{k}_1, \mathbf{k})\nonumber \\
&\quad \times    G_{\mathbf{k}}^{i\epsilon_n} G_{\mathbf{q}-\mathbf{k}}^{i\epsilon_n - i\omega_m} \Gamma^{\nu ,n}_{\bar{\nu} ,m}(\mathbf{k}, \mathbf{k}_2),
\label{BSE_coupled}
\end{align}
where $\gamma^{\nu\bar{\nu}}_{\bar{\nu}\nu}(\mathbf{k}_1,\mathbf{k}_2)$ and $\gamma^{\nu\nu}_{\bar{\nu}\bar{\nu}}(\mathbf{k}_1,\mathbf{k}_2)$ are the bare disorder correlators defining the zeroth-order Cooperons:
\begin{align}
\label{bare_coop}
\gamma^{\nu \bar{\nu}}_{\bar{\nu}\nu}(\mathbf{k}_1,\mathbf{k}_2) 
&\equiv \langle U^{\nu \bar{\nu}}_{\mathbf{k}_1,\mathbf{k}_2} \, U^{\nu \bar{\nu}}_{-\mathbf{k}_1,-\mathbf{k}_2} \rangle, 
\nonumber\\
\gamma^{\nu\nu}_{\bar{\nu}\bar{\nu}}(\mathbf{k}_1,\mathbf{k}_2) 
&\equiv \langle U^{\nu \nu}_{\mathbf{k}_1,\mathbf{k}_2} \, U^{\bar{\nu} \bar{\nu}}_{-\mathbf{k}_1,-\mathbf{k}_2} \rangle.
\end{align}
These equations incorporate the sum over all intermediate momentum $\mathbf{k}$ and the valley/band indices $\nu=m,n$, and they automatically include all maximally crossed diagrams for intervalley/interband scattering.
To obtain the ansatz for the full (dressed) Cooperon $\Gamma$, we adopt an iterative ansatz starting from the bare cooperons in Eq.~(\ref{bare_coop}).

\begin{figure*}
    \centering
    \includegraphics[width=2.1\columnwidth]{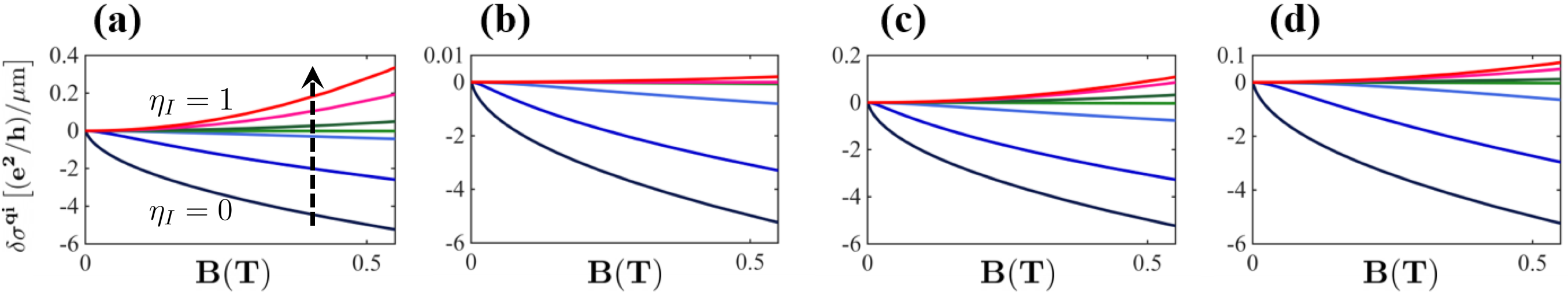}
    \caption{Magnetoconductivity for pseudospin $s=3/2$ fermions for different scattering channel. (a) Intervalley-intraband scattering for $s' = 3/2$  (b) Intravalley-Interband scattering (c) Intervalley-Interband scattering (d) Intervalley-intraband scattering for $s' = 1/2$; The black curve correspond to the $\eta_I = 0$ case and arrow in the direction of increasing $\eta_I$}
    \label{plot}
\end{figure*}
 
\begin{figure}
    \centering
    \includegraphics[width=\columnwidth]{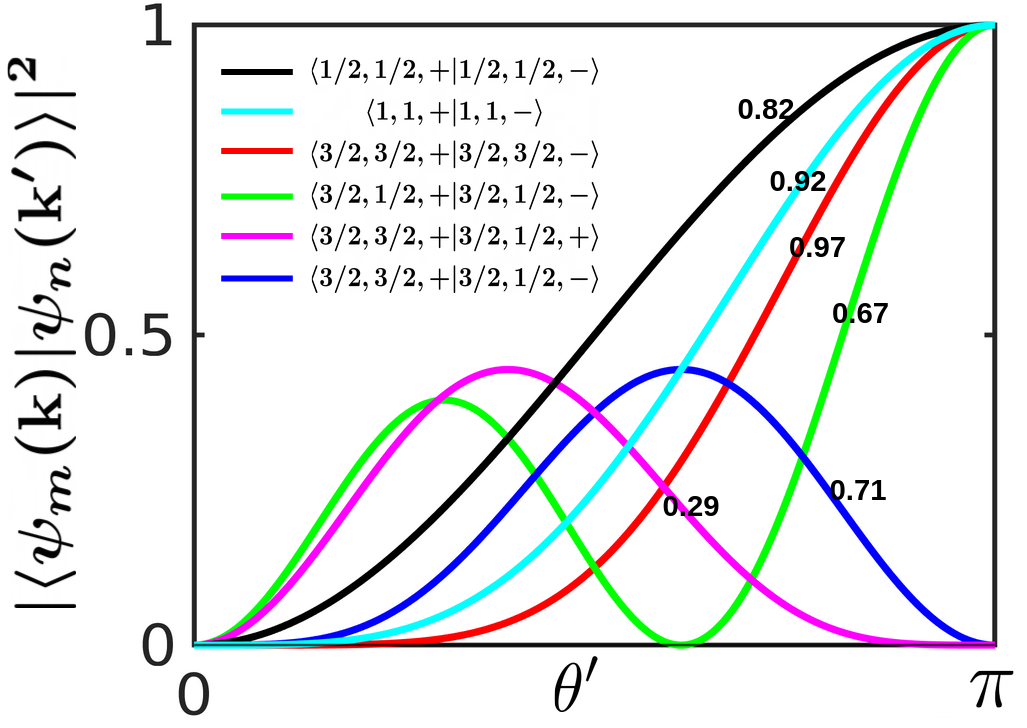}
    \caption{Modulus square of overlap between wavefunction of some higher pseudospin-$s$ fermions within same/different bands $s'$ and same/different valleys ($+$ or $-$), for a fixed azimuthal angle $\phi$ as a function of the difference of incoming $\theta$ and outgoing $\theta^{\prime}$ polar angle. In $\psi_m(\mathbf{k})$, the index $m =equiv |i,j,k\rangle$, where $i$ represents the pseudospin of the fermions, $j$ represents the band index, and $k$ represents the valley of the Weyl fermions. The number on each plot is the backscattering probability $\mathcal{P}_{r's'}^{\lambda\lambda'}$ } 
    \label{back_scatt_plot}
\end{figure}
In practice, we perform the following steps:
(i)  Substitute the bare cooperons $\gamma^{m,n}_{n,m}$ and $\gamma^{n,n}_{m,m}$ into the right-hand side of Eq.~(\ref{BSE_coupled}). The solution $\Gamma^{m,n,0}_{n,m}$, $\Gamma^{n,n,0}_{m,m}$ then just reproduces the bare terms in Eq.~(\ref{BSE_coupled}).  
(ii) Insert the zeroth-order solution into the Bethe–Salpeter integral. This yields a first-order correction $\Gamma^{m,n,1}_{n,m}$, $\Gamma^{n,n,1}_{m,m}$ that contains additional terms coming from one extra impurity scattering. 
This is equivalent to evaluating the $\mathbf{k}$-sum in Eq.~(\ref{BSE_coupled}) with the bare vertices, which produces new angular and band-structure factors not present in the zeroth order.
(iii) Insert $\Gamma^{m,n,1}_{n,m}$, $\Gamma^{n,n,1}_{m,m}$ back into Eq.~(\ref{BSE_coupled}) to compute $\Gamma^{m,n,2}_{n,m}$, $\Gamma^{n,n,2}_{m,m}$, and so on. Therefore, each iteration generates higher-order corrections in the scattering series. After the second iteration, the set of angular structures in $\Gamma^{m,n,p}_{n,m}$, $\Gamma^{n,n,p}_{m,m}$ closes. In other words, the ansatz obtained at first order already includes all linearly independent tensor structures needed. 

While the general formalism to treat the coupled Bethe-Salpeter equations is relegated to Appendix~\ref{interbandcalculation}, here we restrict ourselves to the analysis of $s=3/2$ with two valleys (denoted by $|+\rangle$ and $|-\rangle$). This is also physically motivated by compounds such as {CoSi}~\cite{xu2020optical} that host $s=3/2$ fermions in three spatial dimensions.
Starting from an initial state in the upper band of the $+$ valley, denoted as $\lvert m \rangle = \lvert 3/2, 3/2, + \rangle$, disorder-induced transitions are classified into four distinct channels (see Eq.~\ref{Eq_scatt_channel}): (i) intervalley-intraband scattering that preserves the band index but changes the valley index, i.e., $\lvert m \rangle \to \lvert 3/2, 3/2, - \rangle$, (ii) intravalley-interband scattering, $\lvert m \rangle \to \lvert 3/2, 1/2, + \rangle$, which preserves valley index but changes the band-index; (iii) combined intervalley-interband scattering, $\lvert m \rangle \to \lvert 3/2, 1/2, - \rangle$, which alters both valley and the band-index, and (iv) $|m\rangle\to |m\rangle$, which scatters in the same valley and band index. 
%An analogous set of scattering channels exists for the lower-band initial state $\lvert 3/2, 1/2, + \rangle$ as well. 
%We characterize these processes by the parameter $\eta_I = \tau/\tau_I$, where the total scattering rate is $\tau^{-1} = \tau_{0}^{-1} + \tau_{I}^{-1}$, with $\tau_0$ and $\tau_I$ representing the intravalley and intervalley/interband scattering times, respectively (see Appendix~\ref{interbandcalculation}). 
The scattering time for $| m\rangle \to |m\rangle$ is considered to be $\tau_0$, while the scattering time to any of the other channels is considered to be $\tau_I$. We define $\eta_I = \tau/\tau_I$, the strength of interband/intervalley scattering, where $\tau^{-1} =\tau_0^{-1}+\tau_I^{-1}$ is the total scattering time. Note that in our analysis, we always consider the intravalley-intraband channel to be present and consider only one of the remaining three channels at a time in order to individually study their effect.
\begin{equation}
\left| \tfrac{3}{2}, \tfrac{3}{2}, + \right\rangle 
\;\to\;
\begin{cases}
\left| \tfrac{3}{2}, \tfrac{3}{2}, - \right\rangle \; ; \; \tau_I \\
\left| \tfrac{3}{2}, \tfrac{1}{2}, + \right\rangle \; ; \; \tau_I \\
\left| \tfrac{3}{2}, \tfrac{1}{2}, - \right\rangle \; ; \; \tau_I \\
\left| \tfrac{3}{2}, \tfrac{3}{2}, + \right\rangle \; ; \; \tau_0
\end{cases}
\label{Eq_scatt_channel}
\end{equation}

(i) \textit{Intervalley-intraband scattering:} In the limit $\eta_I\to 0$ (only intraband–intravalley scattering), the magnetoconductivity scales as $\sigma\propto\sqrt{B}$ independent of the band index (black curves in Fig.~\ref{plot}).  As $\eta_I$ increases, the weak-antilocalization peak is suppressed (Fig.~\ref{plot}(a)).  In fact, as $\eta_I\to1$, the magnetoconductivity changes sign at a critical value $\eta_I^C\approx0.58$, signaling a crossover from WAL to weak localization. Such a weak crossover has also been obtained in conventional Weyl fermions~\cite{lu2015weak}. 
We intuitively understand this crossover by examining the backscattering probability:
\begin{equation}
\mathcal{P}_{r's'}^{\lambda\lambda'} 
= 1 - \frac{\int\limits_0^{\pi/2}{\bigl|\langle \psi_{r'}^\lambda(k,\theta,\phi)\mid \psi_{s'}^{\lambda'}(k,\theta'-\theta,\pi+\phi)\rangle\bigr|^2}d\theta'}{\int\limits_0^{\pi}{\bigl|\langle \psi_{r'}^\lambda(k,\theta,\phi)\mid \psi_{s'}^{\lambda'}(k,\theta'-\theta,\pi+\phi)\rangle\bigr|^2}d\theta'},
\end{equation}
where $r',s'$ label the band indices and $\lambda,\lambda'$ label the valley indices.  Figure~\ref{back_scatt_plot} shows the overlap $|\langle \psi_{r'}^\lambda(\mathbf{k})\mid \psi_{s'}^{\lambda'}(\mathbf{k}')\rangle|^2$ as a function of the scattering angle $\theta'-\theta$ (with $\phi'=\pi-\phi$).  In channel-1 (red curve), we find a high backscattering probability $\mathcal{P}_{r's'}^{\lambda\lambda'} \approx0.97$ for $s'=r'=3/2$ with $\lambda\neq\lambda'$. 
\begin{table*}[t]
    \centering
    \small
    \renewcommand{\arraystretch}{1.3} % Adjust row height for better readability
    \setlength{\tabcolsep}{6pt} % Adjust column spacing
    
    \begin{tabular}{l c c c c}
        \toprule
       Channel & {$\eta_I^C$} & {$\mathcal{P}_{r's'}^{\lambda\lambda'}$} & {WAL$\to$WL behavior} \\
       \midrule
        Intervalley-intraband ($s'=r'=3/2$, $\lambda\neq\lambda'$) & 0.58 & 0.97 & Strong interference, early crossover \\
        \midrule
         Intervalley-interband ($s'=3/2, r'=1/2$, $\lambda\neq\lambda'$) & 0.63 & 0.71 & Moderate interference and crossover \\
                      \midrule
       Intervalley-intraband ($s'=r'=1/2$, $\lambda\neq\lambda'$) & 0.65 & 0.67 & Moderate interference and crossover \\
         \midrule
        Intravalley-interband ($s'=3/2, r'=1/2$, $\lambda=\lambda'$) & 0.91 & 0.29 & Weak interference, delayed crossover \\
       %\midrule
       %Intervalley-intraband ($s=1/2$, $s'=r'=1/2$, $\lambda\neq\lambda'$) & 0.71 & 0.82 & Moderate interference and crossover \\
        \bottomrule
    \end{tabular}
    \caption{Critical intervalley/interband strength $\eta_I^C$ where the crossover from negative (WAL) to positive magnetoconductance (WL) occurs, and the corresponding backscattering probability in each channel for pseudospin-$3/2$ fermions.}\label{tab:channels}
\end{table*}

(ii) \textit{Intravalley-interband scattering:} A similar trend is seen for intravalley-interband scattering (Fig.~\ref{plot}(b)).  Increasing $\eta_I$ again suppresses WAL, and the magnetoconductivity crosses zero at $\eta_I^C=0.91$, indicating a transition to WL.  The higher crossover value can be associated with a corresponding lower $\mathcal{P}_{r's'}^{\lambda\lambda'} \approx0.29$ (pink curve in Fig.~\ref{back_scatt_plot}). 
(iii) \textit{Intervalley-interband scattering:} The intervalley-interband case (Fig.~\ref{plot}(c)) follows the same qualitative behavior.  As $\eta_I$ increases, WAL is suppressed and a WL crossover occurs near $\eta_I^C\approx0.63$.  The corresponding backscattering probability is $\mathcal{P}_{r's'}^{\lambda\lambda'}\approx0.71$ (blue curve in Fig.~\ref{back_scatt_plot}). 

(iv) \textit{Band $s'=\tfrac{1}{2}$ channel:} Additionally, we also consider the $s'=1/2$ band under intervalley-intraband scattering (Fig.~\ref{plot}(d)).  Here $\tau_0$ describes intravalley scattering within the $|3/2,1/2,+\rangle$ state and $\tau_I$ the intervalley transition $|3/2,1/2,+\rangle\to|3/2,1/2,-\rangle$.  Increasing $\eta_I$ again suppresses the WAL peak and induces a crossover at $\eta_I^C=0.65$, with  the corresponding $\mathcal{P}_{r's'}^{\lambda\lambda'} \approx0.67$.  The behavior in this channel closely parallels that observed for the $s'=3/2$ channels. {We also apply our formalism to the pseudospin-$\tfrac{1}{2}$ and pseudospin-$1$ systems to analyze the effect of intervalley scattering. The resulting magnetoconductivity is found to be in agreement with previous studies~\cite{lu2015weak, miao2023weak}. See the Fig.~\ref{spin1and1/2}
in Appendix~\ref{spin1and1/2_head}.}

A correlation emerges between the critical scattering strength $\eta_I^C$ and the corresponding backscattering probability (see Table~\ref{tab:channels}). Specifically, channels with a larger backscattering probability exhibit a smaller crossover value $\eta_I^C$, indicating that a relatively weak intervalley/interband scattering is sufficient to drive the system from weak antilocalization (WAL) to weak localization (WL). Conversely, when the backscattering probability is suppressed, a stronger scattering strength (larger $\eta_I^C$) is required to induce the crossover.
The observed trend points toward an approximate inverse relation between $\eta_I^C$ and the backscattering probability: 
\begin{align}
\eta_I^C \propto \frac{1}{\mathcal{P}},
\end{align}
which we propose as a conjecture based on our results, rather than a rigorously derived relation. 
While calculating $\eta_I^C$ for higher pseudospins is technically demanding, the above relation provides insight into the expected behavior as $\mathcal{P}$ is relatively easier to evaluate. Based on the evaluation of $\mathcal{P}$ one may estimate $\eta_I^C$ for higher half-integer pseudospins.
Overall, the crossover behavior is controlled by the competition between symmetry-protected suppression of backscattering (favoring WAL) and disorder-induced mixing between valleys and bands (restoring WL), with $\eta_I^C$ serving as a measure of this balance.

%%%%%%%%%%%%%%%%%%%%%%%%%%%%%%%%%%%%%%%%%%%%%%%%%%%%%%%%%%%%%%
\section{Summary and Outlook}\label{sec7}
In this work, we have developed a unified theory of semiclassical transport and quantum interference for three-dimensional chiral fermions with arbitrary pseudospin-$s$. A central element of our approach is that the disorder is treated at the level of a general matrix structure in pseudospin space, yielding compact arbitrary-$s$ expressions for scattering and vertex renormalization, from which the scalar-disorder universality follows as a controlled limit. While the Drude response is strongly pseudospin dependent, reflecting a geometric suppression of backscattering and a systematic enhancement of transport with increasing $s$, the leading quantum-interference correction retains a universal magnitude identical to that of conventional diffusive metals and conventional Weyl-Dirac fermions, with its sign determined solely by the parity of $2s$, placing half-integer (integer) pseudospins in the symplectic (orthogonal) class. Therefore, only a single diffusive mode survives in the long-wavelength limit despite the enlarged internal Hilbert space. At the same time, this universality is fragile, because interband and intervalley scattering render the Cooperon intrinsically multichannel, and we show explicitly for $s=3/2$ that channel mixing suppresses weak antilocalization and drives a crossover toward localization, with a scale set by the underlying backscattering probability. These findings place multifold fermions within a coherent localization framework that connects band geometry, symmetry class, and disorder, while opening directions for future work, including the role of electron-electron interactions, anisotropy, and realistic disorder, as well as experimental tests in chiral topological materials such as CoSi, RhSi, and AlPt, where magnetotransport may directly probe the predicted pseudospin-dependent interference physics.

%%%%%%%%%%%%%%%%%%%%%%%%%%%%%%%%%%%%%%%%%%%%%%%%%%%%%%%%%%%%%%
\textit{Acknowledgments}: G.S. was funded by ANRF-SERB Core Research Grant CRG/2023/005628. A.G. was funded by IIT Delhi. The authors thank Ekta and Bhanu Pratap Singh for useful discussions.

\bibliography{biblio.bib}
\clearpage
\begin{widetext}

\appendix 
\input{supplementary}

\end{widetext}
\end{document}

%% file: supplementary.tex
\renewcommand\thesection{\Roman{section}}
\renewcommand\thesubsection{\arabic{subsection}}

\appendix
\appendix

\section*{Appendix Contents}
\noindent
\begin{itemize}
    \item \hyperref[Sec_App_A_1]{Appendix A1: Hamiltonian and the Wigner-$D$ functions}
    \item \hyperref[Sec_App_A_2]{Appendix A2: Scattering time in the presence of general disorder $\mathcal{M}$}
    \item \hyperref[AppA_vel_correct]{Appendix A3: Vertex corrections and Kubo conductivity in the presence of general disorder $\mathcal{M}$}
    \item \hyperref[appen.sect-2]{Appendix A4: Conductivity correction from quantum interference for general pseudospin-$s$ for scalar disorder}
    \item \hyperref[Sec_App_A_5]{Appendix A5: Explicit results for $s=1/2$ and $s=1$}
    \item \hyperref[Sec_App_B_1]{Appendix B1: Eigenvalues and eigenvectors for pseudospin-3/2}
    \item \hyperref[Sec_App_B_2]{Appendix B2: Scattering time for pseudospin-3/2}
    \item \hyperref[Sec_App_B_3]{Appendix B3: Velocity correction for pseudospin-3/2}
    \item \hyperref[Sec_App_B_4]{Appendix B4: Conductivity correction with both interband and intervalley scattering for pseudospin-3/2}
    \item \hyperref[Sec_App_B_5]{Appendix B5: Calculation for upper band for pseudospin-3/2}
    \item \hyperref[Sec_App_B_6]{Appendix B6: Calculation for lower band for pseudospin-3/2}
    \item \hyperref[spin1and1/2_head]{Appendix B7: Effect of intervalley scattering on the pseudospin-1/2, and 1}
\end{itemize}
\section{General formalism}
\label{AppA}
%%%%%%%%%%%%%%%%%%%%%%%%%%%%%%%%%%%%%%%%%%%%%%%%%%%%%%%%%%%%%%
\subsection{Hamiltonian and the Wigner-$D$ functions}\label{Sec_App_A_1}
We consider a system with arbitrary pseudospin~$s$ in which the spin degrees of freedom are linearly coupled to the particle momentum. The corresponding effective Hamiltonian is given by
\begin{equation}
H_s(\mathbf{k}) = \hbar v\,\mathbf{k}\cdot\mathbf{S},
\label{eq:H_A}
\end{equation}
where $\mathbf{S}=(S_x,S_y,S_z)$ denotes the spin-$s$ representation of the SU(2) generators
and $v$ is a characteristic velocity. The eigenspectrum of the Hamiltonian is determined using the rotational invariance 
\begin{equation}
E_{s'}(\mathbf{k}) = \hbar v\, s' k, \qquad s' = -s,-s{+}1,\ldots,s,
\label{eq:Ebands_A}
\end{equation}
and the eigenstates of $H_s$ are helicity eigenstates. The spectrum thus consists of $2s{+}1$ linearly dispersing bands that intersect at a single
degeneracy point at $\mathbf{k}=0$. For integer values of $s$, a dispersionless (flat) band
appears at zero energy corresponding to $s'=0$, whereas for half-integers $s$ the spectrum is symmetric about zero energy, and the helicity eigenstates are
\begin{equation}
\ket{\mathbf k;s,s'}=R(\hat{\mathbf k})\ket{s,s'}
   =\sum_{m=-s}^{s} D^{s}_{m \textcolor{black}{s'} }(\phi,\theta,0)\ket{s,m},
\label{eq:wf_A}
\end{equation}
with \(D^{s}_{m s'}(\phi,\theta,0)=e^{-im\phi}d^{s}_{m s'}(\theta)\). The reduced Wigner functions \(d^{s}_{m s'}(\theta)\) obey the orthogonality relations
\begin{align*}
    \sum_{m=-s}^{s}d^{s}_{m s'}(\theta)d^{s}_{m s''}(\theta)=\delta_{s's''},\\ 
\sum_{s'=-s}^{s}d^{s}_{m s'}(\theta)d^{s}_{m' s'}(\theta)=\delta_{m m'}.
\label{eq:dorth_A}
\end{align*}
The closed form of the reduced Wigner functions is:
%\begin{widetext}
\begin{equation}
d^s_{m s'}(\theta)
=\sum_{p}(-1)^{p+m'-m}
\frac{\sqrt{(s{+}m)!(s{-}m)!(s{+}m')!(s{-}m')!}}
{(s{+}m'{-}p)!\,p!\,(m{-}m'{+}p)!\,(s{-}m{-}p)!}
\big(\cos\tfrac{\theta}{2}\big)^{2s+m-m'-2p}
\big(\sin\tfrac{\theta}{2}\big)^{2p+m'-m},
\label{eq:A1_A}
\end{equation}
%\end{widetext}
and for the diagonal entries $m=m'=s'$ , the compact `Pascal' form
\begin{equation}
d^{s}_{s's'}(\theta)
=\sqrt{\binom{2s}{s{+}s'}}\,
\big(\cos\tfrac{\theta}{2}\big)^{s+s'}\big(\sin\tfrac{\theta}{2}\big)^{\,s-s'}.
\label{eq:A2}
\end{equation}
{We provide below the explicit polynomials \(|d^{s}_{s's'}(\theta)|^2\) up to $s=2$ (in $c\equiv\cos\theta$).}
\begin{align*}
s=\tfrac12:~ & s'=\tfrac12:~ |d|^2=\tfrac12(1+c). \\[3pt]
s=1:~ & s'=\pm1:~ |d|^2=\tfrac14(1+2c+c^2),\qquad s'=0:~ |d|^2=c^2. \\[3pt]
s=\tfrac32:~ & s'=\tfrac32:~ |d|^2=\big(\tfrac{1+c}{2}\big)^3=\tfrac18(1+3c+3c^2+c^3),\\
             & s'=\tfrac12:~ |d|^2=\textcolor{black}{\frac{(3c-1)^2}{4}\cdot\frac{1+c}{2}}
              . \\[3pt]
s=2:~ & s'=2:~ |d|^2=\big(\tfrac{1+c}{2}\big)^4=\tfrac1{16}(1+4c+6c^2+4c^3+c^4),\\
      & s'=1:~ |d|^2=\textcolor{black}{\Big[\tfrac14(4c - 2)\cdot(\tfrac{1+c}{2})^2\Big]^2}\\
      & s'=0:~ |d|^2=\tfrac14(3c^2-1)^2=\tfrac14(9c^4-6c^2+1).
\end{align*}
\textit{Density of States:} The density of states (DOS) at the Fermi surface is calculated as 
\begin{align}
    N_{F} &= \int \frac{d^3k}{(2\pi)^3} \delta(E_F - \epsilon_\mathbf{k'}) \nonumber\\
    &= \int_0^{2\pi} \frac{d\phi}{2\pi} \int_0^\pi \frac{\sin \theta \, d\theta}{2\pi} \int_0^\infty \frac{k^2 \, dk}{2\pi} \delta(E_F - \epsilon_\mathbf{k'})\nonumber\\
    &= \frac{4\pi}{8\pi^3}\int_0^\infty k^2 \, dk \delta(E_F - \epsilon_\mathbf{k'})\label{A13}
\end{align}
Since DOS varies by band. In general, we consider the dispersion relation in Eq.~\ref{eq:Ebands} as 
\begin{align*}
     \epsilon_{\mathbf{k}} &= s' \hbar \vartheta k\\
    d\epsilon_{\mathbf{k}} &= s' \hbar \vartheta  dk
\end{align*}
Put the value of $\epsilon_\mathbf{k}$ and $d\epsilon_{\mathbf{k}}$ in Eq.~\ref{A13},
\begin{align}
    N_F^{s'} &= \frac{1}{2\pi^2}\int \frac{\epsilon_{\mathbf{k}}^2 d\epsilon_{\mathbf{k}}}{(s' \hbar \vartheta )^3} \delta(E_F - \epsilon_\mathbf{k'})\nonumber \\
   N_F^{s'} &= \frac{E_F^2}{2\pi^2(s' \hbar \vartheta )^3}\label{ADOS}    
\end{align}  
\subsection{Scattering time in the presence of general disorder $\mathcal{M}$}\label{Sec_App_A_2}
The spinor overlap between the states $\mathbf{k}$ and $\mathbf{k'}$, reflecting the rotation of one helicity state into another, is defined as:

\begin{align}
W_{s''s'}(\theta,\phi)
   =\left|\langle \mathbf{k}', s'' |\mathcal {M} | \mathbf{k}, s' \rangle\right|^2,
   \label{eq:kernal}
\end{align}
where $\mathcal{M}$ is the impurity matrix, and $|{\mathbf k,s,s'}\rangle$ is a helicity eigenstate, given in Eq.~\ref{eq:wf}.  
The overlap can be written in terms of the Wigner $D$-matrix as~\cite{sakurai2020modern}
\begin{equation}
\langle \mathbf{k}', s''| \mathcal{M} |\mathbf{k}, s' \rangle
=
\sum_{m_1,m_2}
D^{s *}_{m_1 s''}(\hat{\mathbf k}')
\, {M}_{m_1 m_2} \,
D^{s}_{m_2 s'}(\hat{\mathbf k})
\label{eq:general_overlap}
\end{equation}
where $M_{m' m} = \langle s, m' |\mathcal{M} | s, m \rangle$.
The scattering rate for $|\mathbf{k}, s'\rangle$ in the first-Born approximation is 
\begin{equation}
    \frac{1}{\tau_{s'}(\mathbf{k})} =  \frac{2\pi}{\hbar} n_i u_0^2\sum_{s''} \int \frac{d^3 k'}{(2\pi)^3} W_{s'' s'}(\mathbf{k}', \mathbf{k}) \delta(\epsilon_{s''}(k') - \epsilon_{s'}(k)).
\end{equation}
Since our system is isotropic, this becomes
\begin{equation}
    \frac{1}{\tau_{s'}} = \frac{2\pi}{\hbar} n_i u_0^2 \sum_{s''} N_F^{(s'')} \int \frac{d\Omega_{\mathbf{k}'}}{4\pi} \left| \langle (\mathbf{\hat{k}}', s'' | \mathcal{M} | \mathbf{\hat{k}}, s' \rangle \right|^2.
    \label{eq:isotropic_scatt_time}
\end{equation}
Using Eq.~\ref{eq:general_overlap}, we obtain 
\begin{equation}
\frac{1}{\tau_{s'}} = \frac{2\pi}{\hbar} n_i u_0^2 \sum_{s''} N_F^{(s'')} \int \frac{d\Omega_{\mathbf{k}'}}{4\pi} \left| \sum_{m', m} D^{s*}_{m' s''}(\mathbf{\hat{k}}') M_{m' m} D^s_{m s'}(\mathbf{\hat{k}}) \right|^2
\end{equation}
\begin{align}
\frac{1}{\tau_{s'}}
&=
\frac{2\pi}{\hbar} n_i u_0^2
\sum_{s''} N_F^{(s'')}
\int \frac{d\Omega_{\mathbf k'}}{4\pi}
\left|
\sum_{m_1,m_2}
D^{s*}_{m_1 s''}(\hat{\mathbf k}')
\,M_{m_1m_2}\,
D^s_{m_2 s'}(\hat{\mathbf k})
\right|^2 \nonumber\\
&=
\frac{2\pi}{\hbar} n_i u_0^2
\sum_{s''} N_F^{(s'')}
\sum_{m_1,m_2}\sum_{n_1,n_2}
M_{m_1m_2}M^*_{n_1n_2}
D^s_{m_2s'}(\hat{\mathbf k})
D^{s*}_{n_2s'}(\hat{\mathbf k})
\int \frac{d\Omega_{\mathbf k'}}{4\pi}
D^{s*}_{m_1 s''}(\hat{\mathbf k}')
D^s_{n_1 s''}(\hat{\mathbf k}').
\label{eq:tau_expand}
\end{align}
Now, using the angular orthogonality relation
\begin{equation}
\int \frac{d\Omega}{4\pi}
D^{s*}_{m s''}(\Omega)\,
D^{s}_{n s''}(\Omega)
=
\frac{\delta_{mn}}{2s+1},
\label{eq:D_orth_sphere}
\end{equation}
the \(m_1,n_1\) sums collapse and we get
\begin{equation}
\frac{1}{\tau_{s'}}
=
\frac{2\pi}{\hbar} n_i u_0^2
\frac{1}{2s+1}
\sum_{s''} N_F^{(s'')}
\sum_{m_1}\sum_{m_2,n_2}
M_{m_1m_2}M^*_{m_1n_2}
D^s_{m_2s'}(\hat{\mathbf k})
D^{s*}_{n_2s'}(\hat{\mathbf k}).
\label{eq:tau_after_orth}
\end{equation}
Defining $
(\mathcal{M}^\dagger \mathcal{M} )_{n_2m_2}
=
\sum_{m_1} M^*_{m_1n_2}M_{m_1m_2},
$, we get
\begin{equation}
\frac{1}{\tau_{s'}}
=
\frac{2\pi}{\hbar} n_i u_0^2
\frac{1}{2s+1}
\sum_{s''} N_F^{(s'')}
\sum_{m_2,n_2}
D^s_{m_2s'}(\hat{\mathbf k})
(\mathcal{M}^\dagger \mathcal{M} )_{n_2m_2}
D^{s*}_{n_2s'}(\hat{\mathbf k}).
\label{eq:tau_component_final}
\end{equation}
Equivalently, we may write
\begin{equation}
\frac{1}{\tau_{s'}}
=
\frac{2\pi}{\hbar} n_i u_0^2
\frac{1}{2s+1}
\sum_{s''} N_F^{(s'')}
\langle{\hat{\mathbf k},s'} |\mathcal{M}^\dagger \mathcal{M} |{\hat{\mathbf k},s'}\rangle.
\label{eq:tau_compact_final}
\end{equation}
Since the system is isotropic, the scattering time should not depend on the direction $\hat{\mathbf{k}}$ and hence we may choose $|\hat{\mathbf k},s'\rangle=|s,s'\rangle$.
Then Eq.~\eqref{eq:tau_compact_final}
reduces to
\begin{equation}
\frac{1}{\tau_{s'}}
=
\frac{2\pi}{\hbar} n_i u_0^2
\frac{1}{2s+1}
\sum_{s''} N_F^{(s'')}
\langle s,s'|\mathcal M^\dagger \mathcal M|s,s'\rangle .
\label{eq:tau_zaxis_start_app}
\end{equation}
We now expand the impurity matrix in irreducible tensor (or polarization) operators~\cite{varshalovich1988quantum},
\begin{equation}
\mathcal M=\sum_{L=0}^{2s}\sum_{Q=-L}^{L} M_{LQ}\,T_Q^{(L)},
\label{eq:M_tensor_expand_app}
\end{equation}
where \(M_{LQ}\) are the corresponding tensor amplitudes. Using~\cite{varshalovich1988quantum} 
\begin{equation}
\langle s,m'|T_Q^{(L)}|s,m\rangle
=
\sqrt{\frac{2L+1}{2s+1}}\,
C^{\,s\,m'}_{\,s\,m,\;L\,Q},
\label{eq:T_matrix_element_app}
\end{equation}
where 
\begin{align}
C^{\,s\, m'}_{\,s\, m,\; L\, Q}
=
(-1)^{s - L + m'}\,
\sqrt{2s+1}\,
\begin{pmatrix}
s & L & s \\
m & Q & -m'
\end{pmatrix}
\end{align}
is the Wigner-3j symbol. 
Using $T_Q^{(L)\dagger}=(-1)^Q T_{-Q}^{(L)}$~\cite{varshalovich1988quantum},
we obtain
\begin{align}
\langle s,s'|\mathcal M^\dagger \mathcal M|s,s'\rangle
&=
\sum_{LQ}\sum_{L'Q'}
M_{LQ}^*M_{L'Q'}
\langle s,s'|T_Q^{(L)\dagger}T_{Q'}^{(L')}|s,s'\rangle
\nonumber\\
&=
\sum_{LQ}\sum_{L'Q'}
M_{LQ}^*M_{L'Q'}(-1)^Q
\langle s,s'|T_{-Q}^{(L)}T_{Q'}^{(L')}|s,s'\rangle .
\end{align}
Now, inserting a complete set of $|s,m\rangle$ states,
\begin{equation}
\langle s,s'|T_{-Q}^{(L)}T_{Q'}^{(L')}|s,s'\rangle
=
\sum_{m=-s}^{s}
\langle s,s'|T_{-Q}^{(L)}|s,m\rangle
\langle s,m|T_{Q'}^{(L')}|s,s'\rangle .
\label{eq:slection}
\end{equation}
Using the Clebsch-Gordan selection rules, we have 
\(m=s'+Q=s'+Q'\), hence \(Q'=Q\). Therefore,
\begin{equation}
\langle s,s'|\mathcal M^\dagger \mathcal M|s,s'\rangle
=
\sum_{Q}\sum_{L,L'}
M_{LQ}^*M_{L'Q}\,
\mathcal K_{LL'}^{(s,s';Q)},
\end{equation}
where
\begin{equation}
\mathcal K_{LL'}^{(s,s';Q)}
=
\frac{\sqrt{(2L+1)(2L'+1)}}{2s+1}\,
C^{\,s,\;s'+Q}_{\,s,\;s',\;L,\;Q}\,
C^{\,s,\;s'+Q}_{\,s,\;s',\;L',\;Q}.
\label{eq:KLL_def_app}
\end{equation}
Hence, the scattering rate becomes
\begin{equation}
\frac{1}{\tau_{s'}}
=
\frac{2\pi}{\hbar} n_i u_0^2
\frac{1}{(2s+1)^2}
\sum_{s''} N_F^{(s'')}
\sum_{Q}\sum_{L,L'}
\sqrt{(2L+1)(2L'+1)}\,
M_{LQ}^*M_{L'Q}\,
C^{\,s,\;s'+Q}_{\,s,\;s',\;L,\;Q}\,
C^{\,s,\;s'+Q}_{\,s,\;s',\;L',\;Q}.
\label{eq:tau_tensor_final_app}
\end{equation}
This can be further simplified to
\begin{equation}
\frac{1}{\tau_{s'}}
=
\frac{2\pi}{\hbar} n_i u_0^2
\frac{1}{(2s+1)^2}
\sum_{s''} N_F^{(s'')}
\sum_Q |\mathcal A_{s'}(Q)|^2,
\end{equation}
where 
\begin{equation}
\mathcal A_{s'}(Q)
=
\sum_L \sqrt{2L+1}\,M_{LQ}\,
C^{\,s,\;s'+Q}_{\,s,\;s',\;L,\;Q},
\label{eq:ASQ}
\end{equation}
%%%%%%%%%%%%%%%%%%%%%%%%%%%%%%%%%%%%%%%%%%%%%%%%%%%%%%%%%%%%%%
We may now apply the above result to some specific cases:
\subsubsection*{(i) Scalar disorder: $\mathcal{M}=\mathcal{I}_{2s+1\times 2s+1}$}
For scalar disorder, only the $L=0$, $Q=0$ component is present:
\begin{equation}
M_{00} \neq 0, \qquad M_{LQ}=0 \ \text{otherwise}.
\end{equation}
Thus,
\begin{equation}
\mathcal A_{s'}(Q)
=
\delta_{Q,0}\,M_{00}\,
C^{\,s,\,s'}_{\,s,\,s',\,0,\,0}.
\end{equation}
Using $
C^{\,s,\,s'}_{\,s,\,s',\,0,\,0} = 1$,
we obtain
\begin{equation}
\mathcal A_{s'}(0)=M_{00}=1, \qquad \mathcal A_{s'}(Q\neq 0)=0.
\end{equation}
Therefore,
\begin{equation}
\frac{1}{\tau_{s'}}
=
\frac{2\pi}{\hbar} n_i u_0^2
\frac{1}{(2s+1)}
\sum_{s''} N_F^{(s'')}.
\end{equation}
%%%%%%%%%%%%%%%%%%%%%%%%%%%%%%%%%%%%%%%%%%%%%%%%%%%%%%%%%%%%%%
\subsubsection*{(ii) $s=1/2$, and $\mathcal{M}=a_0 \mathcal{I}+\mathbf{a}\cdot{\sigma}$}
We obtain
\begin{equation}
\mathcal A_{s'}(Q)
=
M_{00}\,C^{\,\tfrac12,\,s'}_{\,\tfrac12,\,s',\,0,\,0}\,\delta_{Q,0}
+
\sqrt{3}\,M_{1Q}\,C^{\,\tfrac12,\,s'+Q}_{\,\tfrac12,\,s',\,1,\,Q}.
\end{equation}

For $s'=\tfrac12$, allowed values are $Q=0,-1$:
\begin{align}
\mathcal A_{+}(0)
&=
M_{00}
+
\sqrt{3}\,M_{10}\,
C^{\,\tfrac12,\,\tfrac12}_{\,\tfrac12,\,\tfrac12,\,1,\,0}, \\
\mathcal A_{+}(-1)
&=
\sqrt{3}\,M_{1,-1}\,
C^{\,\tfrac12,\,-\tfrac12}_{\,\tfrac12,\,\tfrac12,\,1,\,-1}.
\end{align}

For $s'=-\tfrac12$, allowed values are $Q=0,+1$:
\begin{align}
\mathcal A_{-}(0)
&=
M_{00}
+
\sqrt{3}\,M_{10}\,
C^{\,\tfrac12,\,-\tfrac12}_{\,\tfrac12,\,-\tfrac12,\,1,\,0}, \\
\mathcal A_{-}(+1)
&=
\sqrt{3}\,M_{1,+1}\,
C^{\,\tfrac12,\,\tfrac12}_{\,\tfrac12,\,-\tfrac12,\,1,\,+1}.
\end{align}
For $s=\tfrac12$, the relevant Clebsch-Gordan coefficients are
\begin{equation}
C^{\,\tfrac12,\,\tfrac12}_{\,\tfrac12,\,\tfrac12,\,1,\,0}
=\frac{1}{\sqrt3},
\qquad
C^{\,\tfrac12,\,-\tfrac12}_{\,\tfrac12,\,\tfrac12,\,1,\,-1}
=\sqrt{\frac{2}{3}},
\end{equation}
\begin{equation}
C^{\,\tfrac12,\,-\tfrac12}_{\,\tfrac12,\,-\tfrac12,\,1,\,0}
=-\frac{1}{\sqrt3},
\qquad
C^{\,\tfrac12,\,\tfrac12}_{\,\tfrac12,\,-\tfrac12,\,1,\,1}
=-\sqrt{\frac{2}{3}}.
\end{equation}
The scattering rates are:
\begin{equation}
\frac{1}{\tau_{+}}
=
\frac{2\pi}{\hbar} n_i u_0^2
\frac{1}{4}
\sum_{s''} N_F^{(s'')}
\left(
|M_{00}+M_{10}|^2
+
2|M_{1,-1}|^2
\right),
\end{equation}
\begin{equation}
\frac{1}{\tau_{-}}
=
\frac{2\pi}{\hbar} n_i u_0^2
\frac{1}{4}
\sum_{s''} N_F^{(s'')}
\left(
|M_{00}-M_{10}|^2
+
2|M_{1,+1}|^2
\right).
\end{equation}
Using the relations for \(s=\tfrac12\):,
\begin{equation}
T^{(0)}_0=\frac{\mathcal I}{\sqrt2},
\qquad
T^{(1)}_0=\frac{\sigma_z}{\sqrt2},
\qquad
T^{(1)}_{1}=-\sigma_+,
\qquad
T^{(1)}_{-1}=\sigma_-,
\end{equation}
the disorder matrix
can be written as
\begin{equation}
\mathcal M
=
\sqrt2\,a_0\,T^{(0)}_0
+
\sqrt2\,a_z\,T^{(1)}_0
-
(a_x-i a_y)\,T^{(1)}_{1}
+
(a_x+i a_y)\,T^{(1)}_{-1}.
\end{equation}
Therefore, the scattering rates become
\begin{equation}
\frac{1}{\tau_+}
=
\frac{2\pi}{\hbar}n_i u_0^2\frac12
\sum_{s''}N_F^{(s'')}
\left[
|a_0+a_z|^2+|a_x+i a_y|^2
\right],
\end{equation}
\begin{equation}
\frac{1}{\tau_-}
=
\frac{2\pi}{\hbar}n_i u_0^2\frac12
\sum_{s''}N_F^{(s'')}
\left[
|a_0-a_z|^2+|a_x-i a_y|^2
\right].
\end{equation}
%%%%%%%%%%%%%%%%%%%%%%%%%%%%%%%%%%%%%%%%%%%%%%%%%%%%%%%%%%%%%%
\subsubsection*{(iii) $s=1$, $\mathcal{M}=S_x$}

The operator $S_x$ transforms as a rank-$1$ tensor and can be written as
\begin{equation}
S_x = \frac{S_+ + S_-}{2} = T^{(1)}_{-1} - T^{(1)}_{1},
\end{equation}
so that the only nonzero tensor coefficients are
\begin{equation}
M_{1,-1}=1, \qquad M_{1,1}=-1, \qquad M_{1,0}=0.
\end{equation}

Using
\begin{equation}
\mathcal A_{s'}(Q)
=
\sqrt{3}\,M_{1Q}\,
C^{\,1,\,s'+Q}_{\,1,\,s',\,1,\,Q},
\end{equation}
the nonvanishing amplitudes are obtained for $Q=\pm1$. Evaluating the relevant Clebsch-Gordan coefficients, one finds
\begin{equation}
|\mathcal A_{+1}(-1)|^2 = \frac{3}{2}, \quad
|\mathcal A_{0}(\pm1)|^2 = \frac{3}{2}, \quad
|\mathcal A_{-1}(+1)|^2 = \frac{3}{2}.
\end{equation}

Substituting into the general expression for the scattering rate, we obtain
\begin{equation}
\frac{1}{\tau_{+1}}
=
\frac{2\pi}{\hbar} n_i u_0^2
\frac{1}{6}
\sum_{s''} N_F^{(s'')},
\qquad
\frac{1}{\tau_{0}}
=
\frac{2\pi}{\hbar} n_i u_0^2
\frac{1}{3}
\sum_{s''} N_F^{(s'')},
\qquad
\frac{1}{\tau_{-1}}
=
\frac{2\pi}{\hbar} n_i u_0^2
\frac{1}{6}
\sum_{s''} N_F^{(s'')}.
\end{equation}

%%%%%%%%%%%%%%%%%%%%%%%%%%%%%%%%%%%%%%%%%%%%%%%%%%%%%%%%%%%%%%
\subsubsection*{(iv) $s=1$, $\mathcal{M}=S_y$}
Similarly, the operator $S_y$ can be written as
\begin{equation}
S_y = \frac{S_+ - S_-}{2i} = i\bigl(T^{(1)}_{1} + T^{(1)}_{-1}\bigr),
\end{equation}
and thus
\begin{equation}
M_{1,1}=i, \qquad M_{1,-1}=i, \qquad M_{1,0}=0.
\end{equation}
The amplitudes $\mathcal A_{s'}(Q)$ differ from the $S_x$ case only by phases, and therefore their moduli are identical. Consequently, the scattering rates are unchanged:
\begin{equation}
\frac{1}{\tau_{+1}}
=
\frac{2\pi}{\hbar} n_i u_0^2
\frac{1}{6}
\sum_{s''} N_F^{(s'')},
\qquad
\frac{1}{\tau_{0}}
=
\frac{2\pi}{\hbar} n_i u_0^2
\frac{1}{3}
\sum_{s''} N_F^{(s'')},
\qquad
\frac{1}{\tau_{-1}}
=
\frac{2\pi}{\hbar} n_i u_0^2
\frac{1}{6}
\sum_{s''} N_F^{(s'')}.
\end{equation}

\subsection{Vertex corrections and Kubo conductivity for general disorder $\mathcal{M}$}\label{AppA_vel_correct}
%%%%%%%%%%%%%%%%%%%%%%%%%%%%%%%%%%%%%%%%%%%%%%%%%%%%%%%%%%%%%%
For the short-range disorder potential, the renormalized vertex for helicity~$s'$ is found by solving the ladder equation, 
\begin{align}
\Gamma_{s',\beta}(\hat{\mathbf k})
=
v_{s',\beta}(\hat{\mathbf k}) \quad
+
n_i u_0^2
\sum_{s''}
\mathcal J_{s''}
\int\frac{d\Omega'}{4\pi}\,
W_{s's''}(\hat{\mathbf k},\hat{\mathbf k}')
\,
\Gamma_{s'',\beta}(\hat{\mathbf k}'),
\label{eq:BS_general}
\end{align}
with
\begin{equation}
\mathcal J_{s'}
=
\int \frac{k^{d-1}dk}{(2\pi)^d}\,
G^R_{s'}(\mathbf k,E_F)G^A_{s'}(\mathbf k,E_F)
=
\frac{2\pi}{\hbar}N_F^{(s')}\tau_s'.
\label{eq:Jlambda_general}
\end{equation}
Within the same Born-plus-ladder approximation, the self-energy and vertex are therefore built from the same impurity correlator, so the construction is conserving and compatible with the Ward identity~\cite{rammer1991quantum,mahan2013many}.
For a generic matrix disorder $\mathcal{M}$, the spinor overlap $W_{s's''}(\hat{\mathbf k}',\hat{\mathbf k})$, Eq.~\ref{eq:kernal}, is not purely a function of the relative angle, and the dressed vertex is not characterized by a single scalar renormalization factor. The natural generalization is to expand the vertex in spherical harmonics,
\begin{equation}
\Gamma_{s',\beta}(\hat{\mathbf k})
=
\sum_{\ell m}\eta^{(\beta)}_{s';\ell m}\,
Y_{\ell m}(\hat{\mathbf k}).
\label{eq:Gamma_harmonic_expansion}
\end{equation}
The bare velocity transforms as a vector under rotations and therefore resides entirely in the $l=1$ angular momentum sector; this follows from the fact that the components of $\hat{k}$ are linear combinations of the spherical harmonics $Y_{1m}$.
Equation~\eqref{eq:BS_general} becomes:
\begin{equation}
\eta^{(\beta)}_{s';\ell m}
=
\eta^{(\beta,0)}_{s';\ell m}
+
\sum_{s''\ell'm'}
\mathcal K^{\ell m,\ell' m'}_{s's''}\,
\eta^{(\beta)}_{s'';\ell'm'},
\label{eq:eta_matrix_general}
\end{equation}
with kernel
\begin{align}
\mathcal K^{\ell m,\ell' m'}_{s's''}
&=
n_i u_0^2\,\mathcal J_{s''}
\int\!\frac{d\Omega}{4\pi}\frac{d\Omega'}{4\pi}\,
Y_{\ell m}^{*}(\hat{\mathbf k})\,
W_{s's''}(\hat{\mathbf k},\hat{\mathbf k}')
\,Y_{\ell' m'}(\hat{\mathbf k}'),
\label{eq:Kmatrix_general}
\end{align}
and 
\begin{equation}
\eta^{(\beta,0)}_{s';\ell m}
=
\int \frac{d\Omega}{4\pi}\, v_{s',\beta}(\hat{\mathbf k})
Y^*_{\ell m}(\hat{\mathbf k}).
\end{equation}
Thus, the general vertex renormalization is a matrix
\begin{equation}
\boldsymbol{\eta}^{(\beta)}
=
\left(\mathds 1-\hat{\mathcal K}\right)^{-1}
\boldsymbol{\eta}^{(\beta,0)}.
\label{eq:eta_general_solution}
\end{equation}
Since our system is isotropic, the dressed vertex remains parallel to the bare velocity and only the $\ell=1$ harmonic survives. Then one may write
\begin{equation}
\Gamma_{s',\beta}(\hat{\mathbf k})
=
\eta_{s'}\,v_{s',\beta}(\hat{\mathbf k}),
\label{eq:eta_scalar_ansatz}
\end{equation}
and Eq.~\eqref{eq:BS_general} reduces to
\begin{equation}
\eta_{s'}
=
1+
n_i u_0^2 \mathcal J_{s'}
\
 \int \frac{d\Omega'}{4\pi}\,
W_{s's'}(\hat{\mathbf k}',\hat{\mathbf k})\,\cos\theta'
\,\eta_{s'}.
\label{eq:eta_isotropic_start}
\end{equation}
Using Eq.~ \ref{eq:isotropic_scatt_time}, this can be written compactly as
\begin{equation}
\eta_{s'}
=
\frac{1}{1-\mu_{s'}},
\qquad
\mu_{s'}
\equiv
\frac{
\displaystyle \int \frac{d\Omega'}{4\pi}\,
W_{s's'}(\hat{\mathbf k}',\hat{\mathbf k})\,\cos\theta'
}{
\displaystyle \int \frac{d\Omega'}{4\pi}\,
W_{s's'}(\hat{\mathbf k}',\hat{\mathbf k})
}.
\label{eq:eta_mu_general}
\end{equation}
From Eq.~\ref{eq:general_overlap},
\begin{equation}
\left|
\langle \mathbf{k}', s' | \mathcal{M} |\mathbf{k}, s' \rangle
\right|^2
=
\sum_{m_1,m_2}\sum_{n_1,n_2}
M_{m_1m_2}M^*_{n_1n_2}
D^s_{m_2s'}(\hat{\mathbf k})
D^{s*}_{n_2s'}(\hat{\mathbf k})
D^{s*}_{m_1 s''}(\hat{\mathbf k}')
D^s_{n_1 s''}(\hat{\mathbf k}').
\end{equation}
The angular integral is 
\begin{align}
\int \frac{d\Omega_{\mathbf{k}'}}{4\pi}
\left|
\langle \mathbf{k}', s' | \mathcal{M} |\mathbf{k}, s' \rangle
\right|^2
&=
\sum_{m_1,m_2}\sum_{n_1,n_2}
M_{m_1m_2}M^*_{n_1n_2}
D^s_{m_2s'}(\hat{\mathbf k})
D^{s*}_{n_2s'}(\hat{\mathbf k})
\int \frac{d\Omega_{\mathbf k'}}{4\pi}
D^{s*}_{m_1 s''}(\hat{\mathbf k}')
D^s_{n_1 s''}(\hat{\mathbf k}')
\end{align} 
Now, we used the angular orthogonality relation Eq.~\ref{eq:D_orth_sphere}
\begin{align}
\int \frac{d\Omega_{\mathbf{k}'}}{4\pi}
\left|
\langle \mathbf{k}', s' |  \mathcal{M} |\mathbf{k}, s' \rangle
\right|^2
&=
\frac{1}{2s+1}\sum_{m_2,n_2}\sum_{m_1}
M_{m_1m_2}M^*_{m_1n_2}
D^s_{m_2s'}(\hat{\mathbf k})
D^{s*}_{n_2s'}(\hat{\mathbf k})
\end{align}
By a similar calculation as in section-1, see Eq.~\ref{eq:tau_component_final},
\begin{align}
\int \frac{d\Omega_{\mathbf{k}'}}{4\pi}
\left|
\langle \mathbf{k}', s' |  \mathcal{M} |\mathbf{k}, s' \rangle
\right|^2
&=
\frac{1}{2s+1}
\langle{\hat{\mathbf k},s'} |\mathcal{M}^\dagger \mathcal{M} |{\hat{\mathbf k},s'}\rangle.
\end{align}
\begin{align}
\int \frac{d\Omega_{\mathbf{k}'}}{4\pi}
\left|
\langle \mathbf{k}', s' |  \mathcal{M} |\mathbf{k}, s' \rangle
\right|^2 \cos\theta'
&=
\sum_{m_1,m_2}\sum_{n_1,n_2}
M_{m_1m_2}M^*_{n_1n_2}
D^s_{m_2s'}(\hat{\mathbf k})
D^{s*}_{n_2s'}(\hat{\mathbf k})
\int \frac{d\Omega_{\mathbf k'}}{4\pi}
D^{s*}_{m_1 s''}(\hat{\mathbf k}')
D^s_{n_1 s''}(\hat{\mathbf k}')\cos\theta'
\end{align} 
By using the identity mentioned below,
\begin{equation}
\int \frac{d\Omega}{4\pi}
D^{s*}_{m s'} D^{s}_{n s'} \cos\theta
=
\frac{m\, s'}{s(s+1)(2s+1)}\,\delta_{mn}.
\end{equation}
\begin{align}
\int \frac{d\Omega_{\mathbf{k}'}}{4\pi}
\left|
\langle \mathbf{k}', s' | M | \mathbf{k}, s' \rangle
\right|^2
&=
 \frac{s'}{s(s+1)(2s+1)}\sum_{m_2,n_2}\sum_{m_1}
m_{1} M_{m_1m_2}M^*_{m_1n_2}
D^s_{m_2s'}(\hat{\mathbf k})
D^{s*}_{n_2s'}(\hat{\mathbf k})
\end{align}
The numerator can be written as
\begin{align}
\sum_{m_2,n_2}\sum_{m_1}
m_{1} M_{m_1m_2}M^*_{m_1n_2}
D^s_{m_2s'}(\hat{\mathbf k})
D^{s*}_{n_2s'}(\hat{\mathbf k})\nonumber
&=
\sum_{m_2,n_2}
\left[
\sum_{m_1}
m_1\, M_{m_1 m_2} M^*_{m_1 n_2}
\right]
D^s_{m_2 s'}(\hat{\mathbf k})
D^{s*}_{n_2 s'}(\hat{\mathbf k})\nonumber
\\
&=
\sum_{m_2,n_2}
\langle s,n_2 |
\mathcal{M}^\dagger
\left(
\sum_{m_1} m_1\, |s,m_1\rangle \langle s,m_1|
\right)
\mathcal{M}
| s,m_2 \rangle
D^s_{m_2 s'}(\hat{\mathbf k})
D^{s*}_{n_2 s'}(\hat{\mathbf k})\nonumber\\
&=
\sum_{m_2,n_2}
\langle s,n_2 | \mathcal{M}^\dagger S_z \mathcal{M} | s,m_2 \rangle
D^s_{m_2 s'}(\hat{\mathbf k})
D^{s*}_{n_2 s'}(\hat{\mathbf k})\nonumber\\
&=
\langle s,s' |
R^\dagger(\hat{\mathbf k})
\mathcal{M}^\dagger S_z \mathcal{M}
R(\hat{\mathbf k})
| s,s' \rangle \nonumber \\
&=
\langle \mathbf{k}, s' |
\mathcal{M}^\dagger S_z \mathcal{M}
| \mathbf{k}, s' \rangle,
\end{align}
Since the system is isotropic, the scattering time should not depend on the direction $\hat{\mathbf{k}}$ and hence we may choose $|\hat{\mathbf k},s'\rangle=|s,s'\rangle$, from Eq.~\ref{eq:eta_mu_general}
\begin{equation}
\mu_{s'}
=
\frac{s'}{s(s+1)}
\,
\frac{
\langle s,s'|\mathcal{M}^\dagger S_z \mathcal{M}|s,s'\rangle
}{
\langle s,s'|\mathcal{M}^\dagger \mathcal{M}|s,s'\rangle
}.
\end{equation}
We thus obtain
\begin{equation}
\langle s,s'|\mathcal{M}^\dagger S_z \mathcal{M}|s,s'\rangle
=
\sum_{L,Q}\sum_{L',Q'}
M_{LQ}^*\,M_{L'Q'}\,(-1)^Q\,
\langle s,s'|T^{(L)}_{-Q} S_z T^{(L')}_{Q'}|s,s'\rangle.
\label{eq:num}
\end{equation}
Since $|s,s'\rangle$ is an eigenstate of $S_z$, and for irreducible tensor operators,
\begin{equation}
[S_z, T^{(l')}_{Q'}] = Q'\,T^{(l')}_{Q'} \quad S_z |s,s'\rangle = s' |s,s'\rangle,
\label{eq:tensor_identity}
\end{equation}
By using Eq.~\ref{eq:tensor_identity}, we can write
\begin{align}
\langle s,s'|T^{(L)}_{-Q} S_z T^{(L')}_{Q'}|s,s'\rangle
&=
s'\,\langle s,s'|T^{(L)}_{-Q}T^{(L')}_{Q'}|s,s'\rangle
 + \langle s,s'|T^{(L)}_{-Q}\,[S_z,T^{(L')}_{Q'}]|s,s'\rangle,\nonumber
\\
\quad 
&= (s' + Q')\,
\langle s,s'|T^{(L)}_{-Q}T^{(L')}_{Q'}|s,s'\rangle.
\end{align}
Substituting back in Eq.~\ref{eq:num}, we obtain
\begin{equation}
\langle s,s'|\mathcal{M}^\dagger S_z \mathcal{M}|s,s'\rangle
=
\sum_{L,Q}\sum_{L',Q'}
M_{LQ}^*\,M_{L'Q'}\,(-1)^Q\,(s'+Q')\,
\langle s,s'|T^{(L)}_{-Q}T^{(L')}_{Q'}|s,s'\rangle
\end{equation}
Using the properties of irreducible tensor operators,
\begin{equation}
T^{(L')}_{Q'} |s,s'\rangle \propto |s,s'+Q'\rangle,
\end{equation}
and subsequently
\begin{equation}
T^{(L)}_{-Q} |s,s'+Q'\rangle \propto |s,s'+Q'-Q\rangle,
\end{equation}
the overlap with $\langle s,s'|$ is nonzero only if
\begin{equation}
s' = s' + Q' - Q \quad \Rightarrow \quad Q'=Q.
\end{equation}
Thus, only equal-$Q$ sectors contribute, and we obtain
\begin{equation}
\langle s,s'|\mathcal{M}^\dagger S_z \mathcal{M}|s,s'\rangle
=
\sum_{L,L'}\sum_Q
M_{LQ}^* M_{L'Q} (-1)^Q (s'+Q)\,
\langle s,s'|T^{(L)}_{-Q}T^{(L')}_{Q}|s,s'\rangle
\end{equation}
Similarly as in Eq.~\ref{eq:slection}, inserting a complete set of $|s,m\rangle$ states,
\begin{equation}
\langle s,s'|T^{(L)}_{-Q}T^{(L')}_{Q}|s,s'\rangle
=
\sum_{m}
\langle s,s'|T^{(L)}_{-Q}|s,m\rangle
\langle s,m|T^{(L')}_{Q}|s,s'\rangle.
\end{equation}
From the selection rule, $m = s'+Q$ contributes, so
\begin{equation}
\langle s,s'|T^{(L)}_{-Q}T^{(L')}_{Q}|s,s'\rangle
=
\langle s,s'|T^{(L)}_{-Q}|s,s'+Q\rangle
\langle s,s'+Q|T^{(L')}_{Q}|s,s'\rangle.
\end{equation}
Substituting back, we obtain
\begin{equation}
\langle s,s'|\mathcal{M}^\dagger S_z \mathcal{M}|s,s'\rangle
=
\sum_Q (s'+Q)\,
\left|
\sum_L M_{LQ}\,
\langle s,s'+Q|T^{(L)}_Q|s,s'\rangle
\right|^2
\end{equation}
and,
\begin{equation}
\langle s,s'|\mathcal{M}^\dagger \mathcal{M}|s,s'\rangle
=
\sum_Q
\left|
\sum_L M_{LQ}\,
\langle s,s'+Q|T^{(L)}_Q|s,s'\rangle
\right|^2
\end{equation}
From Eq.~\ref{eq:ASQ}
\begin{equation}
\mathcal A_{s'}(Q)
=
\sum_L \sqrt{2L+1}\,M_{LQ}\,
C^{\,s,\;s'+Q}_{\,s,\;s',\;L,\;Q},
\end{equation}
Then
\begin{align}
\langle s,s'|\mathcal{M}^\dagger \mathcal{M}|s,s'\rangle
&= \sum_Q |\mathcal A_{s'}(Q)|^2,\\
\langle s,s'|\mathcal{M}^\dagger S_z \mathcal{M}|s,s'\rangle
&= \sum_Q (s'+Q)\,|\mathcal A_{s'}(Q)|^2.
\end{align}
Thus,
\begin{equation}
\mu_{s'}
=
\frac{s'}{s(s+1)}
\,
\frac{
\sum_Q (s'+Q)\,|\mathcal A_{s'}(Q)|^2
}{
\sum_Q |\mathcal  A_{s'}(Q)|^2
}
\end{equation}
Define
\begin{equation}
 P_{s'}(Q)
=
\frac{|\mathcal A_{s'}(Q)|^2}{\sum_{Q'} |\mathcal A_{s'}(Q)|^2},
\qquad
\sum_Q  P_{s'}(Q) = 1.
\end{equation}
Then
\begin{equation}
\mu_{s'}
=
\frac{s'}{s(s+1)}
\left(
s' + \sum_Q Q\, P_{s'}(Q)
\right)
\end{equation}
%%%%%%%%%%%%%%%%%%%%%%%%%%%%%%%%%%%%%%%%%%%%%%%%%%%%%%%%%%%%%%
We may now apply the above result to some specific cases. 

\textit{Case:} 1 For scalar disorder, only the $L=0$, $Q=0$ component is present
\begin{equation}
M_{00} \neq 0, \qquad M_{LQ}=0 \ \text{otherwise}.
\end{equation}
Thus,
\begin{equation}
\mathcal A_{s'}(Q)
=
\delta_{Q,0}\,M_{00}\,
C^{\,s,\,s'}_{\,s,\,s',\,0,\,0}.
\end{equation}
Using $
C^{\,s,\,s'}_{\,s,\,s',\,0,\,0} = 1$,
we obtain
\begin{equation}
\mathcal A_{s'}(0)=M_{00}=1, \qquad \mathcal A_{s'}(Q\neq 0)=0.
\end{equation}
then
\begin{equation}
\mu_{s'}
=
\frac{s'^2}{s(s+1)} \qquad 
\eta_{s'}
=
\frac{1}{1-\frac{s'^2}{s(s+1)}},
\end{equation}
\textit{Case:} 2 For arbitrary spin $s$, the operator $S_z$ can be written as a rank-1 irreducible tensor:
\begin{equation}
S_z = \sqrt{s(s+1)}\, T^{(1)}_0.
\end{equation}
Thus, the only nonzero tensor coefficient is
\begin{equation}
M_{1,0} = \sqrt{s(s+1)},
\qquad
M_{LQ}=0 \quad \text{otherwise}.
\end{equation}
The quantity $\mathcal A_{s'}(Q)$ is defined as
\begin{equation}
\mathcal A_{s'}(Q)
=
\sum_L \sqrt{2L+1}\,M_{LQ}\,
C^{\,s,\;s'+Q}_{\,s,\;s',\;L,\;Q}.
\end{equation}
Since only $L=1$, $Q=0$ contributes, we obtain
\begin{equation}
\mathcal A_{s'}(0)
=
\sqrt{3}\,\sqrt{s(s+1)}\,
C^{\,s,\;s'}_{\,s,\;s',\;1,\;0}.
\end{equation}
Using the general Clebsch-Gordan identity
\begin{equation}
C^{\,s,m}_{\,s,m;\,1,0}
=
\frac{m}{\sqrt{s(s+1)}},
\end{equation}
we find
\begin{equation}
\mathcal A_{s'}(0)
=
\sqrt{3}\, s'.
\end{equation}
Since only $Q=0$ contributes, the denominator becomes
\begin{equation}
\sum_Q |\mathcal A_{s'}(Q)|^2
=
3\, s'^2,
\end{equation}
and the numerator is
\begin{equation}
\sum_Q (s'+Q)\,|\mathcal A_{s'}(Q)|^2
=
s' \cdot 3\, s'^2
=
3\, s'^3.
\end{equation}
The transport factor is given by
\begin{equation}
\mu_{s'}
=
\frac{s'}{s(s+1)}
\,
\frac{
\sum_Q (s'+Q)\,|\mathcal A_{s'}(Q)|^2
}{
\sum_Q |\mathcal A_{s'}(Q)|^2
}.
\end{equation}
Substituting the above expressions, we obtain
\begin{equation}
\mu_{s'}
=
\frac{s'}{s(s+1)}
\cdot
\frac{3 s'^3}{3 s'^2}
=
\frac{s'^2}{s(s+1)}
\end{equation}

\subsection{Conductivity correction from quantum interference for general pseudospin-\texorpdfstring{$s$}{s} for scalar disorder}\label{appen.sect-2}
Here we will assume the presence of only scalar disorder($\mathcal{M}=\mathcal{I}$), and only consider a single band. In the next section, we will discuss the general interband-intervalley case as well. The bare Cooperon, an overlap of the time-reversed trajectory, is defined as 
\begin{align}\label{A14}
     \Gamma^{0}_{{\kn_1}{\kn_2}} &= \langle U_{\kn_{1},\kn_{2}}U_{-\kn_{1},-\kn_{2}}\rangle\\
      \Gamma^{0}_{{\kn_1}{\kn_2}} &= {n_0 u_0^2}\sum_{m,n,p} C^0_\mathrm{mnp}e^{i(m\theta_1 + n\theta_2)}   e^{p(i\phi_1 + i\phi_2)}\nonumber
\end{align}
the bare Cooperon, has an angular structure \(\Gamma^{0}_{{\kn_1}{\kn_2}}\propto e^{p(i\phi_1 - i\phi_2))} \), where $p \in ({0,1,....,4s+1})$, $s$ is the pseudospin of particle
. There are several channels for backscattering. Now the time evolution of the Cooperon obeys a Bethe–Salpeter (BSE) equation analogous to that of the vertex function but in the particle–particle channel.

In polar coordinates, the Bethe-Salpeter equation
\begin{align}\label{ABSE}
     \Gamma_{{\kn_1}{\kn_2}} &=  \Gamma^{0}_{{\kn_1}{\kn_2}} + \int_0^{2\pi} \frac{d\phi}{2\pi} \int_0^\pi \frac{\sin \theta \, d\theta}{2\pi} \int_0^\infty \frac{k^2 \, dk}{2\pi} \Gamma^{0}_{{\kn_1}{\kn}}G^R_{{\kn}}G^A_{\mathbf{q}-{\kn}} \Gamma_{{\kn}{\kn_2}}
\end{align}
where $\Gamma_{{\kn_1}{\kn_2}}$ is the vertex function, which we have to find; initially, we take an ansatz for a vertex function.
\begin{align}\label{agamma}
     \Gamma_{{\kn_1}{\kn_2}} &= n_0 u_0^2\sum_{m,n,p} C_\mathrm{mnp}e^{i(m\theta_1 + n\theta_2)}   e^{p(i\phi_1 + i\phi_2)}
\end{align}
the $k$ integration is independent of $\Gamma_{{\kn_1}{\kn_2}}$ and $\Gamma^{0}_{{\kn_1}{\kn_2}}$ because they are the function of $\theta$ and $\phi$, so we can evaluate $k$ integration by contour integration method and becomes
\begin{equation}
    \int_0^\infty \frac{k^2 \, dk}{2\pi} G^R_{{\kn}}G^A_{\mathbf{q}-{\kn}} = \frac{2\pi^2 N_F \tau}{\hbar}[1 - i\tau v_F q\cos{\theta} - \tau^2 v_F^2 q^2\cos^2{\theta}]
\end{equation}
Put back the value of $k$ integration in Eq.~\ref{ABSE} 
\begin{align*}
    n_0 u_0^2\sum_{m,n,p} \Gamma_\mathrm{mnp}e^{i(m\theta_1 + n\theta_2)}   e^{p(i\phi_1 + i\phi_2)} &=  n_0 u_0^2\sum_{m,n,p} \Gamma^0_\mathrm{mnp}e^{i(m\theta_1 + n\theta_2)}   e^{p(i\phi_1 + i\phi_2)} + \frac{2\pi^2 N_F \tau}{\hbar} (n_0u_0^2)^2
     \\
    &\quad \intsh  \Bigg[ 1 - i\tau v_F q \cos{\theta} - \tau^2 v_F^2 q^2 \cos^2{\theta} \Bigg] \\
    &\quad \sum_{i,j,k} \Gamma^0_{\mathrm{ijk}} e^{i(i\theta_1 + j\phi_1 + k\theta + j\phi)}  \sum_{r,s,p} \Gamma_\mathrm{rsp} e^{i(r\theta + s\phi + p\theta_2 + s\phi_2)}
\end{align*}
\begin{align}\label{BSEmatrix}
   \sum_{m,n,p} \Gamma_\mathrm{mnp}e^{i(m\theta_1 + n\theta_2)}   e^{p(i\phi_1 + i\phi_2)} &=  \sum_{m,n,p} \Gamma^0_\mathrm{mnp}e^{i(m\theta_1 + n\theta_2)}   e^{p(i\phi_1 + i\phi_2)} + \frac{2\pi^2 N_F \tau}{\hbar} n_0u_0^2\nonumber
     \\
    &\quad \intsh  \Bigg[ 1 - i\tau v_F q \cos{\theta} - \tau^2 v_F^2 q^2 \cos^2{\theta} \Bigg] \nonumber\\
    &\quad \sum_{i,j,k} \Gamma^0_{\mathrm{ijk}} e^{i(i\theta_1 + j\phi_1 + k\theta + j\phi)}  \sum_{r,s,q} \Gamma_\mathrm{rsq} e^{i(r\theta + s\phi + q\theta_2 + s\phi_2)}
\end{align}
By using the identity $ \int_0^{2\pi} d\phi e^{i(j\phi + s\phi)} = 2\pi\delta_{j,s}$, the Eq.~\ref{BSEmatrix} becomes 
\begin{align}
   \sum_{m,n,p} \Gamma_\mathrm{mnp}e^{i(m\theta_1 + n\theta_2)}   e^{p(i\phi_1 + i\phi_2)} &=  \sum_{m,n,p} \Gamma^0_\mathrm{mnp}e^{i(m\theta_1 + n\theta_2)}   e^{p(i\phi_1 + i\phi_2)} + \frac{2\pi^2 N_F \tau}{\hbar} n_0u_0^2\nonumber
     \\
    &\quad \int_0^{\pi} \frac{\sin \theta d\theta}{2\pi} \Bigg[ 1 - i\tau v_F q \cos{\theta} - \tau^2 v_F^2 q^2 \cos^2{\theta} \Bigg] \nonumber\\
   &\quad \sum_{i,j,k,r,q} \Gamma^0_{\mathrm{ijk}} e^{i(i\theta_1 + j\phi_1 + k\theta )} \Gamma_\mathrm{rjq} e^{i(r\theta  + q\theta_2 + j\phi_2)}
\end{align}
\begin{align}
   \sum_{m,n,p} \Gamma_\mathrm{mnp}e^{i(m\theta_1 + n\theta_2)}   e^{p(i\phi_1 + i\phi_2)} &=  \sum_{m,n,p} \Gamma^0_\mathrm{mnp}e^{i(m\theta_1 + n\theta_2)}   e^{p(i\phi_1 + i\phi_2)} + \frac{2\pi^2 N_F \tau}{\hbar} n_0u_0^2\nonumber
     \\
    &\quad  \sum_{k,r}\int_0^{\pi}\frac{\sin\theta d\theta}{2\pi} \Bigg[ 1 - i\tau v_F q \cos{\theta} - \tau^2 v_F^2 q^2 \cos^2{\theta} \Bigg](e^{i(k\theta + r\theta)}) \nonumber\\
   &\quad \sum_{i,j,q} \Gamma^0_{\mathrm{ijk}}\Gamma_\mathrm{rjq} e^{i(i\theta_1 + j\phi_1 )}  e^{i( q\theta_2 + j\phi_2)}
\end{align}
For every channel $p$, we find a coupled equation. We can express these coupled equations in matrix form 
\begin{equation}
     \Bigg[\Gamma\Bigg] =  
     \Bigg[\Gamma^0\Bigg] +\Bigg[\Gamma^0\Bigg]\Bigg[\Phi\Bigg]\Bigg[\Gamma\Bigg]
\end{equation}
where $\Phi$-matrix is whose elements are defined as 
\begin{equation}
    \phi_\mathrm{k,r} = \int_0^{\pi}\frac{\sin\theta d\theta}{2\pi}\Bigg[ 1 - i\tau v_F q \cos{\theta} - \tau^2 v_F^2 q^2 \cos^2{\theta} \Bigg]e^{i(r+k)\theta }
\end{equation}
 
After solving the BSE, we get a value of the coefficient \(C_\mathrm{mnp}\) for each channel in the form 
\begin{equation}\label{answergamma}
    C_\mathrm{mnp} = \frac{1}{q^2 + Q^2_{mnp}}
\end{equation}
where it \(Q_{mnp}^2\) is non-zero except for the case when \(p= 2s, m=n=0\). Only one \(C_\mathrm{m=n=0, p = 2s}\) channel is divergent in the diffusive limit \({q}\rightarrow0\), and the full Cooperon acquires the form for spin-\(s\) with helicity \(s'\),  
\begin{equation}
\Gamma(\mathbf q) =
 \frac{e^{p(i\phi_1 - i\phi_2)}}{D_{s'} q^2 }
 ,
\label{eq:ACooperon_diff}
\end{equation}
where\[
D_{s'} = \frac{v^2\tau_{s'}\eta_{s'}}{3}.
\]The conductivity contribution from the intraband-intravalley cooperons is given by 
\begin{equation}\label{Aconduccorr}
    \sigma^{qi}_0 = \sigma_{a1} + 2\sigma_{a2}, 
\end{equation}
where
\begin{equation}
    \sigma_{\mathrm{a_1}} = \frac{e^2 \hbar}{2\pi} \sum_{\mathbf{q}} \Gamma_{\mathbf{k}, \mathbf{q}-\mathbf{k}} \sum_{\mathbf{k}} G_{\mathbf{k}}^R \tilde{v}_{\mathbf{k}}^x G_{\mathbf{k}}^A G_{\mathbf{q}-\mathbf{k}}^R \tilde{v}_{\mathbf{q}-\mathbf{k}}^x G_{\mathbf{q}-\mathbf{k}}^A,\nonumber
\end{equation}
and
\begin{equation}
    \sigma_{\mathrm{a_2}} = \frac{e^2 \hbar}{2\pi} \sum_{\mathbf{q}} \Gamma_{\mathbf{k_1}, \mathbf{q}-\mathbf{k_1}} \sum_{\mathbf{k_1}} \tilde{v}_{\mathbf{k_1}}^x \tilde{v}_{\mathbf{q}-\mathbf{k_1}}^x G_{\kn}^R G_{\mathbf{k_1}}^R G_{\mathbf{q}-\mathbf{k}}^R G_{\mathbf{q}-\mathbf{k_1}}^R G_{\mathbf{k}}^A G_{\mathbf{q}-\mathbf{k_1}}^A \langle U_{\mathbf{k},\mathbf{k_1}} U_{\mathbf{q}-\mathbf{k},\mathbf{q}-\mathbf{k_1}} \rangle.\nonumber
\end{equation}
The $\sigma_{\mathrm{a_1}}$ is calculated as 
\begin{equation}
    \sigma_{\mathrm{a_1}} = \frac{e^2\hbar}{2\pi}\sum_\mathbf{q}\Gamma_{\kn,\mathbf{q}-\kn}\sum_\kn G^R_\kn\widetilde{v}^x_\kn G^A_\kn G^R_{\mathbf{q}-\kn}\widetilde{v}^x_{\mathbf{q}-\kn}G^A_{\mathbf{q}-\kn}
\end{equation}
In the limit $\mathbf{q}\to 0$
\begin{equation}
     \Gamma_{\kn,\mathbf{q}-\kn} =\Gamma_{\kn,-\kn}
\end{equation}
 As we find the $\Gamma_{\kn_1\kn_2}$ in Eq.~\ref{eq:ACooperon_diff}
\begin{equation}
    \Gamma_{\kn_1\kn_2} = \frac{1}{D_{s'}q^2} e^{ip(\phi_2 - \phi_1)}\nonumber
\end{equation}
\begin{equation}
   \Gamma_{\kn, -\kn} = \frac{1}{D_{s'}q^2} (-1)^{2s}
\end{equation}
Now, we called  $\Gamma_{\kn, -\kn}$= -$\gamma(q)$  
\begin{align}
    \sigma_{\mathrm{a_1}} &= (-1)^{2s}\frac{e^2\hbar}{2\pi}\sum_\mathbf{q} \gamma(q)\sum_\kn G^R_\kn\widetilde{v}^x_\kn G^A_\kn G^R_{\mathbf{q}-\kn}\widetilde{v}^x_{\mathbf{q}-\kn}G^A_{\mathbf{q}-\kn} \nonumber \\
    &= (-1)^{2s}\frac{e^2\hbar}{2\pi}\sum_\mathbf{q} \gamma(q)\isum  G^R_\kn\widetilde{v}^x_\kn G^A_\kn G^R_{\mathbf{q}-\kn}\widetilde{v}^x_{\mathbf{q}-\kn}G^A_{\mathbf{q}-\kn} \nonumber \\
    &= (-1)^{2s}\frac{e^2\hbar}{2\pi}\sum_\mathbf{q} \gamma(q)\intsh \widetilde{v}^x_\kn \widetilde{v}^x_{\mathbf{q}-\kn}\int_0^\infty \frac{k^2 dk}{2\pi}G^R_\kn G^A_\kn G^R_{\mathbf{q}-\kn}G^A_{\mathbf{q}-\kn} \label{Asigmaa1}
\end{align}
From Eq.\ref{eq:Ebands}
\begin{align*}
     E_\kn &= s' \hbar \vartheta k\\
    dE_\kn &=  s' \hbar \vartheta dk
\end{align*}
So, first, we solve the $k$ integration 
\begin{align*}
   \int_0^\infty\frac{k^2 \, dk}{2\pi} G^R_\kn G^A_\kn G^R_{\mathbf{q}-\kn}G^A_{\mathbf{q}-\kn} &= \int_0^\infty\frac{k^2 \, dk}{2\pi}G^R_\kn G^A_\kn G^R_{\mathbf{q}-\kn}G^A_{\mathbf{q}-\kn}\\
   &= \frac{1}{2\pi (s' \hbar \vartheta)^3}\int_0^\infty E^2 dE G^R_\kn G^A_\kn G^R_{\mathbf{q}-\kn}G^A_{\mathbf{q}-\kn} \\
   &= \frac{E^2_F}{2\pi (s' \hbar \vartheta)^3}\int_{-\infty}^\infty  dE G^R_\kn G^A_\kn G^R_{\mathbf{q}-\kn}G^A_{\mathbf{q}-\kn} \\
\end{align*}
By using the relation between the Fermi energy and the density of states (DOS) at the Fermi level mentioned in Eq.~\ref{ADOS}
\begin{align*}
    \int_0^\infty\frac{k^2 \, dk}{2\pi} G^R_\kn G^A_\kn G^R_{\mathbf{q}-\kn}G^A_{\mathbf{q}-\kn}&=\pi N_F\int_{-\infty}^\infty  dE G^R_\kn G^A_\kn G^R_{\mathbf{q}-\kn}G^A_{\mathbf{q}-\kn}\\
    &= \pi N_F\int_{-\infty}^\infty  dE \frac{1}{\omega - \epsilon_k  + i\hbar/2\tau}\frac{1}{\omega - \epsilon_k -i\hbar/2\tau} \\
    &\quad\times   \frac{1}{\omega - \epsilon_k  +\textbf{Q}+ i\hbar/2\tau}\frac{1}{\omega - \epsilon_k+ \textbf{Q} -i\hbar/2\tau} 
\end{align*}
where $\textbf{Q} = \hbar {\mathbf{v}_F\cdot \mathbf{q}}$
\begin{equation}
    \pi N_F\int_{-\infty}^\infty  dE G^R_\kn G^A_\kn G^R_{\mathbf{q}-\kn}G^A_{\mathbf{q}-\kn} =  \pi N_F\frac{4\pi\tau^3}{\textbf{Q}^2\tau^2\hbar + \hbar^3}\nonumber
\end{equation}
As $\mathbf{q}\to 0$ $\implies$ $\mathbf{Q}\to 0$,
\begin{equation}
    \pi N_F\int_{-\infty}^\infty  dEG^R_\kn G^A_\kn G^R_{\mathbf{q}-\kn}G^A_{\mathbf{q}-\kn} =   \frac{4\pi^2 N_F\tau^3}{  \hbar^3}
\end{equation}
Put the value of $k$ integration in Eq.~\ref{Asigmaa1}
\begin{align}
    \sigma_{\mathrm{a_1}}
    &= (-1)^{2s}\frac{e^2\hbar}{2\pi}\sum_\mathbf{q} \gamma(q)\intsh \widetilde{v}^x_\kn \widetilde{v}^x_{\mathbf{q}-\kn}\int_0^\infty \frac{k^2 dk}{2\pi}G^R_\kn G^A_\kn G^R_{\mathbf{q}-\kn}G^A_{\mathbf{q}-\kn}\nonumber\\
    &= (-1)^{2s}\frac{e^2\hbar}{2\pi}\frac{4\pi^2 N_F\tau^3}{  \hbar^3}\sum_\mathbf{q} \gamma(q)\intsh \widetilde{v}^x_\kn \widetilde{v}^x_{\mathbf{q}-\kn}\nonumber\\
    &=(-1)^{2s}\frac{e^2\hbar}{2\pi}\frac{4\pi^2 N_F\tau^3}{  \hbar^3}\sum_\mathbf{q} \gamma(q)\intsh \eta v_\kn^x \eta v_{\mathbf{q}-\kn}^x\nonumber
\end{align}
We are performing integration in polar coordinates, so the velocity in polar coordinates is 
\begin{align}
    v_\kn^x &= v_F\sin{\theta}\cos{\phi}\nonumber\\
    {v}_{-\kn}^x &= - v_F\sin{\theta}\cos{\phi}\nonumber
\end{align}
put the values of $v_\kn^x$ and ${v}_{\mathbf{q}-\kn}^x$ and integrate over $\theta$ and $\phi$
\begin{align*}
\sigma_{\mathrm{a_1}}
&= (-1)^{2s}\frac{e^2 \hbar}{2\pi}\frac{4\pi^2 N_F \tau^3}{\hbar^3}
\sum_{\mathbf{q}} \gamma(q)
\int d\Omega \, \eta \, v_{\mathbf{k}}^x \, \eta \, v_{\mathbf{q}-\mathbf{k}}^x \\
&= (-1)^{2s+1}\frac{e^2 \hbar}{2\pi}\frac{4\pi^2 N_F \tau^3}{\hbar^3}
\sum_{\mathbf{q}} \gamma(q)
\int d\Omega \, \eta^2 \, v_F^2 \sin\theta \cos\phi \, \sin\theta \cos\phi \\
&= (-1)^{2s+1}\frac{e^2 \hbar}{2\pi}\frac{4\pi^2 N_F \tau^3}{\hbar^3}
\eta^2 v_F^2
\sum_{\mathbf{q}} \gamma(q)
\int d\Omega \, \sin^2\theta \cos^2\phi
\end{align*}
\begin{equation}
   \intsh \sin{\theta}\cos{\phi}  \sin{\theta}\cos{\phi} = \frac{ 1}{3\pi}\nonumber 
\end{equation}
\begin{equation}
   \sigma_{\mathrm{a_1}} = (-1)^{2s+1} \frac{e^2\hbar}{2\pi}\frac{4\pi^2 N_F\tau^3}{  \hbar^3}\eta^2 v_F^2\frac{1}{3\pi}\sum_\q \gamma(q) \nonumber
\end{equation}
\begin{equation}
   {\sigma_{\mathrm{a_1}} =(-1)^{2s+1} \frac{2\eta^2 v_F^2 N_F\tau^3 e^2}{ 3 \hbar^2}\sum_\q \gamma(q)} 
\end{equation}
The $\mathbf{\sigma}_{a_2}$ is calculated as 
\begin{equation}
\sigma_{\mathrm{a_2}} = \frac{e^2 \hbar}{2 \pi} \sum_{\mathbf{q}} \Gamma_{\mathbf{k_1}, \mathbf{q} - \mathbf{k}} \sum_{\mathbf{k}} \sum_{\mathbf{k_1}} \tilde{v}_{\mathbf{k}}^x \tilde{v}_{\mathbf{q} - \mathbf{k_1}}^x G^R_{\mathbf{k}} G^R_{\mathbf{k_1}}G^A_{\mathbf{k}}  G^R_{\mathbf{q} - \mathbf{k}} 
 G^R_{\mathbf{q} - \mathbf{k_1}} G^A_{\mathbf{q} - \mathbf{k_1}} \left\langle U_{\mathbf{k}, \mathbf{k_1}} U_{\mathbf{q} - \mathbf{k}, \mathbf{q} - \mathbf{k_1}} \right\rangle \nonumber .
\end{equation}
The $\left\langle U_{\mathbf{k}, \mathbf{k_1}} U_{\mathbf{q} - \mathbf{k}, \mathbf{q} - \mathbf{k_1}} \right\rangle $ is independent of $k$, by contour integration method we can evaluate $k$ integration as,
\begin{align}
  \int_0^\infty \frac{k^2 dk}{2\pi} G^R_{\mathbf{k}} G^A_{\mathbf{k}} G^R_{\mathbf{q}-\mathbf{k}} &=\pi N_F \int_{-\infty}^{\infty} d\epsilon_{\mathbf{k}} \frac{1}{\left( \omega - \epsilon_{\mathbf{k}} + \frac{i \hbar}{2\tau} \right)} \frac{1}{\left( \omega - \epsilon_{\mathbf{k}} - \frac{i \hbar}{2\tau} \right)} \frac{1}{\left( \omega - \epsilon_{\mathbf{k}} + \hbar \mathbf{v_F} \cdot \mathbf{q} + \frac{i \hbar}{2\tau} \right)}\nonumber \\
  \int_0^\infty \frac{k^2 dk}{2\pi} G^R_{\mathbf{k}} G^A_{\mathbf{k}} G^R_{\mathbf{q}-\mathbf{k}} &=  \frac{2\pi^2 N_F \tau^2}{ i\hbar^2} \nonumber
\end{align}
Similarly,
\begin{align}
    \int_0^\infty \frac{k_{1}^2 dk_1}{2\pi} G^R_{\mathbf{k_1}} G^R_{\mathbf{q} -\mathbf{k_1}} G^A_{\mathbf{q}-\mathbf{k_1}} &=  \frac{2\pi^2 N_F \tau^2}{ i\hbar^2} \nonumber
\end{align}
put the value of $k$-integration,
\begin{equation}
\sigma_{\mathrm{a_2}} = -\frac{e^2 \hbar}{2 \pi}  \frac{4\pi^4 N_F^2 \tau^4}{ \hbar^4}\sum_{\mathbf{q}} \Gamma_{\mathbf{k_1}, \mathbf{q} - \mathbf{k}} \intsh \tilde{v}_{\mathbf{k}}^x \tilde{v}_{\mathbf{q} - \mathbf{k_1}}^x  
\intshone \left\langle U_{\mathbf{k}, \mathbf{k_1}} U_{\mathbf{q} - \mathbf{k}, \mathbf{q} - \mathbf{k_1}} \right\rangle.
\end{equation}
We are performing integration in polar coordinates, so the velocity in polar coordinates is 
\begin{align*}
    v_\kn^x &= v_F\sin{\theta}\cos{\phi}\\
    {v}_{\q-\kn}^x &=  -v_F\sin{\theta}\cos{\phi}
\end{align*}
Put the value of $\widetilde{v}_\kn^x$ and $\widetilde{v}_{\q-\kn}^x$ and perform integration 
\begin{equation}
\sigma_{\mathrm{a_2}} = \frac{e^2 \hbar}{2 \pi} \frac{4\pi^4 N_F^2 \tau^4}{ \hbar^4}\sum_{\mathbf{q}} \Gamma_{\mathbf{k_1}, \mathbf{q} - \mathbf{k}} \intsh \eta^2 v_F^2\sin^2{\theta}\cos^2{\phi}  
 \intshone \left\langle U_{\mathbf{k}, \mathbf{k_1}} U_{\mathbf{q} - \mathbf{k}, \mathbf{q} - \mathbf{k_1}} \right\rangle.
\end{equation}
\begin{equation}\label{A15}
\sigma_{\mathrm{a_2}} = \frac{\eta^2 v_F^2 e^2 \hbar}{2 \pi} \frac{4\pi^4 N_F^2 \tau^4}{ \hbar^4}\sum_{\mathbf{q}} \Gamma_{\mathbf{k_1}, \mathbf{q} - \mathbf{k}} \intsh \sin^2{\theta}\cos^2{\phi}    \intshone \left\langle U_{\mathbf{k}, \mathbf{k_1}} U_{\mathbf{q} - \mathbf{k}, \mathbf{q} - \mathbf{k_1}} \right\rangle.
\end{equation}
from Eq.~\ref{eq:ACooperon_diff}
\begin{equation}
    \Gamma_{\kn_1\kn_2} = \frac{1}{D_{s'} q^2 } e^{2si(\phi_2 - \phi_1)}\nonumber
\end{equation}
In the limit $\mathbf{q}\to 0$
\begin{align}
    \Gamma_{\mathbf{k_1}, \mathbf{q} - \mathbf{k}} &= \Gamma_{\mathbf{k_1},  -\mathbf{k}}\nonumber
\end{align}
 Since $e^{2si(\pi +\phi -\phi_1)} = (-1)^{2s}e^{2si(\phi -\phi_1)}$ 
 \begin{align*}
    \Gamma_{\mathbf{k_1}, -\mathbf{k}} 
    &=  \gamma(q)e^{2si(\pi +\phi -\phi_1)}\\
    &=  (-1)^{2s}\gamma(q)e^{2si(\phi -\phi_1)}
\end{align*}
In the limit $\mathbf{q}\to 0$  
\begin{align*}
 \left\langle U_{\mathbf{k}, \mathbf{k_1}} U_{\mathbf{q} - \mathbf{k}, \mathbf{q} - \mathbf{k_1}} \right\rangle &= \left\langle U_{\mathbf{k}, \mathbf{k_1}} U_{ - \mathbf{k},- \mathbf{k_1}} \right\rangle  = \Gamma^{0}_{\kn ,\kn_1} 
\end{align*}
We define $\Gamma^{0}_{\kn ,\kn_1}$ in Eq.~\ref{A14},
\begin{equation}
  \Gamma^{0}_{{\mathbf{k}_1}{\mathbf{k}_2}} =\frac{(2s+1)\hbar}{2\pi N_F \tau}\sum_{r,s,p} \Gamma^{0}_\mathrm{rsp}e^{i(r\theta_1 + p\phi_1 + s\theta_2 + p\phi_2)}  
\end{equation}
Put the value of $\Gamma^{0}_{{\kn_1}{\kn_2}}$ and $ \Gamma_{\mathbf{k_1}, -\mathbf{k}}$ in Eq.~\ref{A15}, 
\begin{equation}
 I_{a2} = \intsh \sin^2{\theta}\cos^2{\phi}  
 \intshone \left\langle U_{\mathbf{k}, \mathbf{k_1}} U_{\mathbf{q} - \mathbf{k}, \mathbf{q} - \mathbf{k_1}} \right\rangle e^{2si(\phi -\phi_1)}
\end{equation}
we get
\begin{equation}
\sigma_{\mathrm{a_2}} = (-1)^{2s}\frac{\eta^2 v_F^2 e^2 \hbar}{2 \pi} \frac{4\pi^4 N_F^2 \tau^4}{ \hbar^4}  (I_{a2})\sum_{\mathbf{q}} \gamma(q)
\end{equation}  
The total conductivity correction from intraband-intravalley Cooperon is given by 
\begin{equation}
  \sigma_0^{qi} =\mathbf{ \sigma}_{\mathrm{a_1}} +2\mathbf{\sigma}_{\mathrm{a_2}} \nonumber.
\end{equation}
%%%%%%%%%%%%%%%%%%%%%%%%%%%%%%%%%%%%%%%%%%%%%%%%%%%%%%%%%%%%%%
\subsection{Explicit results for $s = \frac{1}{2}$, and $s=1$    }\label{Sec_App_A_5}
%%%%%%%%%%%%%%%%%%%%%%%%%%%%%%%%%%%%%%%%%%%%%%%%%%%%%%%%%%%%%%

To demonstrate the general framework developed above, we now evaluate
the localization corrections explicitly for pseudospin–$\tfrac12$, pseudospin–1 fermions, and pseudospin–$\tfrac32$.  These three cases
represent the canonical Dirac–Weyl triply degenerate (Maxwell fermion or pseudospin-1) semimetals~\cite {zhu2017emergent}, and serve as the first members of the pseudospin sequence.

\subsection*{A. Pseudospin-\texorpdfstring{$\tfrac12$}{1/2}: Dirac and Weyl fermions}

For $s=\tfrac12$, the low-energy Hamiltonian
\(H=\hbar v\,\mathbf S\!\cdot\!\mathbf k
 =\hbar v\,\bm\sigma\!\cdot\!\mathbf k/2\)
produces two bands $s'=\pm\tfrac12$.  The corresponding spinors
are
\[
\ket{\mathbf k,\pm} =
 \begin{pmatrix}
  \cos\frac{\theta}{2}\\
  \pm e^{i\phi}\sin\frac{\theta}{2}
 \end{pmatrix},
\]
with $|d^{1/2}_{1/2,1/2}(\theta)|^2=\cos^2(\theta/2)$.
Using Eq.~(\ref{eq:tau_result}), we obtain
\[
\frac{\tau^{\mathrm{tr}}}{\tau}
 = \frac23,
\qquad
\eta_{1/2,1/2} = \frac{3}{2}.
\]
The diffusion constant and Drude conductivity follow as
\[
D_{\tfrac12} = \frac{v^2\tau_{\tfrac12}\eta_{\tfrac12}}{d},\qquad
\sigma_0 = e^2 N_F D_{\tfrac12},
\]
in agreement with the known Dirac results ($d$ denoting the dimensionality).~\cite{lu2015weak}

Because $\mathcal T^2=-1$, the system belongs to the symplectic class
and exhibits weak antilocalization (WAL).  The quantum interference correction for one valley of Weyl fermions takes the form
\begin{equation}
\sigma^{\mathrm{qi}} = \frac{e^2}{h}\frac{1}{\pi^2}\left(\frac{1}{\ell}-\frac{1}{\ell_\phi}\right),
\label{eq:s12_ds}
\end{equation}
with the positive sign reflecting destructive interference between
time-reversed trajectories.
In the presence of a perpendicular magnetic field, the familiar Hikami–Larkin–Nagaoka formula yields a positive magnetoconductivity
cusp characterized by $\alpha_{1/2}=+1/2$.~\cite{lu2015weak}
\subsection*{B. Pseudospin-1: Triply degenerate fermions}
For $s=1$, the Hamiltonian
\(H=\hbar v\,\mathbf S\!\cdot\!\mathbf k\)
has three bands: $s'=0,\pm1$.
The eigenvectors are the spin-1 rotation matrices
\(\ket{\mathbf k,s'} = D^{1}_{m s'}(\phi,\theta,0)\ket{m}\),
with overlaps
\(|d^1_{1,1}(\theta)|^2=\tfrac14(1+\cos\theta)^2\),
\(|d^1_{0,0}(\theta)|^2=\tfrac12\sin^2\theta\),
and
\(|d^1_{-1,-1}(\theta)|^2=\tfrac14(1-\cos\theta)^2\).
From Eq.~(\ref{eq:tau_result}) one finds
\[
\frac{\tau^{\mathrm{tr}}_{\pm1}}{\tau_{\pm1}}=\frac12,
\qquad
\frac{\tau^{\mathrm{tr}}_0}{\tau_0}=1,
\]
and from Eq.~(\ref{eq:eta_scalar_general_1})
\[
\eta_{1,\pm1}=2,\qquad
\eta_{1,0}=1.
\]
The diffusion constants are therefore
\[
D_{\pm1} = \frac{v^2\tau_{\pm1}\eta_{\pm1}}{2d},\qquad
D_0 = \frac{v^2\tau_0}{d}.
\]
because $\mathcal T^2=+1$ for integer $s$, all three channels belong to the orthogonal class and produce weak localization with
\(\zeta_1=-1\).
The total correction is dominated by the outer branches, yielding
\begin{equation}
\sigma^{\mathrm{qi}} = -\frac{e^2}{h}\frac{1}{\pi^2}\left(\frac{1}{\ell}-\frac{1}{\ell_\phi}\right)
\label{eq:s1_ds}
\end{equation}
with a negative sign opposite to that of Eq.~(\ref{eq:s12_ds}).
The evaluation of the Cooperon matrix yields the magnetoconductivity cusp
is negative with $\alpha_1=-1/2$~\cite{miao2023weak}, consistent with
the change in the symmetry class.
%%%%%%%%%%%%%%%%%%%%%%%%%%%%%%%%%%%%%%%%%%%%%%%%%%%%%%%%%%%%%%
\section{Explicit calculation for pseudospin-\texorpdfstring{$3/2$}{3/2}}\label{interbandcalculation}
%%%%%%%%%%%%%%%%%%%%%%%%%%%%%%%%%%%%%%%%%%%%%%%%%%%%%%%%%%%%%%
\subsection{Eigenvalues and eigenvectors for pseudospin-3/2}~\label{Sec_App_B_1}
The linearized $\mathbf{k}\cdot\mathbf{p}$ Hamiltonian for the pseudospin-${3}/{2}$ is, 
          $$
H(\mathbf{k})=\hbar \vartheta\begin{pmatrix}
-\frac{3}{2} k \cos\theta & \frac{\sqrt{3}}{2} e^{-i\phi} k \sin\theta & 0 & 0 & 0 \\
\frac{\sqrt{3}}{2} e^{i\phi} k \sin\theta & \frac{1}{2} k \cos\theta & e^{-i\phi} k \sin\theta & 0 & 0 \\
0 & e^{i\phi} k \sin\theta & -\frac{1}{2} k \cos\theta & \frac{\sqrt{3}}{2} e^{-i\phi} k \sin\theta & 0 \\
0 & 0 & \frac{\sqrt{3}}{2} e^{i\phi} k \sin\theta & -\frac{3}{2} k \cos\theta
\end{pmatrix}
$$
where $\mathbf{k}(k_x,k_y,k_z)$ is the wave vector, $\hbar$ is the reduced Planck's constant, and $\vartheta$ is the factor with the dimensionality of velocity. 
 The eigenvalues of the Hamiltonian of pseudospin-${3}/{2}$ are $\epsilon_\mathbf{k}/(\hbar \vartheta) = \{\pm \frac{1}{2} \mathrm{k}, \pm\frac{3}{2} \mathrm{k} \}$. The pseudospin system ${3}/{2}$ has two conduction bands and two valence bands, and the four bands degenerate at $\mathbf{k}=0$.
We focus solely on the properties of the conduction bands, as the properties of the valence bands are identical. In our calculation, there are two conduction bands that have a dispersion relation,
\begin{align}
    \epsilon_{\mathbf{k}} &= \frac{\vartheta \hbar k}{2} = \frac{\vartheta \hbar \sqrt{k_x^2 + k_y^2 + k_z^2}}{2},\nonumber \\
     \epsilon_{\mathbf{k}} &= \frac{3 \vartheta \hbar k}{2} = \frac{3 \vartheta \hbar \sqrt{k_x^2 + k_y^2 + k_z^2}}{2},\label{12}
\end{align}

where \(\mathbf{k}\) is measured from a four-fold degenerate point. Therefore,
The Hamiltonian of Weyl Fermions with pseudo spin $S=3/2$ is written as:
\begin{align}
H(\mathbf{k}) = \sum_{\chi} \chi \hbar  \vartheta \mathbf{k}\cdot\mathbf{S}
\end{align}
Dirac cones are split into two identical Weyl cones, with the carrier having an extra quantum number denoted by $\chi$ called the chirality of the particles. In Weyl fermions, there is a valley of particles with opposite chirality. Since the valleys have particles with opposite chirality, we denote these as "+" and "-" valleys.
In the "+" valley, the spinor wave function of the conduction band corresponds to $\epsilon_{\mathbf{k}} = \frac{\hbar \vartheta k}{2}$, in the future we denote these bands in Dirac notation as $|m \rangle$ $\equiv$ $|i,j,k\rangle$, where $i$ represents the pseudospin of the fermions, $j$ represents the band index, $k$ represents the valley of the Weyl fermions.

\begin{align}
    |3/2, 1/2, + \rangle_{\mathbf{k}} &= 
    \left(
    \begin{array}{c}
       -\frac{\sqrt{3}}{4}  \csc\frac{\theta}{2} \sin^2\theta e^{-3 i \phi} \\
\frac{1}{2} \cos\frac{\theta}{2}\left(-1 + 3 \cos\theta\right)  e^{-2 i \phi} \\
\frac{1}{2} \left(1 + 3 \cos\theta\right) \sin\frac{\theta}{2} e^{-i \phi} \\
\frac{\sqrt{3}}{2}  \sin\frac{\theta}{2} \sin\theta
    \end{array}
    \right) {e^{i \mathbf{k} \cdot \mathbf{r}}}, 
\end{align}

where \(\theta\) and \(\varphi\) are the wave vector angles, \(\tan \varphi \equiv k_y / k_x\), and \(\cos \theta \equiv k_z / k\). The spinor wave function of the conduction band corresponds to $\epsilon_{\mathbf{k}} = \frac{3 \hbar \vartheta k}{2}$, and in the future we denote this band in the Dirac notation as $|+,\mathrm{U}, \mathbf{k} \rangle $, which is

\begin{align}
    |3/2, 3/2, + \rangle_{\mathbf{k}} &= 
    \left(
    \begin{array}{c}
         \cos^3\frac{\theta}{2} e^{-3 i \phi} \\
\frac{\sqrt{3}}{4}   \csc\frac{\theta}{2} \sin^2\theta e^{-2 i \phi} \\
\frac{\sqrt{3}}{2}   \sin\frac{\theta}{2} \sin\theta e^{-i \phi} \\
\sin^3\frac{\theta}{2}
    \end{array}
    \right) {e^{i \mathbf{k} \cdot \mathbf{r}}}. 
\end{align}
In the "-" valley, the spinor wave function of the conduction of both bands can be found as ($\theta \to \pi-\theta$ and $\phi \to \pi+\phi$), 
\begin{align}
    |3/2, 1/2, -\rangle_{\mathbf{k}} &= 
    \left(
    \begin{array}{c}
       -\frac{\sqrt{3}}{2}  \sin\frac{\theta}{2} \sin\theta e^{3 i \phi} \\
\frac{-1}{2} \sin\frac{\theta}{2}\left(1 + 3 \cos\theta\right)  e^{2 i \phi} \\
\frac{1}{2} \left(-1 + 3 \cos\theta\right) \cos\frac{\theta}{2} e^{i \phi} \\
\frac{\sqrt{3}}{2}  \cos\frac{\theta}{2} \sin\theta
    \end{array}
    \right) {e^{i \mathbf{k} \cdot \mathbf{r}}}, 
\end{align}

\begin{align}
    |3/2, 3/2, - \rangle_{\mathbf{k}} &= 
    \left(
    \begin{array}{c}
         -\sin^3\frac{\theta}{2} e^{3 i \phi} \\
\frac{\sqrt{3}}{2}   \sin\frac{\theta}{2} \sin\theta e^{2 i \phi} \\
-\frac{\sqrt{3}}{2}   \cos\frac{\theta}{2} \sin\theta e^{i \phi} \\
\cos^3\frac{\theta}{2}
    \end{array}
    \right) {e^{i \mathbf{k} \cdot \mathbf{r}}}. 
\end{align}
The Density of states (DOS) at the Fermi surface is calculated as 
\begin{align}
    N_{F} &= \int \frac{d^3k}{(2\pi)^3} \delta(E_F - \epsilon_\mathbf{k'}) \nonumber\\
    &= \int_0^{2\pi} \frac{d\phi}{2\pi} \int_0^\pi \frac{\sin \theta \, d\theta}{2\pi} \int_0^\infty \frac{k^2 \, dk}{2\pi} \delta(E_F - \epsilon_\mathbf{k'})\nonumber\\
    &= \frac{4\pi}{8\pi^3}\int_0^\infty k^2 \, dk \delta(E_F - \epsilon_\mathbf{k'})\label{13}
\end{align}
Since DOS is different for each band. In general, we consider the dispersion relation in Eq.~\ref{12} as 
\begin{align*}
     \epsilon_{\mathbf{k}} &= C_m \hbar \vartheta k\\
    d\epsilon_{\mathbf{k}} &= C_m \hbar \vartheta  dk
\end{align*}
where $C_m \in \{\frac{1}{2},\frac{3}{2}\}$ ($\frac{3}{2}$ correspond to upper band and $\frac{1}{2}$ correspond to lower band ). Put the value of $\epsilon_\mathbf{k}$ and $d\epsilon_{\mathbf{k}}$ in Eq.~\ref{13},
\begin{align}
    N_F^m &= \frac{1}{2\pi^2}\int \frac{\epsilon_{\mathbf{k}}^2 d\epsilon_{\mathbf{k}}}{(C_m \hbar \vartheta )^3} \delta(E_F - \epsilon_\mathbf{k'})\nonumber \\
   N_F^m &= \frac{E_F^2}{2\pi^2(C_m \hbar \vartheta )^3}\label{DOS}    
\end{align}  
The DOS depends on the factor $C_m $ of the dispersion relation. For lower band the $C_m = \frac{1}{2}$, then
\begin{equation}
    {N_F = \frac{8 E_F^2}{2\pi^2(\hbar \vartheta )^3}}\nonumber
\end{equation}
For upper band $C_m = \frac{3}{2}$, then
\begin{equation}
    {N_F = \frac{8 E_F^2}{27\times 2\pi^2(\hbar \vartheta )^3}}\nonumber
\end{equation}
The dispersion relation of the lower and upper bands in the "-" valley is the same as in the "+" valley,  so the density of states is also equal.

\subsection{Scattering time for pseudospin-3/2}~\label{Sec_App_B_2}

In general, the scattering time is calculated by the Fermi golden rule as 
\begin{equation}\label{fgr}
\frac{1}{\tau}=\frac{2\pi}{\hbar}\sum_{\mathbf{k}'}\langle U_{\mathbf{k}\mathbf{k'}} U_{\mathbf{k'}\mathbf{k}}\rangle\delta(E_F-\epsilon_\mathbf{k'})
\end{equation}
where $U_{\mathbf{k}\mathbf{k'}} = \langle \mathbf{k}|U(\mathbf{r})|\mathbf{k'}\rangle$ represents the transition amplitude form initial state $|\mathbf{k}^{'}\rangle$ to final state $|\mathbf{k}\rangle$. We aim to study the effect of intervalley and interband scattering on the weak localization (WL)/ weak antilocalization (WAL) for multifold Weyl fermion. When two independent scattering processes are involved, the total scattering time $\tau$ is typically determined using Matthiessen's rule as
\begin{equation}\label{MR}
    \frac{1}{\tau} = \frac{1}{\tau_0} + \frac{1}{\tau_I} 
\end{equation}
where $\tau_0$ and $\tau_{I}$ are the scattering times corresponding to the independent scattering processes.

In our case, $\tau_0$ represents the scattering time associated with intraband and intravalley scattering, that is, where the band index and valley index are the same before and after scattering, whereas $\tau_{I}$ denotes the scattering time corresponding to interband or intervalley scattering, that is, where either the band index and valley index or both change after the scattering. This includes scenarios where the particle scatters within the same valley but transitions to a different dispersion relation(different band), scatters to a different valley while maintaining the same dispersion relation, or undergoes scattering to a different valley with a different dispersion relation. In general, the scattering amplitude between the initial and final states is calculated as 
\begin{equation}
     U^{m,n}_{\mathbf{k},\mathbf{k}^{'}} = \langle n |U(\mathbf{r})| m\rangle_{\mathbf{k},\mathbf{k}^{'}}
\end{equation}
where $m$ denotes the initial state of the particle, whereas $n$ denotes the final state of the particle, i.e., (particle is scattered from $|m\rangle \to |n\rangle$), and $U(\mathbf{r})$ is the elastic scattering potential,  which is defined as 
\begin{equation}
    U(\mathbf{r}) = \sum_i u_{0/\mathrm{I}}^i\delta(\mathbf{r}-\mathbf{R_i})\nonumber
\end{equation}
Where \( u_0^i \) is the scattering strength of the random potential that conserved the valley and band index after scattering, at position \( \mathbf{R}_i \), whereas \( u_{\mathrm{I}}^i \) is the scattering strength of the random potential at position \( \mathbf{R}_i \) responsible for the scattering where either the band index and valley index or both change after the scattering.
\begin{equation}
           U^{m,n}_{\mathbf{k},\mathbf{k}^{'}} 
           = \int d\mathbf{r}  e^{i\mathbf{k}\cdot\mathbf{r}}\sum_i u^i_0\delta(\mathbf{r}-\mathbf{R}_i)e^{-i\mathbf{k'}\cdot\mathbf{r}} \langle m, \mathbf{k}|n,\mathbf{k'}\rangle \nonumber
\end{equation}
let  $\Upsilon$ = $\langle m|n,\rangle$$_{\mathbf{k},\mathbf{k}^{'}}$ and  $\Upsilon^*$ = $\langle n |m \rangle$$_{\mathbf{k},\mathbf{k}^{'}}$, $^*$ denotes the complex conjugate
\begin{equation}
      U^{m,n}_{\mathbf{k}\mathbf{k'}}= \int d\mathbf{r}\sum_i u^i_0\delta(\mathbf{r}-\mathbf{R}_i) e^{i \mathbf{r}\cdot(\mathbf{k}-\mathbf{k'})}\Upsilon\nonumber 
\end{equation}
\begin{equation}
     U^{m,n}_{\mathbf{k}\mathbf{k'}}=   \sum_i u^i_0e^{i \mathbf{R}_i\cdot(\mathbf{k}-\mathbf{k'})}\Upsilon
\end{equation}
\\
In general, impurity correlation is calculated as 
\begin{align*}
   \langle|U^{m,n}_{\mathbf{kk}'}|^2\rangle &= \langle U^{m,n}_{\mathbf{kk}'}(U^{m,n}_{\mathbf{kk}'})^*\rangle\\
   &= \sum_i\sum_ju^i_0u^j_0e^{i \mathbf{R}_i\cdot(\mathbf{k}-\mathbf{k}')}e^{-i \mathbf{R}_j\cdot(\mathbf{k}'-\mathbf{k})}\Upsilon\Upsilon^*
\end{align*}
where $\langle...\rangle$ is averaging over impurity.
Upon averaging, only the terms with $i = j$ contribute, while all other terms cancel out. Consequently, the impurity correlation after averaging simplifies to, 
\begin{equation}
   \langle|U^{m,n}_{\mathbf{kk}'}|^2\rangle = {n_0 u_{0/\mathrm{I}}^2}\Upsilon\Upsilon^*
\end{equation}
where $n_0$ is impurity concentration and $u_{0/\mathrm{I}}$ is average elastic impurity strength. To calculate the total scattering time, we need to calculate $\tau_0$ and $\tau_{I}$, which are calculated from Eq.~\ref{fgr}. In a multifold Weyl fermion, different types of scattering processes are involved. So, first, we do all the calculations abstractly, and later, we discuss the results of each type of scattering. 

 The  $\tau_0$ is calculated as 
\begin{align}\label{intrascatt1}
    \frac{1}{\tau_0}&=\frac{2\pi}{\hbar}\sum_{\kn'}\langle U^{m,m}_{\mathbf{k}\mathbf{k'}} U^{m,m}_{\mathbf{k'}\mathbf{k}}\rangle\delta(E_F-\epsilon_\mathbf{k'}^m)\nonumber\\
    &=\frac{2\pi}{\hbar}\int_0^{2\pi} \frac{d\phi}{2\pi} \int_0^\pi \frac{\sin \theta \, d\theta}{2\pi} \int_0^\infty \frac{k^2 \, dk}{2\pi} \langle U^{m,m}_{\mathbf{k}\mathbf{k'}} U^{m,m}_{\mathbf{k'}\mathbf{k}}\rangle\delta(E_F-\epsilon_\mathbf{k'}^m)\nonumber\\
    &= \frac{2\pi}{\hbar}  \int_0^\infty \frac{k^2 \, dk}{2\pi} \delta(E_F-\epsilon_\mathbf{k'}^m)\int_0^{2\pi}\frac{d\phi}{2\pi} \int_0^\pi \frac{\sin \theta \, d\theta}{2\pi}\langle U^{++}_{\mathbf{k}\mathbf{k'}} U^{++}_{\mathbf{k'}\mathbf{k}}\rangle\nonumber\\
     &= \frac{2\pi}{\hbar} *I_{T1} * I_{T2}
\end{align}
where 
\begin{align}\label{intrascattint1}
    I_{T1} &= \int_0^\infty \frac{k^2 \, dk}{2\pi} \delta(E_F - \epsilon_\mathbf{k}^m)\nonumber  \\
   I_{T2} &= \int_0^{2\pi}\frac{d\phi}{2\pi} \int_0^\pi \frac{\sin \theta \, d\theta}{2\pi}\langle U^{m,m}_{\mathbf{k}\mathbf{k'}} U^{m,m}_{\mathbf{k'}\mathbf{k}}\rangle
\end{align} 
The  $\tau_I$ is calculated as 
\begin{align}\label{intrascatt2}
    \frac{1}{\tau_I}&=\frac{2\pi}{\hbar}\sum_{k'}\langle U^{m,n}_{\mathbf{k}\mathbf{k'}} U^{m,n}_{\mathbf{k'}\mathbf{k}}\rangle\delta(E_F-\epsilon_\mathbf{k'}^n)\nonumber\\
    &=\frac{2\pi}{\hbar}\int_0^{2\pi} \frac{d\phi}{2\pi} \int_0^\pi \frac{\sin \theta \, d\theta}{2\pi} \int_0^\infty \frac{k^2 \, dk}{2\pi} \langle U^{m,n}_{\mathbf{k}\mathbf{k'}} U^{m,n}_{\mathbf{k'}\mathbf{k}}\rangle\delta(E_F-\epsilon_\mathbf{k'}^n)\nonumber\\
    &= \frac{2\pi}{\hbar}  \int_0^\infty \frac{k^2 \, dk}{2\pi} \delta(E_F-\epsilon_\mathbf{k'}^n)\int_0^{2\pi}\frac{d\phi}{2\pi} \int_0^\pi \frac{\sin \theta \, d\theta}{2\pi}\langle U^{m,n}_{\mathbf{k}\mathbf{k'}} U^{m,n}_{\mathbf{k'}\mathbf{k}}\rangle\nonumber\\
    &=\frac{2\pi}{\hbar} *I_{T3} * I_{T4}
\end{align}
where 
\begin{align}\label{intrascattint2}
     I_{T3} &= \int_0^\infty \frac{k^2 \, dk}{2\pi} \delta(E_F - \epsilon_\mathbf{k}^n)\nonumber \\
    I_{T4} &= \int_0^{2\pi}\frac{d\phi}{2\pi} \int_0^\pi \frac{\sin \theta \, d\theta}{2\pi}\langle U^{m,n}_{\mathbf{k}\mathbf{k'}} U^{m,n}_{\mathbf{k'}\mathbf{k}}\rangle  
\end{align}

\subsection{Velocity correction for pseudospin-3/2}~\label{Sec_App_B_3}
The iterative equation can find the velocity correction.
\begin{equation}\label{vc}
   \widetilde{ v}_\kn^i = v_\kn^i + \sum_{\kp} G^R_{\kp}G^A_{\kp}\langle U_{\kn\kp}U_{\kp\kn}\rangle \widetilde{ v}_{\kp}^i
\end{equation}
where $i\in \{x,y,z\}$, $G^{R/A}$ is reduced/advanced Green's function, $v_{\mathbf{k}}^i$ is bare velocity, and $\widetilde{v}_{\mathbf{k}}^i$ is corrected velocity by the disorder scattering. In polar coordinates, Eq.~\ref{vc} becomes, 
\begin{equation}
   \widetilde{ v}_\kn^z = v_\kn^z + \int_0^{2\pi} \frac{d\phi'}{2\pi} \int_0^\pi \frac{\sin \theta' \, d\theta'}{2\pi} \int_0^\infty \frac{k'^2 \, dk'}{2\pi} G^R_{\kp}G^A_{\kp}\langle U_{\kn\kp}U_{\kp\kn}\rangle \widetilde{ v}_{\kp}^z
\end{equation}
The retarded (R) and advanced (A) Green's functions
\begin{equation}
    G^{R/A}_{\kn}(\omega) = \frac{1}{\omega - \epsilon_{\kn}\pm\frac{i\hbar}{2\tau\nonumber}}
\end{equation}
\begin{align*}
   \int_0^\infty\frac{k'^2 \, dk'}{2\pi} G^R_{\kp}G^A_{\kp} &= \int_0^\infty\frac{k'^2 \, dk'}{2\pi}\frac{1}{\omega - \epsilon_\kp  + i\hbar/2\tau}\times\frac{1}{\omega - \epsilon_\kp -    i\hbar/2\tau} \\
   &= \frac{1}{2\pi (C_{\alpha}\hbar v)^3}\int_0^\infty E^2 dE \frac{1}{\omega - \epsilon_\kp  + i\hbar/2\tau}\times\frac{1}{\omega - \epsilon_\kp -    i\hbar/2\tau} \\
   &= \frac{E^2_F}{2\pi (C_{\alpha}\hbar v)^3}\int_{-\infty}^\infty  dE \frac{1}{\omega - \epsilon_\kp  + i\hbar/2\tau}\times\frac{1}{\omega - \epsilon_\kp -i\hbar/2\tau} \\
\end{align*}
We evaluate this integral using the contour integration method, which gives
\begin{align}
    \int_{-\infty}^\infty  dE \frac{1}{\omega - \epsilon_\kp  + i\hbar/2\tau}\times\frac{1}{\omega - \epsilon_\kp -i\hbar/2\tau} &= \frac{2\pi\tau}{\hbar}\nonumber
\end{align}
and from Eq.~\ref{DOS}
\begin{align}
    \frac{E^2_F}{2\pi (C_m \hbar \vartheta)^3} =\pi N_F^{m}    
\end{align}
\begin{equation}
    {  \int_0^\infty\frac{k'^2 \, dk'}{2\pi} G^R_{\kp}G^A_{\kp} = \frac{2\pi^2 N_F^{m} \tau}{\hbar}}
\end{equation}
\begin{equation}\label{6}
   \widetilde{ v}_\kn^z = v_\kn^z + \frac{2\pi^2 N_F^{m} \tau}{\hbar}\int_0^{2\pi} \frac{d\phi'}{2\pi} \int_0^\pi \frac{\sin \theta' \, d\theta'}{2\pi} \langle U_{\kn\kp}U_{\kn'\kn}\rangle \widetilde{ v}_{\kn'}^z 
\end{equation}
and 
\begin{align}
\langle U_{\mathbf{k}, \mathbf{k}'} U_{\mathbf{k}', \mathbf{k}} \rangle &= \langle U_{\mathbf{k}, \mathbf{k}'}^{m,m} U_{\mathbf{k}', \mathbf{k}}^{m,m} \rangle + \langle U_{\mathbf{k}, \mathbf{k}'}^{m,n} U_{\mathbf{k}', \mathbf{k}}^{n,m} \rangle \nonumber ,
\end{align}
Put the value of $ \langle U_{\kn\kn'}U_{\kn'\kn}\rangle$ in Eq.~\ref{vc}
\begin{equation}
   \widetilde{ v}_\kn^i = v_\kn^i + \sum_{\kp} G^R_{\kp}G^A_{\kp}\langle U_{\mathbf{k}, \mathbf{k}'}^{m,m} U_{\mathbf{k}', \mathbf{k}}^{m,m} \rangle + \sum_{\kp} G^R_{\kp}G^A_{\kp} \langle U_{\mathbf{k}, \mathbf{k}'}^{m,n} U_{\mathbf{k}', \mathbf{k}}^{n,m} \rangle \widetilde{ v}_{\kp}^i
\end{equation}
In the above expression, there is  $\sum_{\kp}$, where $\mathbf{k'}$ is the state after the scattering, so in the second term in RHS, the ${\kp}$ belongs to the state denoted by $|m\rangle $  whereas in the third term in RHS, ${\kp}$ belong to the state denoted by $|n\rangle $.
\begin{align}
    \widetilde{ v}_\kn^z &= v_\kn^z + {\int_0^{2\pi} \frac{d\phi'}{2\pi} \int_0^\pi \frac{\sin \theta' \, d\theta'}{2\pi} \int_0^\infty \frac{k'^2 \, dk'}{2\pi} G^R_{\kn'}G^A_{\kn'}\langle U^{m,m}_{\kn\kn'}U^{m,m}_{\kn'\kn}\rangle \widetilde{ v}_{\kn'}^z} \nonumber\\
    & \quad + {\int_0^{2\pi} \frac{d\phi'}{2\pi} \int_0^\pi \frac{\sin \theta' \, d\theta'}{2\pi} \int_0^\infty \frac{k'^2 \, dk'}{2\pi} G^R_{\kn'}G^A_{\kn'}\langle U^{m,n}_{\kn\kn'}U^{n,m}_{\kn'\kn}\rangle \widetilde{ v}_{\kn'}^z}\nonumber
\end{align}
In polar coordinates,
\begin{align*}
    v_\kn^z &= C_m \vartheta \cos{\theta}\\
    \widetilde{v}_\kn^z &= \eta C_m \vartheta \cos{\theta}\\
    \widetilde{v}_\kn^z &= \eta  v_\kn^z
\end{align*}
where $\eta$ is the velocity correction factor.
\begin{align}
    \widetilde{ v}_\kn^z &= v_\kn^z + \frac{2\pi^2 N^m_F \tau}{\hbar}\int_0^{2\pi} \frac{d\phi'}{2\pi} \int_0^\pi \frac{\sin \theta' \, d\theta'}{2\pi} \langle U^{m,m}_{\kn\kn'}U^{m,m}_{\kn'\kn}\rangle \eta \vartheta C_{m} \cos{\theta'}  \nonumber \\
   & \quad + \frac{2\pi^2 N^n_F \tau}{\hbar}\int_0^{2\pi} \frac{d\phi'}{2\pi} \int_0^\pi \frac{\sin \theta' \, d\theta'}{2\pi} \langle U^{m,n}_{\kn\kn'}U^{n,m}_{\kn'\kn}\rangle \eta \vartheta C_{n} \cos{\theta'}\nonumber\\
   &= v_\kn^z + \frac{2\pi^2 N^m_F \tau}{\hbar}\eta \vartheta C_{m} \int_0^{2\pi} \frac{d\phi'}{2\pi} \int_0^\pi \frac{\sin \theta' \, d\theta'}{2\pi} \langle U^{m,m}_{\kn\kn'}U^{m,m}_{\kn'\kn}\rangle \cos{\theta'}  \nonumber \\
   & \quad + \frac{2\pi^2 N^n_F \tau}{\hbar}\eta \vartheta C_{n} \int_0^{2\pi} \frac{d\phi'}{2\pi} \int_0^\pi \frac{\sin \theta' \, d\theta'}{2\pi} \langle U^{m,n}_{\kn\kn'}U^{n,m}_{\kn'\kn}\rangle \cos{\theta'}\nonumber
\end{align}
where 
\begin{align}\label{intvelcor}
    \int_0^{2\pi} \frac{d\phi'}{2\pi} \int_0^\pi \frac{\sin \theta' \, d\theta'}{2\pi} \langle U^{m,m}_{\kn\kn'}U^{m,m}_{\kn'\kn}\rangle \cos{\theta'} &= I_{V1}\nonumber\\
    \int_0^{2\pi} \frac{d\phi'}{2\pi} \int_0^\pi \frac{\sin \theta' \, d\theta'}{2\pi} \langle U^{m,n}_{\kn\kn'}U^{n,m}_{\kn'\kn}\rangle \cos{\theta'} &=I_{V2}
\end{align}

\begin{align}\label{expvelcor}
    \eta{ v}_\kn^z  &= v_\kn^z + \frac{2\pi^2 N^m_F \tau}{\hbar}\eta \vartheta C_{m} I_{V1}  + \frac{2\pi^2 N^n_F \tau}{\hbar}\eta \vartheta C_{n}I_{V2}\nonumber\\
     C_{m} \eta &= C_{m} + \frac{2\pi^2 N^m_F \tau}{\hbar} C_{m} \eta I_{V1}  + \frac{2\pi^2 N^n_F \tau}{\hbar} C_{n}\eta I_{V2}
\end{align}

\subsection{Conductivity with both interband and intervalley scattering for pseudospin-3/2}~\label{Sec_App_B_4}

Let the initial state of the particle \( |m\rangle_{\mathbf{k}} \), and the final state \( |n \rangle_\mathbf{k'} \). When the particle undergoes elastic scattering at the Fermi surface, there are two types of scattering: (i) when the band and valley index do not change, i.e., $|m\rangle_{\mathbf{k}}$ $\to$ $|m\rangle_{\mathbf{k'}}$.  
(ii) The valley or the band index or both will change, i.e., $|m\rangle_{\mathbf{k}}$ $\to$ $|n\rangle_{\mathbf{k'}}$.

Thus, the total conductivity correction due to quantum interference includes both type (i) and type (ii) scattering processes, and is given by.
\begin{equation}\label{16}
    \sigma^{qi} = 2 \sigma^{qi}_{0} + \sigma^{qi}_{I}
\end{equation}
where $\sigma^{qi}_{0}$ is from the Cooperons corresponding to the type (i) process, whereas $\sigma^{qi}_{I}$ is from the Cooperons corresponding to the type (ii) process.

First, we have to solve the Bethe-Salpeter equation for finding the intraband-intravalley Cooperons, i.e., the band and valley index do not change, corresponding to the type (i) process, is given as  
\begin{align}
    \Gamma^m_{{\kn_1}{\kn_2}} &=  \Gamma^{m,0}_{{\kn_1}{\kn_2}} + \sum_{{\kn}} \Gamma^{m,0}_{{\kn_1}{\kn}}G^R_{{\kn}}G^A_{\mathbf{q}-{\kn}} \Gamma^m_{{\kn}{\kn_2}}\nonumber
\end{align}
In polar coordinates, the Bethe-Salpeter equation becomes
\begin{align}\label{BSE}
     \Gamma^m_{{\kn_1}{\kn_2}} &=  \Gamma^{m,0}_{{\kn_1}{\kn_2}} + \int_0^{2\pi} \frac{d\phi}{2\pi} \int_0^\pi \frac{\sin \theta \, d\theta}{2\pi} \int_0^\infty \frac{k^2 \, dk}{2\pi} \Gamma^{m,0}_{{\kn_1}{\kn}}G^R_{{\kn}}G^A_{\mathbf{q}-{\kn}} \Gamma^m_{{\kn}{\kn_2}}
\end{align}
where $\Gamma^m_{{\kn_1}{\kn_2}}$ is the vertex function, which we have to find, and $\Gamma^{m,0}_{{\kn_1}{\kn_2}}$ is the bare Cooperon, which is defined as,
\begin{align}\label{14}
     \Gamma^{m,0}_{{\kn_1}{\kn_2}} &= \langle U^{m,m}_{\kn_{1},\kn_{2}}U^{m,m}_{-\kn_{1},-\kn_{2}}\rangle\\
      \Gamma^{m,0}_{{\kn_1}{\kn_2}} &= {n_0 u_0^2}\sum_{r,s,p,q} \Gamma^0_\mathrm{mnpq}e^{i(r\theta_1 + s\phi_1 + p\theta_2 + q\phi_2)}\nonumber
\end{align}
\begin{align}
     \Gamma^{m,0}_{{\kn_1}{\kn_2}} &=\frac{4(1-\eta_I)\hbar}{2\pi N_F^m \tau}\sum_{r,s,p,q} \Gamma^0_\mathrm{rspq}e^{i(r\theta_1 + s\phi_1 + p\theta_2 + q\phi_2)}\nonumber
\end{align}
where $\eta_{I} = \tau/\tau_{I}$ measures the relative strength of the intervalley/interband scattering, which is zero in the absence of type-(ii) scattering process, and we can recover bare intraband-intravalley Cooperons. 

Anstaz for a Bethe-Salpeter equation is
\begin{align}
     \Gamma^m_{{\kn_1}{\kn_2}} &=\frac{4\hbar}{2\pi N_F^m \tau}\sum_{r,s,p,q} \Gamma_\mathrm{rspq}e^{i(r\theta_1 + s\phi_1 + p\theta_2 + q\phi_2)}\nonumber
\end{align}

the $k$ integration is independent of $\Gamma^m_{{\kn_1}{\kn_2}}$ and $\Gamma^{m,0}_{{\kn_1}{\kn_2}}$ because they are the function of $\theta$ and $\phi$, so we can evaluate $k$ integration by contour integration method and becomes
\begin{equation}
    \int_0^\infty \frac{k^2 \, dk}{2\pi} G^R_{{\kn}}G^A_{\mathbf{q}-{\kn}} = \frac{2\pi^2 N_F^m \tau}{\hbar}[1 - i\tau v_F q\cos{\theta} - \tau^2 v_F^2 q^2\cos^2{\theta}]
\end{equation}
Here $v_F = C_{m} \vartheta$. Put back the value of $k$ integration in Eq.~\ref{BSE} 
\begin{align*}
    \gone &= (1-\eta_I)\gzero + \frac{2\pi^2 N_F^m \tau(1-\eta_I)}{\hbar} 
     \\
    &\quad \intsh  \Bigg[ 1 - i\tau v_F q \cos{\theta} - \tau^2 v_F^2 q^2 \cos^2{\theta} \Bigg]\Bigg(\frac{4\hbar}{2\pi\tau N_F^m} \Bigg)^2 \\
    &\quad \sum_{i,j,k,l} \Gamma^0_{\mathrm{ijkl}} e^{i(i\theta_1 + j\phi_1 + k\theta + l\phi)}  \sum_{r,s,p,q} \Gamma_\mathrm{rspq} e^{i(r\theta + s\phi + p\theta_2 + q\phi_2)}
\end{align*}
\begin{align*}
    \sum_{r,s,p,q}\Gamma_\mathrm{rspq}e^{i(r\theta_1 + s\phi_1 + p\theta_2 + q\phi_2)} &= (1-\eta_I)\sum_{r,s,p,q}\Gamma^{m,0}_\mathrm{rspq}e^{i(r\theta_1 + s\phi_1 + p\theta_2 + q\phi_2)}+\frac{2\pi^2 N_F^m \tau}{\hbar}\Bigg(\frac{4\hbar(1-\eta_I)}{2\pi N_F^m \tau} \Bigg)\\
    &\quad \intsh\Bigg[ 1 - i\tau v_F q \cos{\theta} - \tau^2 v_F^2 q^2 \cos^2{\theta} \Bigg]\\
    &\sum_{i,j,k,l}\Gamma^0_\mathrm{{ijkl}} e^{i(i\theta_1 + j\phi_1 + k\theta + l\phi)}  \sum_{m,n,p,q} \Gamma_\mathrm{rspq} e^{i(r\theta + s\phi + p\theta_2 + q\phi_2)}\\
    \sum_{r,s,p,q}\Gamma_\mathrm{rspq}e^{i(r\theta_1 + s\phi_1 + p\theta_2 + q\phi_2)} &= (1-\eta_I)\Gamma^0_\mathrm{rspq}e^{i(r\theta_1 + s\phi_1 + p\theta_2 + q\phi_2)}+ {4(1-\eta_I)\pi}\sum_{i,j,p,q}\Bigg[ 1 - i\tau v_F q \cos{\theta} - \tau^2 v_F^2 q^2 \cos^2{\theta} \Bigg]\\
    &\quad\sum_{k,l,m,n}\intsh\Gamma^0_\mathrm{ijkl} e^{i(i\theta_1 + j\phi_1 + p\theta_2 + q\phi_2)}   \Gamma_\mathrm{rspq} e^{i((r+k)\theta + (s+q)\phi}
\end{align*}
This equation can be written in matrix form
\begin{equation}\label{MBSE}
     \Bigg[\Gamma\Bigg] =  (1-\eta_I)\Bigg[\Gamma^0\Bigg] +\Bigg[\Gamma^0\Bigg]\Bigg[\Phi\Bigg]\Bigg[\Gamma\Bigg]
\end{equation}
where $\Phi$ is a matrix whose elements are defined as 
\begin{equation}
    \phi_\mathrm{rspq} = {4(1-\eta_I)\pi}\intsh\Bigg[ 1 - i\tau v_F q \cos{\theta} - \tau^2 v_F^2 q^2 \cos^2{\theta} \Bigg]e^{i((m+k)\theta + (n+q)\phi)}
\end{equation}
after solving Eq.~\ref{MBSE}, we get 
\begin{equation}
    \Gamma^m_{\kn_1\kn_2} = \gamma^m(q) e^{3i(\phi_2 - \phi_1)}
\end{equation}
where 
\begin{equation}\label{18}
    \gamma(q) = \frac{4\hbar}{2\pi N_F^m v_F^2 \tau^3}\frac{\chi^m}{\lambda^2 + q^2 }
\end{equation}
\begin{equation}\label{gammaanstaz}
    \Gamma^m_{\kn_1\kn_2} = \frac{4\hbar}{2\pi N_F^m v_F^2 \tau^3}\frac{\chi^m}{\lambda^2 + q^2 } e^{3i(\phi_2 - \phi_1)}
\end{equation}
The conductivity contribution from the intraband-intravalley cooperons is given by 
\begin{equation}\label{conduccorr}
    \sigma^{qi}_0 = \sigma_{a1} + 2\sigma_{a2}, 
\end{equation}
where
\begin{equation}
    \sigma_{\mathrm{a_1}} = \frac{e^2 \hbar}{2\pi} \sum_{\mathbf{q}} \Gamma_{\mathbf{k}, \mathbf{q}-\mathbf{k}} \sum_{\mathbf{k}} G_{\mathbf{k}}^R \tilde{v}_{\mathbf{k}}^x G_{\mathbf{k}}^A G_{\mathbf{q}-\mathbf{k}}^R \tilde{v}_{\mathbf{q}-\mathbf{k}}^x G_{\mathbf{q}-\mathbf{k}}^A,\nonumber
\end{equation}
and
\begin{equation}
    \sigma_{\mathrm{a_2}} = \frac{e^2 \hbar}{2\pi} \sum_{\mathbf{q}} \Gamma_{\mathbf{k_1}, \mathbf{q}-\mathbf{k_1}} \sum_{\mathbf{k_1}} \tilde{v}_{\mathbf{k_1}}^x \tilde{v}_{\mathbf{q}-\mathbf{k_1}}^x G_{\kn}^R G_{\mathbf{k_1}}^R G_{\mathbf{q}-\mathbf{k}}^R G_{\mathbf{q}-\mathbf{k_1}}^R G_{\mathbf{k}}^A G_{\mathbf{q}-\mathbf{k_1}}^A \langle U_{\mathbf{k},\mathbf{k_1}} U_{\mathbf{q}-\mathbf{k},\mathbf{q}-\mathbf{k_1}} \rangle.\nonumber
\end{equation}
The $\sigma_{\mathrm{a_1}}$ is calculated as 
\begin{equation}
    \sigma_{\mathrm{a_1}} = \frac{e^2\hbar}{2\pi}\sum_\mathbf{q}\Gamma_{\kn,\mathbf{q}-\kn}\sum_\kn G^R_\kn\widetilde{v}^x_\kn G^A_\kn G^R_{\mathbf{q}-\kn}\widetilde{v}^x_{\mathbf{q}-\kn}G^A_{\mathbf{q}-\kn}
\end{equation}
In the limit $\mathbf{q}\to 0$
\begin{equation}
     \Gamma_{\kn,\mathbf{q}-\kn} =\Gamma_{\kn,-\kn}
\end{equation}
 As we find the $\Gamma_{\kn_1\kn_2}$ in Eq.\ref{gammaanstaz}
\begin{equation}
    \Gamma_{\kn_1\kn_2} = \frac{4\hbar}{2\pi N_F v_F^2 \tau^3}\frac{\chi^m}{\lambda^2 + q^2 } e^{i3(\phi_2 - \phi_1)}\nonumber
\end{equation}
\begin{equation}
   \Gamma_{\kn, -\kn} =  \frac{4\hbar}{2\pi N_F v_F^2 \tau^3}\frac{\chi^m}{\lambda^2 + q^2 } (-1)
\end{equation}
Now, we called  $\Gamma_{\kn, -\kn}$= -$\gamma(q)$  
\begin{align}
    \sigma_{\mathrm{a_1}} &= -\frac{e^2\hbar}{2\pi}\sum_\mathbf{q} \gamma(q)\sum_\kn G^R_\kn\widetilde{v}^x_\kn G^A_\kn G^R_{\mathbf{q}-\kn}\widetilde{v}^x_{\mathbf{q}-\kn}G^A_{\mathbf{q}-\kn} \nonumber \\
    &= -\frac{e^2\hbar}{2\pi}\sum_\mathbf{q} \gamma(q)\isum  G^R_\kn\widetilde{v}^x_\kn G^A_\kn G^R_{\mathbf{q}-\kn}\widetilde{v}^x_{\mathbf{q}-\kn}G^A_{\mathbf{q}-\kn} \nonumber \\
    &= -\frac{e^2\hbar}{2\pi}\sum_\mathbf{q} \gamma(q)\intsh \widetilde{v}^x_\kn \widetilde{v}^x_{\mathbf{q}-\kn}\int_0^\infty \frac{k^2 dk}{2\pi}G^R_\kn G^A_\kn G^R_{\mathbf{q}-\kn}G^A_{\mathbf{q}-\kn} \label{sigmaa1}
\end{align}
From Eq.\ref{12}
\begin{align*}
     E_\kn &= C _m \hbar \vartheta k\\
    dE_\kn &=  C _m \hbar \vartheta dk
\end{align*}
So, first, we solve the $k$ integration 
\begin{align*}
   \int_0^\infty\frac{k^2 \, dk}{2\pi} G^R_\kn G^A_\kn G^R_{\mathbf{q}-\kn}G^A_{\mathbf{q}-\kn} &= \int_0^\infty\frac{k^2 \, dk}{2\pi}G^R_\kn G^A_\kn G^R_{\mathbf{q}-\kn}G^A_{\mathbf{q}-\kn}\\
   &= \frac{1}{2\pi (C _m \hbar \vartheta)^3}\int_0^\infty E^2 dE G^R_\kn G^A_\kn G^R_{\mathbf{q}-\kn}G^A_{\mathbf{q}-\kn} \\
   &= \frac{E^2_F}{2\pi (C _m \hbar \vartheta)^3}\int_{-\infty}^\infty  dE G^R_\kn G^A_\kn G^R_{\mathbf{q}-\kn}G^A_{\mathbf{q}-\kn} \\
\end{align*}
By using the relation between the Fermi energy and the density of states (DOS) at the Fermi level mentioned in Eq.~\ref{DOS}
\begin{align*}
    \int_0^\infty\frac{k^2 \, dk}{2\pi} G^R_\kn G^A_\kn G^R_{\mathbf{q}-\kn}G^A_{\mathbf{q}-\kn}&=\pi N_F\int_{-\infty}^\infty  dE G^R_\kn G^A_\kn G^R_{\mathbf{q}-\kn}G^A_{\mathbf{q}-\kn}\\
    &= \pi N_F\int_{-\infty}^\infty  dE \frac{1}{\omega - \epsilon_k  + i\hbar/2\tau}\frac{1}{\omega - \epsilon_k -i\hbar/2\tau} \\
    &\quad\times   \frac{1}{\omega - \epsilon_k  +\textbf{Q}+ i\hbar/2\tau}\frac{1}{\omega - \epsilon_k+ \textbf{Q} -i\hbar/2\tau} 
\end{align*}
where $\textbf{Q} = \hbar {\mathbf{v}_F\cdot \mathbf{q}}$
\begin{equation}
    \pi N_F\int_{-\infty}^\infty  dE G^R_\kn G^A_\kn G^R_{\mathbf{q}-\kn}G^A_{\mathbf{q}-\kn} =  \pi N_F\frac{4\pi\tau^3}{\textbf{Q}^2\tau^2\hbar + \hbar^3}\nonumber
\end{equation}
As $\mathbf{q}\to 0$ $\implies$ $\mathbf{Q}\to 0$,
\begin{equation}
    \pi N_F\int_{-\infty}^\infty  dEG^R_\kn G^A_\kn G^R_{\mathbf{q}-\kn}G^A_{\mathbf{q}-\kn} =   \frac{4\pi^2 N_F\tau^3}{  \hbar^3}
\end{equation}
Put the value of $k$ integration in Eq.~\ref{sigmaa1}
\begin{align}
    \sigma_{\mathrm{a_1}}
    &= -\frac{e^2\hbar}{2\pi}\sum_\mathbf{q} \gamma(q)\intsh \widetilde{v}^x_\kn \widetilde{v}^x_{\mathbf{q}-\kn}\int_0^\infty \frac{k^2 dk}{2\pi}G^R_\kn G^A_\kn G^R_{\mathbf{q}-\kn}G^A_{\mathbf{q}-\kn}\nonumber\\
    &= -\frac{e^2\hbar}{2\pi}\frac{4\pi^2 N_F\tau^3}{  \hbar^3}\sum_\mathbf{q} \gamma(q)\intsh \widetilde{v}^x_\kn \widetilde{v}^x_{\mathbf{q}-\kn}\nonumber\\
    &=-\frac{e^2\hbar}{2\pi}\frac{4\pi^2 N_F\tau^3}{  \hbar^3}\sum_\mathbf{q} \gamma(q)\intsh \eta v_\kn^x \eta v_{\mathbf{q}-\kn}^x\nonumber
\end{align}
We are performing integration in polar coordinates, so the velocity in polar coordinates is 
\begin{align}
    v_\kn^x &= v_F\sin{\theta}\cos{\phi}\nonumber\\
    {v}_{-\kn}^x &= - v_F\sin{\theta}\cos{\phi}\nonumber
\end{align}
put the values of $v_\kn^x$ and ${v}_{\mathbf{q}-\kn}^x$ and integrate over $\theta$ and $\phi$
\begin{align*}
    \sigma_{\mathrm{a_1}}&=-\frac{e^2\hbar}{2\pi}\frac{4\pi^2 N_F\tau^3}{  \hbar^3}\sum_\mathbf{q} \gamma(q)\intsh \eta v_\kn^x \eta v_{\mathbf{q}-\kn}^x\\
    &= \frac{e^2\hbar}{2\pi}\frac{4\pi^2 N_F\tau^3}{  \hbar^3}\sum_\q \gamma(q)\intsh \eta^2 v_F\sin{\theta}\cos{\phi}  v_F\sin{\theta}\cos{\phi}\\
     &= \frac{e^2\hbar}{2\pi}\frac{4\pi^2 N_F\tau^3}{  \hbar^3}\eta^2 v_F^2\sum_\q \gamma(q)\intsh \sin{\theta}\cos{\phi}  \sin{\theta}\cos{\phi}
\end{align*}
\begin{equation}
   \intsh \sin{\theta}\cos{\phi}  \sin{\theta}\cos{\phi} = \frac{ 1}{3\pi}\nonumber 
\end{equation}
\begin{equation}
   \sigma_{\mathrm{a_1}} =  \frac{e^2\hbar}{2\pi}\frac{4\pi^2 N_F\tau^3}{  \hbar^3}\eta^2 v_F^2\frac{1}{3\pi}\sum_\q \gamma(q) \nonumber
\end{equation}
\begin{equation}
   {\sigma_{\mathrm{a_1}} = \frac{2\eta^2 v_F^2 N_F\tau^3 e^2}{ 3 \hbar^2}\sum_\q \gamma(q)} 
\end{equation}
The $\mathbf{\sigma}_{a_2}$ is calculated as 
\begin{equation}
\sigma_{\mathrm{a_2}} = \frac{e^2 \hbar}{2 \pi} \sum_{\mathbf{q}} \Gamma_{\mathbf{k_1}, \mathbf{q} - \mathbf{k}} \sum_{\mathbf{k}} \sum_{\mathbf{k_1}} \tilde{v}_{\mathbf{k}}^x \tilde{v}_{\mathbf{q} - \mathbf{k_1}}^x G^R_{\mathbf{k}} G^R_{\mathbf{k_1}}G^A_{\mathbf{k}}  G^R_{\mathbf{q} - \mathbf{k}} 
 G^R_{\mathbf{q} - \mathbf{k_1}} G^A_{\mathbf{q} - \mathbf{k_1}} \left\langle U_{\mathbf{k}, \mathbf{k_1}} U_{\mathbf{q} - \mathbf{k}, \mathbf{q} - \mathbf{k_1}} \right\rangle \nonumber .
\end{equation}
The $\left\langle U_{\mathbf{k}, \mathbf{k_1}} U_{\mathbf{q} - \mathbf{k}, \mathbf{q} - \mathbf{k_1}} \right\rangle $ is independent of $k$, by contour integration method we can evaluate $k$ integration as,
\begin{align}
  \int_0^\infty \frac{k^2 dk}{2\pi} G^R_{\mathbf{k}} G^A_{\mathbf{k}} G^R_{\mathbf{q}-\mathbf{k}} &=\pi N_F \int_{-\infty}^{\infty} d\epsilon_{\mathbf{k}} \frac{1}{\left( \omega - \epsilon_{\mathbf{k}} + \frac{i \hbar}{2\tau} \right)} \frac{1}{\left( \omega - \epsilon_{\mathbf{k}} - \frac{i \hbar}{2\tau} \right)} \frac{1}{\left( \omega - \epsilon_{\mathbf{k}} + \hbar \mathbf{v_F} \cdot \mathbf{q} + \frac{i \hbar}{2\tau} \right)}\nonumber \\
  \int_0^\infty \frac{k^2 dk}{2\pi} G^R_{\mathbf{k}} G^A_{\mathbf{k}} G^R_{\mathbf{q}-\mathbf{k}} &=  \frac{2\pi^2 N_F \tau^2}{ i\hbar^2} \nonumber
\end{align}
Similarly,
\begin{align}
    \int_0^\infty \frac{k_{1}^2 dk_1}{2\pi} G^R_{\mathbf{k_1}} G^R_{\mathbf{q} -\mathbf{k_1}} G^A_{\mathbf{q}-\mathbf{k_1}} &=  \frac{2\pi^2 N_F \tau^2}{ i\hbar^2} \nonumber
\end{align}
put the value of $k$-integration,
\begin{equation}
\sigma_{\mathrm{a_2}} = -\frac{e^2 \hbar}{2 \pi}  \frac{4\pi^4 N_F^2 \tau^4}{ \hbar^4}\sum_{\mathbf{q}} \Gamma_{\mathbf{k_1}, \mathbf{q} - \mathbf{k}} \intsh \tilde{v}_{\mathbf{k}}^x \tilde{v}_{\mathbf{q} - \mathbf{k_1}}^x  
\intshone \left\langle U_{\mathbf{k}, \mathbf{k_1}} U_{\mathbf{q} - \mathbf{k}, \mathbf{q} - \mathbf{k_1}} \right\rangle.
\end{equation}
We are performing integration in polar coordinates, so the velocity in polar coordinates is 
\begin{align*}
    v_\kn^x &= v_F\sin{\theta}\cos{\phi}\\
    {v}_{\q-\kn}^x &=  -v_F\sin{\theta}\cos{\phi}
\end{align*}
Put the value of $\widetilde{v}_\kn^x$ and $\widetilde{v}_{\q-\kn}^x$ and perform integration 
\begin{equation}
\sigma_{\mathrm{a_2}} = \frac{e^2 \hbar}{2 \pi} \frac{4\pi^4 N_F^2 \tau^4}{ \hbar^4}\sum_{\mathbf{q}} \Gamma_{\mathbf{k_1}, \mathbf{q} - \mathbf{k}} \intsh \eta^2 v_F^2\sin^2{\theta}\cos^2{\phi}  
 \intshone \left\langle U_{\mathbf{k}, \mathbf{k_1}} U_{\mathbf{q} - \mathbf{k}, \mathbf{q} - \mathbf{k_1}} \right\rangle.
\end{equation}
\begin{equation}\label{15}
\sigma_{\mathrm{a_2}} = \frac{\eta^2 v_F^2 e^2 \hbar}{2 \pi} \frac{4\pi^4 N_F^2 \tau^4}{ \hbar^4}\sum_{\mathbf{q}} \Gamma_{\mathbf{k_1}, \mathbf{q} - \mathbf{k}} \intsh \sin^2{\theta}\cos^2{\phi}    \intshone \left\langle U_{\mathbf{k}, \mathbf{k_1}} U_{\mathbf{q} - \mathbf{k}, \mathbf{q} - \mathbf{k_1}} \right\rangle.
\end{equation}
from Eq.~\ref{6}
\begin{equation}
    \Gamma_{\kn_1\kn_2} = \frac{4\hbar}{2\pi N_F v_F^2 \tau^3}\frac{\chi
    ^m}{\lambda^2 + q^2 } e^{3i(\phi_2 - \phi_1)}\nonumber
\end{equation}
In the limit $\mathbf{q}\to 0$
\begin{align}
    \Gamma_{\mathbf{k_1}, \mathbf{q} - \mathbf{k}} &= \Gamma_{\mathbf{k_1},  -\mathbf{k}}\nonumber
\end{align}
 Since $e^{3i(\pi +\phi -\phi_1)} = -e^{3i(\phi -\phi_1)}$ 
 \begin{align*}
    \Gamma_{\mathbf{k_1}, -\mathbf{k}} 
    &=  \frac{4\hbar}{2\pi N_F v_F^2 \tau^3}\gamma(q)e^{3i(\pi +\phi -\phi_1)}\\
    &=  -\frac{4\hbar}{2\pi N_F v_F^2 \tau^3}\gamma(q)e^{3i(\phi -\phi_1)}
\end{align*}
In the limit $\mathbf{q}\to 0$  
\begin{align*}
 \left\langle U_{\mathbf{k}, \mathbf{k_1}} U_{\mathbf{q} - \mathbf{k}, \mathbf{q} - \mathbf{k_1}} \right\rangle &= \left\langle U_{\mathbf{k}, \mathbf{k_1}} U_{ - \mathbf{k},- \mathbf{k_1}} \right\rangle  = \Gamma^{m,0}_{\kn ,\kn_1} 
\end{align*}
We define $\Gamma^{m,0}_{\kn ,\kn_1}$ in Eq.~\ref{14},
\begin{equation}
  \Gamma^{m,0}_{{\mathbf{k}_1}{\mathbf{k}_2}} =\frac{4\hbar}{2\pi N_F \tau}\sum_{r,s,p,q} \Gamma^{m,0}_\mathrm{rspq}e^{i(r\theta_1 + s\phi_1 + p\theta_2 + q\phi_2)}  
\end{equation}
Put the value of $\Gamma^{m,0}_{{\kn_1}{\kn_2}}$ and $ \Gamma
^m_{\mathbf{k_1}, -\mathbf{k}}$ in Eq.~\ref{15}, 
\begin{equation}\label{dseconduc}
 I_{a2} = \intsh \sin^2{\theta}\cos^2{\phi}  
 \intshone \left\langle U_{\mathbf{k}, \mathbf{k_1}} U_{\mathbf{q} - \mathbf{k}, \mathbf{q} - \mathbf{k_1}} \right\rangle e^{3i(\phi -\phi_1)}
\end{equation}
we get
\begin{equation}
\sigma_{\mathrm{a_2}} = -\frac{\eta^2 v_F^2 e^2 \hbar}{2 \pi} \frac{4\pi^4 N_F^2 \tau^4}{ \hbar^4} \frac{4\hbar}{2\pi N_F v_F^2 \tau^3} (I_{a2})\sum_{\mathbf{q}} \gamma(q)
\end{equation}  
The total conductivity correction from intraband-intravalley Cooperon is given by 
\begin{equation}
  \sigma_0^{qi} =\mathbf{ \sigma}_{\mathrm{a_1}} +2\mathbf{\sigma}_{\mathrm{a_2}} \nonumber.
\end{equation}
To include the effect of the scattering between the band with a different index i.e., $|m\rangle_{\mathbf{k}}$ $\to$ $|n\rangle_{\mathbf{k'}}$  on the conductivity correction, we need to solve the coupled Bethe-Salpeter equations for finding the, we called as interband/intervalley Cooperons, which depends on the nature of scattering
\begin{align}
\Gamma^{m,n}_{n,m}(\mathbf{k}_1, \mathbf{k}_2) &= \gamma^{m,n}_{n,m}(\mathbf{k}_1, \mathbf{k}_2) + \int_0^{2\pi} \frac{d\varphi}{2\pi} \int_0^{\pi} \frac{d\theta \sin \theta}{2\pi} \nonumber \\
&\quad \times \int_0^{\infty} \frac{dk k^2}{2\pi} \sum_{\nu = m,n} \gamma^{m,\nu}_{n,\bar{\nu}} (\mathbf{k}_1, \mathbf{k}) G_{\mathbf{k}}^{i\epsilon_n}  G_{\mathbf{q}-\mathbf{k}}^{i\epsilon_n - i\omega_m} \Gamma^{\nu ,n}_{\bar{\nu} ,m}(\mathbf{k}, \mathbf{k}_2), \nonumber \\
\Gamma^{n,n}_{m,m}(\mathbf{k}_1, \mathbf{k}_2) &= \gamma^{n,n}_{m,m}(\mathbf{k}_1, \mathbf{k}_2) + \int_0^{2\pi} \frac{d\varphi}{2\pi} \int_0^{\pi} \frac{d\theta \sin \theta}{2\pi} \nonumber \\
&\quad \times \int_0^{\infty} \frac{dk k^2}{2\pi} \sum_{\nu = m,n} \gamma^{n,\nu}_{m,\bar{\nu}} (\mathbf{k}_1, \mathbf{k})  G_{\mathbf{k}}^{i\epsilon_n} G_{\mathbf{q}-\mathbf{k}}^{i\epsilon_n - i\omega_m} \Gamma^{\nu ,n}_{\bar{\nu} ,m}(\mathbf{k}, \mathbf{k}_2),
\end{align}
where $\bar{\nu} = n $ if $\nu = m$ and
\begin{align}
\gamma^{\nu \bar{\nu}}_{ \bar{\nu} \nu} (\mathbf{k}_1, \mathbf{k}_2) &\equiv \langle U^{\nu \bar{\nu}}_{\mathbf{k}_1, \mathbf{k}_2} U^{\nu \bar{\nu}}_{-\mathbf{k}_1, -\mathbf{k}_2} \rangle \nonumber \\
\gamma^{\nu \nu}_{\bar{\nu} \bar{\nu}} (\mathbf{k}_1, \mathbf{k}_2) &\equiv \langle U^{\nu \nu}_{\mathbf{k}_1, \mathbf{k}_2} U^{\bar{\nu} \bar{\nu}}_{-\mathbf{k}_1, -\mathbf{k}_2} \rangle \nonumber 
\end{align}
where $\nu = m, n$, $\eta_I \equiv \tau / \tau_I$, and $\eta_\ast \equiv \tau / \tau_\ast$ measure the correlation between intervalley/interband scattering. After solving the Bethe-Salpeter equation. In general, we get the interband/intervalley Cooperons as,
\begin{align}\label{interbandcoop1}
    \Gamma_{n,m}^{m,n} (\mathbf{k_1}, \mathbf{k_2})  &=\frac{4\hbar}{2\pi\tau N_F^n}\sum_{l=0}^{6} e^{ik(\phi_2 - \phi_1)} (\kappa_{n,m}^{m,n})_k(\theta)\nonumber\\
    \Gamma_{n,n}^{m,m} (\mathbf{k_1}, \mathbf{k_2})  &=\frac{4\hbar}{2\pi\tau N_F^n}\sum_{l=0}^{6} e^{ik(\phi_2 - \phi_1)} (\kappa_{- -}^{+ +})_k(\theta)
\end{align}
\begin{equation}
    \int_0^\infty \frac{k^2 \, dk}{2\pi} G^R_{{\kn}}G^A_{\mathbf{q}-{\kn}} = \frac{2\pi^2 N_F^n \tau}{\hbar}[1 - i\tau v_F q\cos{\theta} - \tau^2 v_F^2 q^2\cos^2{\theta}]
\end{equation}
The conductivity contribution from the interband/intervalley Cooperons is given by 
\[
\sigma_I = 2 \left( \sigma_{c1} + 2 \sigma_{c2} + 2 \sigma_{d2} \right),
\]
where
\begin{align}\label{sigma1}
\sigma_{c1} &= \frac{e^2 \hbar}{2\pi} \sum_{\mathbf{q}} \Gamma^{m,n}_{n,m} (\mathbf{k}, -\mathbf{k}) 
\sum_{\mathbf{k}} \tilde{v}_{\mathbf{k}, m}^x G_{\mathbf{k}, m}^R G_{\mathbf{q}-\mathbf{k}, n}^R 
\tilde{v}_{\mathbf{q}-\mathbf{k}, n}^x G_{\mathbf{q}-\mathbf{k}, n}^A G_{\mathbf{k}, m}^A, \nonumber\\
\sigma_{c2} &= \frac{e^2 \hbar}{2\pi} \sum_{\mathbf{q}} \Gamma^{m,n}_{n,m} (\mathbf{k_1}, -\mathbf{k})
\sum_{\mathbf{k}} \sum_{\mathbf{k_1}} \tilde{v}_{\mathbf{k}, m}^x G_{\mathbf{k}, m}^R G_{\mathbf{k_1}, m}^R G_{\mathbf{q}-\mathbf{k}, n}^R G_{\mathbf{q}-\mathbf{k_1}, n}^R \nonumber\\
& \quad \times \tilde{v}_{\mathbf{q}-\mathbf{k_1}, n}^x G_{\mathbf{q}-\mathbf{k_1}, n}^A G_{\mathbf{k_1}, m}^A 
\langle U_{\mathbf{k}, \kn_1}^{m,m} U_{\mathbf{q}-\mathbf{k}, \mathbf{q}-\kn_1}^{n,n} \rangle, \nonumber\\
\sigma_{d2} &= \frac{e^2 \hbar}{2\pi} \sum_{\mathbf{q}} \Gamma^{n,n}_{m,m} (\mathbf{k_1}, -\mathbf{k}) 
\sum_{\mathbf{k}} \sum_{\mathbf{k_1}} \tilde{v}_{\mathbf{k}, m}^x G_{\mathbf{k}, m}^R G_{\mathbf{k_1}, n}^R G_{\mathbf{q}-\mathbf{k}, n}^R G_{\mathbf{q}-\mathbf{k_1}, m}^R \nonumber\\
& \quad \times \tilde{v}_{\mathbf{q}-\mathbf{k_1}, m}^x G_{\mathbf{q}-\mathbf{k_1}, m}^A G_{\mathbf{k_1}, m}^A 
\langle U_{\mathbf{k}, \kn_1}^{m,n} U_{\mathbf{q}-\mathbf{k}, \mathbf{q}-\mathbf{k}_1}^{n,m} \rangle.
\end{align}
First we calculate $\sigma_{c1}$, which is given as 
\begin{align}
\sigma_{c1} &= \frac{e^2 \hbar}{2\pi} \sum_{\mathbf{q}} \Gamma^{m,n}_{n,m} (\mathbf{k}, -\mathbf{k}) 
 \sum_{\mathbf{k}} \tilde{v}_{\mathbf{k},m}^x G^R_{\mathbf{k},m}  G^R_{\mathbf{q}-\mathbf{k},n} \tilde{v}_{\mathbf{q}-\mathbf{k},n}^x  G^A_{\mathbf{q}-\mathbf{k},n} G^A_{\mathbf{k},m}\nonumber
\end{align}
In polar coordinates
\begin{align}\label{sigmac1eq}
\sigma_{c1} &= \frac{e^2 \hbar}{2\pi} \sum_{\mathbf{q}} \Gamma^{m,n}_{n,m} (\mathbf{k}, -\mathbf{k}) 
 \isum \tilde{v}_{\mathbf{k},m}^x G^R_{\mathbf{k},m}  G^R_{\mathbf{q}-\mathbf{k},n} \tilde{v}_{\mathbf{q}-\mathbf{k},n}^x  G^A_{\mathbf{q}-\mathbf{k},n} G^A_{\mathbf{k},m}
\end{align}
Since $\tilde{v}_{\mathbf{k},m}^x$, and $\tilde{v}_{\mathbf{q}-\mathbf{k},n}^x$ are functions of $\theta$ and $\phi$ , So first we evaluate the $k$ integration, 
\begin{equation}\label{sigmaintc1}
   \mathrm{I}_{c1}= \int_0^{\infty} \frac{k^2 dk}{2\pi}G^R_{\mathbf{k},m}  G^R_{\mathbf{q}-\mathbf{k},n}  G^A_{\mathbf{q}-\mathbf{k},n} G^A_{\mathbf{k},m}
\end{equation} 
For calculating $\sigma_{c1}$, from Eq.~\ref{sigmac1eq}, we need  $\Gamma^{m,n}_{n,m} (\mathbf{k}, -\mathbf{k}) $. If we substitute $\mathbf{k}$, and $-\mathbf{k}$ in place of  $\mathbf{k}_1$, and  $\mathbf{k}_2$ respectively in Eq.~\ref{interbandcoop1}, we get $\Gamma^{m,n}_{n,m} (\mathbf{k}, -\mathbf{k}) $,
\begin{align}
\sigma_{c1} &= \frac{e^2 \hbar}{2\pi} \mathrm{I}_{c1} \sum_{\mathbf{q}} \Gamma^{m,n}_{n,m} (\mathbf{k}, -\mathbf{k}) 
 \intsh \tilde{v}_{\mathbf{k},m}^x  \tilde{v}_{\mathbf{q}-\mathbf{k},n}^x  
\end{align}
\begin{align}
    v_\kn^x &= C_m \vartheta\sin{\theta}\cos{\phi}\nonumber\\
    {v}_{-\kn}^x &=  -C_m \vartheta \sin{\theta}\cos{\phi}\nonumber\\
   \tilde{v}_{\kn}^x &= \eta v_{\kn}^x\nonumber 
\end{align}
\begin{align}
\sigma_{c1} &= \frac{e^2 \hbar}{2\pi} \mathrm{I}_{c1} \sum_{\mathbf{q}}  
 \intsh  \Gamma^{m,n}_{n,m} (\mathbf{k}, -\mathbf{k})\eta_n\eta_m C_m C_n \vartheta^2 \sin^2{\theta}\cos^2{\phi}
\end{align}
where $\eta_{n,m}$ is velocity correction factor of band corresponding to the index $| n \rangle$, and $| m \rangle$ after including interband/intervalley scattering, and 
\begin{equation}\label{sigmaintc2}
    \mathrm{I}_{c2} = \intsh  \Gamma^{m,n}_{n,m} (\mathbf{k}, -\mathbf{k})\sin^2{\theta}\cos^2{\phi}
\end{equation}
then
\begin{align}\label{finalsigmac1}
\sigma_{c1} &= \frac{e^2 \hbar}{2\pi}  \eta^n\eta^m v_F^m v_F^n \times \mathrm{I}_{c1}
  \sum_{\mathbf{q}} \mathrm{I}_{c2}.
\end{align}

\begin{comment}
Calculation of $\sigma_{c2}$,
\begin{align}
\sigma_{c2} &= \frac{e^2 \hbar}{2\pi} \sum_{\mathbf{q}} \Gamma^{+-}_{-+} (\mathbf{k}_1, -\mathbf{k})  \sum_{\mathbf{k}} \sum_{\mathbf{k}_1} \tilde{v}_{\mathbf{k},+}^x G^R_{\mathbf{k},+}  G^R_{\mathbf{k}_1,+}  G^R_{\mathbf{q}-\mathbf{k},-} G^R_{\mathbf{q}-\mathbf{k}_1,-} \nonumber \\
&\times \tilde{v}_{\mathbf{q}-\mathbf{k}_1,-}  G^A_{\mathbf{q}-\mathbf{k}_1,-} G^A_{\mathbf{k},+} \left\langle U^{++}_{\mathbf{k},\mathbf{k}_1} U^{--}_{\mathbf{\mathbf{q}-k}, \mathbf{q}-\mathbf{k}_1} \right\rangle  \nonumber
\end{align} 
In polar coordinates,
\begin{align}
\sigma_{c2} &= \frac{e^2 \hbar}{2\pi} \sum_{\mathbf{q}} \Gamma^{+-}_{-+} (\mathbf{k}_1, -\mathbf{k})  \isum \intshone\int_0^{\infty}\frac{k_1^2 dk_1}{2\pi} \nonumber \\ & \widetilde{v}_{\mathbf{k},+}^x G^R_{\mathbf{k},+}  G^R_{\mathbf{k}_1,+}  G^R_{\mathbf{q}-\mathbf{k},-} G^R_{\mathbf{q}-\mathbf{k}_1,-} 
\widetilde{v}_{\mathbf{q}-\mathbf{k}_1,-}  G^A_{\mathbf{q}-\mathbf{k}_1,-} G^A_{\mathbf{k},+} \left\langle U^{++}_{\mathbf{k},\mathbf{k}_1} U^{--}_{\mathbf{\mathbf{q}-k}, \mathbf{q}-\mathbf{k}_1} \right\rangle
\end{align}
Since $\langle U^{++}_{\mathbf{k},\mathbf{k}_1} U^{--}_{\mathbf{\mathbf{q}-k}, \mathbf{q}-\mathbf{k}_1}\rangle$, $\tilde{v}_{\mathbf{k},+}^x$, and $\tilde{v}_{\mathbf{q}-\mathbf{k},-}^x$ are independent of $k$ . So first we evaluate the $k$ integration, by the method of contour integration. \\
\end{comment}

Calculation of $\sigma_{d2}$,
\begin{align}
\sigma_{d2} &= \frac{e^2 \hbar}{2\pi} \sum_{\mathbf{q}} \Gamma^{n,n}_{m,m} (\mathbf{k}_1, -\mathbf{k})  \sum_{\mathbf{k}} \sum_{\mathbf{k}_1} \tilde{v}_{\mathbf{k},m}^x G^R_{\mathbf{k},m}  G^R_{\mathbf{k}_1,n}  G^R_{\mathbf{q}-\mathbf{k},n} G^R_{\mathbf{q}-\mathbf{k}_1,n} \nonumber \\
&\times \tilde{v}_{\mathbf{q}-\mathbf{k}_1,m}  G^A_{\mathbf{q}-\mathbf{k}_1,m} G^A_{\mathbf{k},m} \left\langle U^{m,n}_{\mathbf{k},\mathbf{k}_1} U^{n,m}_{\mathbf{\mathbf{q}-k}, \mathbf{q}-\mathbf{k}_1} \right\rangle  \nonumber
\end{align} 
In polar coordinates,
\begin{align}\label{sigmad1eq}
\sigma_{d2} &= \frac{e^2 \hbar}{2\pi} \sum_{\mathbf{q}} \Gamma^{n,n}_{m,m} (\mathbf{k}_1, -\mathbf{k})  \isum \intshone\int_0^{\infty}\frac{k_1^2 dk_1}{2\pi} \nonumber \\ & \widetilde{v}_{\mathbf{k},m}^x G^R_{\mathbf{k},m}  G^R_{\mathbf{k}_1,n}  G^R_{\mathbf{q}-\mathbf{k},n} G^R_{\mathbf{q}-\mathbf{k}_1,n} 
 \tilde{v}_{\mathbf{q}-\mathbf{k}_1,m}  G^A_{\mathbf{q}-\mathbf{k}_1,m} G^A_{\mathbf{k},m} \left\langle U^{m,n}_{\mathbf{k},\mathbf{k}_1} U^{n,m}_{\mathbf{\mathbf{q}-k}, \mathbf{q}-\mathbf{k}_1} \right\rangle
\end{align}
Since $\langle U^{m,n}_{\mathbf{k},\mathbf{k}_1} U^{n,m}_{\mathbf{\mathbf{q}-k}, \mathbf{q}-\mathbf{k}_1}\rangle$, $\tilde{v}_{\mathbf{k},m}^x$, and $\tilde{v}_{\mathbf{q}-\mathbf{k},m}^x$ are independent of ${k}$ . So first, we evaluate the $k$ integration by the method of contour integration. In the limit $\mathbf{q} \to 0$ implies $\mathbf{Q} \to 0$,
\begin{align}
   \mathrm{I}_{d1}&= \int_0^\infty\frac{k^2 dk}{2\pi} G^R_{\mathbf{k},m}   G^R_{\mathbf{q}-\mathbf{k},n} G^A_{\mathbf{k},m} 
\nonumber\\
     \mathrm{I}_{d2}&=\int_0^\infty\frac{k_1^2 dk_1}{2\pi}   G^R_{\mathbf{k}_1,n}   G^R_{\mathbf{q}-\mathbf{k}_1,n} G^A_{\mathbf{q}-\mathbf{k}_1,m} 
\nonumber
\end{align}
To calculate $\sigma_{d1}$ from Eq.~\ref{sigmad1eq}, we require \( \Gamma^{n,n}_{m,m} (\mathbf{k}, -\mathbf{k}) \). By substituting \( \mathbf{k} \) and \( -\mathbf{k} \) in place of \( \mathbf{k}_1 \) and \( \mathbf{k}_2 \), respectively, in Eq.~\ref{interbandcoop1}, we obtain \( \Gamma^{m,n}_{n,m} (\mathbf{k}, -\mathbf{k}) \).
As \( \mathbf{q} \to 0 \), the following relation holds:
$ \left\langle U^{m,n}_{\mathbf{k},\mathbf{k}_1} U^{n,m}_{\mathbf{q}-\mathbf{k}, \mathbf{q}-\mathbf{k}_1} \right\rangle = \left\langle U^{m,n}_{\mathbf{k},\mathbf{k}_1} U^{n,m}_{-\mathbf{k}, -\mathbf{k}_1} \right\rangle,$
which is equal to \( \gamma^{m,n}_{n,m} \), as previously defined. Thus, from Eq.~\ref{sigmad1eq}, we can express \( \sigma_{d1} \) accordingly.

\begin{align}
\sigma_{d2} &= \frac{e^2 \hbar}{2\pi} \mathrm{I}_{d1}*\mathrm{I}_{d2}\sum_{\mathbf{q}} \Gamma^{m,n}_{n,m} (\mathbf{k}_1, -\mathbf{k})  \intsh \intshone \nonumber  \widetilde{v}_{\mathbf{k},m}^x  
\widetilde{v}_{\mathbf{q}-\mathbf{k}_1,m} \left\langle U^{m,m}_{\mathbf{k},\mathbf{k}_1} U^{n,n}_{\mathbf{\mathbf{q}-k}, \mathbf{q}-\mathbf{k}_1} \right\rangle\nonumber
\end{align}
\begin{align}
\sigma_{d2} &= \frac{e^2 \hbar}{2\pi} \mathrm{I}_{d1}*\mathrm{I}_{d2}\sum_{\mathbf{q}}   \intsh \intshone \nonumber  (\eta_m)^2 C_m^2 \vartheta^2 \\ & \times \Gamma^{m,n}_{n,m} (\mathbf{k}_1, -\mathbf{k}) \sin{\theta}\cos{\phi}\sin{\theta_1 }\cos{\phi_1 }\left\langle U^{m,m}_{\mathbf{k},\mathbf{k}_1} U^{n,n}_{\mathbf{\mathbf{q}-k}, \mathbf{q}-\mathbf{k}_1} \right\rangle.
\end{align}
\begin{align}
    \mathrm{I}_{d3} = \intsh \intshone \nonumber   \Gamma^{m,n}_{n,m} (\mathbf{k}_1, -\mathbf{k}) \sin{\theta}\cos{\phi}\sin{\theta_1 }\cos{\phi_1 }\left\langle U^{m,m}_{\mathbf{k},\mathbf{k}_1} U^{n,n}_{\mathbf{\mathbf{q}-k}, \mathbf{q}-\mathbf{k}_1} \right\rangle.
\end{align}
\begin{align}\label{finalsigmad1}
\sigma_{d2} &= \frac{e^2 \hbar}{2\pi} \mathrm{I}_{d1}*\mathrm{I}_{d2} (\eta_m)^2 C_m^2 \vartheta^2 \sum_{\mathbf{q}} \mathrm{I}_{d3}   
\end{align}
Until now, we have performed abstract calculations. As discussed in the spin-\(\frac{3}{2}\) system, there are two conduction bands; therefore, we can now carry out the calculations for both bands.
\subsection{Calculation for Upper band for pseudospin-3/2}~\label{Sec_App_B_5}
Since we are considering the upper band in the \( + \) valley, i.e., \( |m\rangle = |3/2, 3/2, +\rangle \), there are three possible types of scattering processes:  
(i) Scattering from the upper band of the \( + \) valley to the upper band of the \( - \) valley, i.e., \( |m\rangle \to |n\rangle = |3/2, 3/2, -\rangle \).  
(ii) Scattering from the upper band of the \( + \) valley to the lower band of the \( + \) valley, i.e., \( |m\rangle \to |n\rangle = |3/2, 1/2, +\rangle \).  
(iii) Scattering from the upper band of the \( + \) valley to the lower band of the \( - \) valley, i.e., \( |m\rangle \to |n\rangle = |3/2, 1/2, -\rangle \).  
We will discuss each case individually.

\textbf{Case-1} Scattering from the upper band of the \( + \) valley to the upper band of the \( - \) valley, that is, \(|n\rangle = |3/2, 3/2, -\rangle \). Since the dispersion is the same for both bands $C_m = \frac{3}{2}$ and $C_m = C_n$, due to this, the DOS are also the same i.e., $\mathrm{N}^{m}_F = \mathrm{N}^{n}_F$. 

\textit{Scattering Time}: To calculate the intravalley and intervalley scattering times, we need to evaluate Eq.~\ref{intrascattint1} and Eq.~\ref{intrascattint2}. In this case, the values are given by  
\( \mathrm{I}_{T1} = \pi N_F^m \),  
\( \mathrm{I}_{T2} = \frac{\pi n_0 u_0^2}{4\pi^2} \),  
\( \mathrm{I}_{T3} = \pi N_F^n \), and  
\( \mathrm{I}_{T4} = \frac{\pi n_0 u_I^2}{4\pi^2} \).
Substituting these values into Eq.~\ref{intrascatt1} and Eq.~\ref{intrascatt2} gives an intra- and inter- scattering time 
\begin{align}\label{scatttimeU1}
    \frac{1}{\tau_0} &= \frac{2\pi}{\hbar} N_F^{m} \frac{n_0 u_0^2}{4}, \nonumber \\
    \frac{1}{\tau_I} &= \frac{2\pi}{\hbar} N_F^{n} \frac{n_0 u_I^2}{4}.
\end{align}
 \textit{Velocity correction}: To calculate a velocity correction, we need the value of $\mathrm{I}_{V1}$ and $\mathrm{I}_{V2}$ from Eq.~\ref{intvelcor} and put those values in Eq.~\ref{expvelcor}. So the value of $\mathrm{I}_{V1}$ and $\mathrm{I}_{V2}$ are 
 \begin{align*}
      \mathrm{I}_{V1}&= \frac{n_0 u_0^2}{4\pi^2}\frac{3\pi\cos\theta}{5}\\
     \mathrm{I}_{V2}&= -\frac{n_0 u_I^2}{4\pi^2}\frac{3\pi\cos\theta}{5}
 \end{align*}
 and from Eq.~\ref{scatttimeU1}
\begin{align*}
    {\frac{4\hbar}{2\pi\tau_{0}} =  N_F^m n_0 u_0^2 }\nonumber\\
    {\frac{4\hbar}{2\pi\tau_{I}} = N_F^n n_0 u_I^2}\nonumber
\end{align*}
after putting all these values in Eq.~\ref{expvelcor}, we get 
\begin{equation}
   {\eta = \frac{5}{2+6\eta_I}}
\end{equation}
\textit{Conductivity correction}: For calculating conductivity correction from Eq.~\ref{conduccorr}, we need to find $\sigma_{a1}$ and $\sigma_{a2}$, where $\sigma_{a1}$ is solely depends on the factor $\gamma(q) = {\chi_0}/({q_0^2 + q^2 })$, which we can find after solving the Bethe-Salpeter equation Eq.~\ref{BSE},
\begin{align*}
    \chi_0(\eta_I) &= \frac{9 (34 + \eta_I) (2 + 3\eta_I) (64 + 21\eta_I)}
{640 (204 + \eta_I (338 + \eta_I (95 + 3\eta_I)))}\\
q^2_0(\eta_I) &= \frac{9 \eta_I (34 + \eta_I) (2 + 3\eta_I) (4 + \eta_I) \eta(\eta_I)}
{10 (1 - \eta_I) (204 + \eta_I (338 + \eta_I (95 + 3\eta_I))) 3 l^2}
\end{align*}
and for calculating $\sigma_{a2}$, we need to evalute the value of integral $\mathrm{I}_{a2}$ from Eq.~\ref{dseconduc}, the value of $\mathrm{I}_{a2} = 1/(20\pi^2)$.Thus from Eq.~\ref{conduccorr}, the conductivity correction is given by 
\begin{align}
    \sigma^{qi}_0(B) &= \frac{16 e^2 \eta[\eta_I]^2}{15 h}\sum_{\mathbf{q}}\frac{\chi_0}{q_0^2 + q^2 }
\end{align}
To calculate the magneto conductivity, we have to replace $\sum_{\mathbf{q}}\frac{\chi_0}{q_0^2 + q^2 }$ in $\sigma^{qi}_0(B)$ by 
\begin{align}
    \Psi_3(B,q_0) &=  \int_0^{1/\ell} \frac{dx}{(2\pi)^2} \bigg[ 
    \psi \left( \frac{\ell_B^2}{\ell^2} \right. \nonumber  \left. + \ell_B^2(q_0^2+ x^2) + \frac{1}{2} \right) 
    - \psi \left( \frac{\ell_B^2}{\ell_\phi^2} \right. \left. + \ell_B^2 (q_0^2+ x^2) + \frac{1}{2} \right)
    \bigg],
\end{align}
where magnetic length is $\ell_{\textit{B}}$ = $\sqrt{{\hbar}/{4e\textit{B}}}$, the magnetic field $B$ is along arbitrary directions. The magnetoconductivity is defined as
\begin{equation}
    \delta\sigma^{qi}(B)=\sigma^{qi}(B)-\sigma^{qi}(0).
\end{equation}

This is the quantum correction in conductivity due to the intravalley Cooperons. Our next task is to find the effect of intervalley cooperons on the magnitude and nature of conductivity.

For finding the effect of intervalley Cooperons on the conductivity correction due to intravalley Cooperons, we need to evaluate $\sigma_{c1}$, $\sigma_{c2}$, and $\sigma_{d1}$ from Eq.~\ref{sigma1}. To find the value of $\sigma_{c1}$, first we need to evaluate $\mathrm{I}_{c1}$,and $\mathrm{I}_{c2}$ from Eq.~\ref{sigmaintc1}, and \ref{sigmaintc2} respectively. In this case, the dispersion relation is the same for the initial and final band; thus, the value of $\mathrm{I}_{c1}$ and $\mathrm{I}_{c2}$ are
\begin{align*}
    \mathrm{I}_{c1}= \frac{4\pi^2 N_F^n\tau^3}{  \hbar^3}
\end{align*}
\begin{align*}
    \mathrm{I}_{c2}= \frac{35 a_0 - 10 a_1 + 5 a_2 - 2 a_3}{630 \pi}
\end{align*}
Put the value of  $\mathrm{I}_{c1}$, and $\mathrm{I}_{c2}$ in Eq.~\ref{finalsigmac1}, we get
\begin{equation}
    \sigma_{c1}=-\frac{4\eta(\eta_I)^2}{315}(35 a_0 - 10 a_1 + 5 a_2 - 2 a_3).
\end{equation}
To find the value of $\sigma_{d1}$, first we need to evaluate $\mathrm{I}_{d1}$, $\mathrm{I}_{d2}$, and $\mathrm{I}_{d3}$ from Eq.~\ref{sigmaintc1}, \ref{sigmaintc2} and , \ref{sigmaintc2} respectively. In this case, the dispersion relation is the same for the initial and final band; thus, the value of $\mathrm{I}_{d1}$,  $\mathrm{I}_{d2}$, and $\mathrm{I}_{d3}$ are
\begin{align*}
    \mathrm{I}_{d1}= \frac{2\pi^2 N_F^n \tau^2}{ i\hbar^2}
\end{align*}
\begin{align*}
    \mathrm{I}_{d2}= \frac{2\pi^2 N_F^n \tau^2}{ i\hbar^2}
\end{align*}
\begin{align*}
    \mathrm{I}_{d3}= -\frac{3 b_0 + 8 b_1 + 11 b_2 + 6 b_3}{784 \pi^2}
\end{align*}
Put the value of  $\mathrm{I}_{d1}$, $\mathrm{I}_{d2}$, and $\mathrm{I}_{d3}$ in Eq.~\ref{finalsigmad1}, we get
\begin{equation}
    \sigma_{d1}=-\frac{\eta_I \times\eta(\eta_I)^2}{49}(3 b_0 + 8 b_1 + 11 b_2 + 6 b_3).
\end{equation}

\textbf{Case-2} Scattering from the upper band of the \( + \) valley to the lower band of the \( + \) valley, that is, \( |n\rangle = |3/2, 1/2, + \rangle\). The values are $C_m =3/2$ and $C_n = 1/2$ for the upper and lower band, respectively. The DOS depends on the factor $C_m$; due to the different values of $C_m$, the bands have different DOS.  

\textit{Scattering Time}: To calculate the intraband-intravalley and interband-intravalley scattering times, we need to evaluate Eqs.~\ref{intrascattint1} and \ref{intrascattint2}. In this case, the values are given by  
\( \mathrm{I}_{T1} = \pi N_F^m \),  
\( \mathrm{I}_{T2} = \frac{\pi n_0 u_0^2}{4\pi^2} \),  
\( \mathrm{I}_{T3} = \pi N_F^n \), and  
\( \mathrm{I}_{T4} = \frac{\pi n_0 u_I^2}{4\pi^2} \).
Substituting these values into Eqs.~\ref{intrascatt1} and \ref{intrascatt2} gives an intra- and inter- scattering time 
\begin{align}\label{scatttimeU2}
    \frac{1}{\tau_0} &= \frac{2\pi}{\hbar} N_F^{m} \frac{n_0 u_0^2}{4}, \nonumber \\
    \frac{1}{\tau_I} &= \frac{2\pi}{\hbar} N_F^{n} \frac{n_0 u_I^2}{4}.
\end{align}
 \textit{Velocity correction}: To calculate a velocity correction, we need the value of $\mathrm{I}_{V1}$ and $\mathrm{I}_{V2}$ from Eq.~\ref{intvelcor} and put those values in Eq.~\ref{expvelcor}. So the value of $\mathrm{I}_{V1}$ and $\mathrm{I}_{V2}$ are 
 \begin{align*}
      \mathrm{I}_{V1}&= \frac{n_0 u_0^2}{4\pi^2}\frac{3\pi\cos\theta}{5}\\
     \mathrm{I}_{V2}&= \frac{n_0 u_I^2}{4\pi^2}\frac{1\pi\cos\theta}{5}
 \end{align*}
 and from Eq.~\ref{scatttimeU2}
\begin{align*}
    {\frac{4\hbar}{2\pi\tau_{0}} =  N_F^{m} n_0 u_0^2 }\nonumber\\
    {\frac{4\hbar}{2\pi\tau_{I}} = N_F^{n}  n_0 u_I^2}\nonumber
\end{align*}
After putting all these values in Eq.~\ref{expvelcor}, we get the velocity correction for the upper band due to including scattering from the upper band to the lower band.
\begin{equation}
   {\eta_m = \frac{15}{6+8\eta_I}}
\end{equation}
Similarly, we can find a velocity correction for the lower band, where the values of $\mathrm{I}_{V1}$ and $\mathrm{I}_{V2}$ are 
 \begin{align*}
      \mathrm{I}_{V1}&= \frac{n_0 u_0^2}{4\pi^2}\frac{\pi\cos\theta}{15}\\
     \mathrm{I}_{V2}&= \frac{n_0 u_I^2}{4\pi^2}\frac{1\pi\cos\theta}{5}
 \end{align*}
 and the velocity correction for the lower band is 
 \begin{equation}
   {\eta_n = \frac{15}{14-8\eta_I}}.
\end{equation}
\textit{Conductivity correction}: For calculating conductivity correction from Eq.~\ref{conduccorr}, we need to find $\sigma_{a1}$ and $\sigma_{a2}$, where $\sigma_{a1}$ is solely depends on the factor $\gamma(q) = {\chi_0}/({q_0^2 + q^2 })$, which we can find after solving the Bethe-Salpeter equation Eq.~\ref{BSE},
\begin{align*}
    \chi_0(\eta_I) &= \frac{9 (34 + \eta_I) (2 + 3\eta_I) (64 + 21\eta_I)}
{640 (204 + \eta_I (338 + \eta_I (95 + 3\eta_I)))}\\
q^2_0(\eta_I) &= \frac{9 \eta_I (34 + \eta_I) (2 + 3\eta_I) (4 + \eta_I) \eta(\eta_I)}
{10 (1 - \eta_I) (204 + \eta_I (338 + \eta_I (95 + 3\eta_I))) 3 l^2}
\end{align*}
and for calculating $\sigma_{a2}$, we need to evalute the value of integral $\mathrm{I}_{a2}$ from Eq.~\ref{dseconduc}, the value of $\mathrm{I}_{a2} = 1/(20\pi^2)$.Thus from Eq.~\ref{conduccorr}, the conductivity correction is given by 
\begin{align}
    \sigma^{qi}_0(B) &= \frac{16 e^2 \eta[\eta_I]^2}{15 h}\sum_{\mathbf{q}}\frac{\chi_0}{q_0^2 + q^2 }
\end{align}
To calculate the magneto conductivity, we have to replace $\sum_{\mathbf{q}}\frac{\chi_0}{q_0^2 + q^2 }$ in $\sigma^{qi}_0(B)$ by 
\begin{align}
    \Psi_3(B,q_0) &=  \int_0^{1/\ell} \frac{dx}{(2\pi)^2} \bigg[ 
    \psi \left( \frac{\ell_B^2}{\ell^2} \right. \nonumber  \left. + \ell_B^2(q_0^2+ x^2) + \frac{1}{2} \right) 
    - \psi \left( \frac{\ell_B^2}{\ell_\phi^2} \right. \left. + \ell_B^2 (q_0^2+ x^2) + \frac{1}{2} \right)
    \bigg],
\end{align}
where magnetic length is $\ell_{\textit{B}}$ = $\sqrt{{\hbar}/{4e\textit{B}}}$, the magnetic field $B$ is along arbitrary directions. The magnetoconductivity is defined as
\begin{equation}
    \delta\sigma^{qi}(B)=\sigma^{qi}(B)-\sigma^{qi}(0).
\end{equation}

This is the quantum correction in conductivity due to the intraband Cooperons. Our next task is to find the effect of interband Cooperons on the magnitude and nature of conductivity.

To find the effect of interband Cooperons on the conductivity correction due to intraband Cooperons, we need to evaluate $\sigma_{c1}$, $\sigma_{c2}$, and $\sigma_{d1}$ from Eq.~\ref{sigma1}. To find the value of $\sigma_{c1}$, we first need to evaluate $\mathrm{I}_{c1}$,and $\mathrm{I}_{c2}$ from Eq.~\ref{sigmaintc1}, and \ref{sigmaintc2} respectively. $\mathrm{I}_{c1}$, and $\mathrm{I}_{c2}$ are
\begin{align*}
    \mathrm{I}_{c1}= \frac{2\pi^2 N_F^n\tau^3}{  \hbar^3}
\end{align*}
\begin{align*}
    \mathrm{I}_{c2} &=
\frac{1}{630 \pi} \Big( 5 a_0 - 12 a_{10} + 3 a_{11} - 3 a_{12} + 2 a_{13} 
+ 63 a_{20} + 21 a_{21} - 33 a_{22} - 21 a_{23} - 15 a_{24} \\ 
\quad &
+ 3 a_{25} - 33 a_{26} - 3 a_{27} + 31 a_{28} + 18 a_{29} 
- 168 a_{30} + 120 a_{31} + 24 a_{32} + 120 a_{33} - 104 a_{34} \\
\quad &
+ 63 a_{40} - 21 a_{41} - 33 a_{42} - 18 a_{43} + 21 a_{44} 
- 15 a_{45} - 3 a_{46} - 12 a_{47} - 33 a_{48} + 3 a_{49} 
\\
\quad &
- 12 a_{50} - 3 a_{51} + 3 a_{52} + 2 a_{53} + 10 a_{66} 
- 12 a_{210} + 31 a_{410} \Big)
\end{align*}
Put the value of  $\mathrm{I}_{c1}$, and $\mathrm{I}_{c2}$ in Eq.~\ref{finalsigmac1}, we get
\begin{equation}
    \sigma_{c1}=-\frac{12 \eta_m\eta_n}{630}\mathrm{I}_{c2}.
\end{equation}
To find the value of $\sigma_{d1}$, we first need to evaluate $\mathrm{I}_{d1}$, $\mathrm{I}_{d2}$, and $\mathrm{I}_{d3}$ from Eq.~\ref{sigmaintc1}, \ref{sigmaintc2} and , \ref{sigmaintc2} respectively. In this case, the dispersion relation is the same for the initial and final band; thus, the value of $\mathrm{I}_{d1}$,  $\mathrm{I}_{d2}$, and $\mathrm{I}_{d3}$ are
\begin{align*}
    \mathrm{I}_{d1}= \frac{\pi^2 N_F^n \tau^2}{ i\hbar^2}
\end{align*}
\begin{align*}
    \mathrm{I}_{d2}= \frac{\pi^2 N_F^n \tau^2}{ i\hbar^2}
\end{align*}
\begin{align*}
    \mathrm{I}_{d3} &= 
\frac{1}{58800 \pi^2} \Big( 42 b_0 - 53 b_{10} - 23 b_{11} - 23 b_{12} - 5 b_{13} 
+ 98 b_{20} + 98 b_{21} - 14 b_{22} + 98 b_{23} + 78 b_{24} + 98 b_{25} \\
\quad &
+ 58 b_{26} - 46 b_{27} + 70 b_{28} + 42 b_{29} - 392 b_{30} + 56 b_{31} 
- 8 b_{32} + 56 b_{33} + 120 b_{34} + 98 b_{40} - 98 b_{41} - 14 b_{42}  \\
\quad &
- 98 b_{43} + 58 b_{44} + 46 b_{45} - 14 b_{46} + 46 b_{47} - 30 b_{48} 
+ 98 b_{49} - 53 b_{50} + 23 b_{51} + 23 b_{52} - 5 b_{53} + 168 b_{66} \\
\quad & 
- 14 b_{210} - 46 b_{211} - 30 b_{212} - 78 b_{410} - 70 b_{411} + 42 b_{412} \Big)
\end{align*}
Put the value of  $\mathrm{I}_{d1}$, $\mathrm{I}_{d2}$, and $\mathrm{I}_{d3}$ in Eq.~\ref{finalsigmad1}, we get
\begin{equation}
    \sigma_{d1}=-\frac{36 \eta_I \times(\eta_m)^2}{58800}\mathrm{I}_{d3}.
\end{equation}

\textbf{Case-3} Scattering from the upper band of the \( + \) valley to the lower band of the \( - \) valley, that is, \( |n\rangle = |3/2, 1/2, - \rangle\). The values are $C_m =3/2$ and $C_n = 1/2$ for the upper and lower band, respectively. The DOS depends on the factor $C_m$; due to the different values of $C_m$, the bands have different DOS.  

\textit{Scattering Time}: To calculate the intraband-intravalley and interband-intervalley scattering times, we need to evaluate Eqs.~\ref{intrascattint1} and \ref{intrascattint2}. In this case, the values are given by  
\( \mathrm{I}_{T1} = \pi N_F^{m} \),  
\( \mathrm{I}_{T2} = \frac{\pi n_0 u_0^2}{4\pi^2} \),  
\( \mathrm{I}_{T3} = \pi N_F^{n} \), and  
\( \mathrm{I}_{T4} = \frac{\pi n_0 u_I^2}{4\pi^2} \).
Substituting these values into Eqs.~\ref{intrascatt1} and \ref{intrascatt2} gives an intra- and inter- scattering time 
\begin{align}\label{scatttimeL1}
    \frac{1}{\tau_0} &= \frac{2\pi}{\hbar} N_F^{m} \frac{n_0 u_0^2}{4}, \nonumber \\
    \frac{1}{\tau_I} &= \frac{2\pi}{\hbar} N_F^{n} \frac{n_0 u_I^2}{4}.
\end{align}
 \textit{Velocity correction}: To calculate a velocity correction, we need the value of $\mathrm{I}_{V1}$ and $\mathrm{I}_{V2}$ from Eq.~\ref{intvelcor} and put those values in Eq.~\ref{expvelcor}. So the value of $\mathrm{I}_{V1}$ and $\mathrm{I}_{V2}$ are 
 \begin{align*}
      \mathrm{I}_{V1}&= \frac{n_0 u_0^2}{4\pi^2}\frac{3\pi\cos\theta}{5}\\
     \mathrm{I}_{V2}&= -\frac{n_0 u_I^2}{4\pi^2}\frac{1\pi\cos\theta}{5}
 \end{align*}
 and from Eq.~\ref{scatttimeL1}
\begin{align*}
    {\frac{4\hbar}{2\pi\tau_{0}} =  N_F^{m} n_0 u_0^2 }\nonumber\\
    {\frac{4\hbar}{2\pi\tau_{I}} = N_F^{n}  n_0 u_I^2}\nonumber
\end{align*}
After putting all these values in Eq.~\ref{expvelcor}, we get the velocity correction for the upper band due to including scattering from the upper band to the lower band.
\begin{equation}
   {\eta_m = \frac{15}{6+10\eta_I}}
\end{equation}
Similarly, we can find a velocity correction for the lower band, where the values of $\mathrm{I}_{V1}$ and $\mathrm{I}_{V2}$ are 
 \begin{align*}
      \mathrm{I}_{V1}&= \frac{n_0 u_0^2}{4\pi^2}\frac{\pi\cos\theta}{15}\\
     \mathrm{I}_{V2}&= -\frac{n_0 u_I^2}{4\pi^2}\frac{1\pi\cos\theta}{5}
 \end{align*}
 and the velocity correction for the lower band is 
 \begin{equation}
   {\eta_n = \frac{15}{14+10\eta_I}}.
\end{equation}
\textit{Conductivity correction}: For calculating conductivity correction from Eq.~\ref{conduccorr}, we need to find $\sigma_{a1}$ and $\sigma_{a2}$, which we can find after solving the Bethe-Salpeter equation Eq.~\ref{BSE},
\begin{align*}
    \chi_0(\eta_I) &= \frac{9 (34 + \eta_I) (2 + 3\eta_I) (64 + 21\eta_I)}
{640 (204 + \eta_I (338 + \eta_I (95 + 3\eta_I)))}\\
q^2_0(\eta_I) &= \frac{9 \eta_I (34 + \eta_I) (2 + 3\eta_I) (4 + \eta_I) \eta(\eta_I)}
{10 (1 - \eta_I) (204 + \eta_I (338 + \eta_I (95 + 3\eta_I))) 3 l^2}
\end{align*}
and for calculating $\sigma_{a2}$, we need to evaluate the value of integral $\mathrm{I}_{a2}$ from Eq.~\ref{dseconduc}, the value of $\mathrm{I}_{a2} = 1/(20\pi^2)$.Thus from Eq.~\ref{conduccorr}, the conductivity correction is given by 
\begin{align}
    \sigma^{qi}_0(B) &= \frac{16 e^2 \eta[\eta_I]^2}{15 h}\sum_{\mathbf{q}}\frac{\chi_0}{q_0^2 + q^2 }
\end{align}
To calculate the magneto conductivity, we have to replace $\sum_{\mathbf{q}}\frac{\chi_0}{q_0^2 + q^2 }$ in $\sigma^{qi}_0(B)$ by 
\begin{align}
    \Psi_3(B,q_0) &=  \int_0^{1/\ell} \frac{dx}{(2\pi)^2} \bigg[ 
    \psi \left( \frac{\ell_B^2}{\ell^2} \right. \nonumber  \left. + \ell_B^2(q_0^2+ x^2) + \frac{1}{2} \right) 
    - \psi \left( \frac{\ell_B^2}{\ell_\phi^2} \right. \left. + \ell_B^2 (q_0^2+ x^2) + \frac{1}{2} \right)
    \bigg],
\end{align}
where magnetic length is $\ell_{\textit{B}}$ = $\sqrt{{\hbar}/{4e\textit{B}}}$, the magnetic field $B$ is along arbitrary directions. The magnetoconductivity is defined as
\begin{equation}
    \delta\sigma^{qi}(B)=\sigma^{qi}(B)-\sigma^{qi}(0).
\end{equation}

This is the quantum correction in conductivity due to the intraband Cooperons. Our next task is to find the effect of interband Cooperons on the magnitude and nature of conductivity.

To find the effect of interband Cooperons on the conductivity correction due to intraband Cooperons, we need to evaluate $\sigma_{c1}$, $\sigma_{c2}$, and $\sigma_{d1}$ from Eq.~\ref{sigma1}. To find the value of $\sigma_{c1}$, we first need to evaluate $\mathrm{I}_{c1}$,and $\mathrm{I}_{c2}$ from Eq.~\ref{sigmaintc1}, and \ref{sigmaintc2} respectively. $\mathrm{I}_{c1}$, and $\mathrm{I}_{c2}$ are
\begin{align*}
    \mathrm{I}_{c1}= \frac{2\pi^2 N_F^n\tau^3}{  \hbar^3}
\end{align*}
\begin{align*}
    \mathrm{I}_{c2} &=
\frac{1}{630 \pi} \Big( 10 a_0 - 30 a_{10} + 15 a_{11} - 15 a_{12} + 10 a_{13} 
+ 12 a_{20} - 3 a_{21} + 3 a_{22} - 2 a_{23} - 144 a_{30} + 16 a_{31} \\
\quad &
+ 12 a_{40} + 3 a_{41} - 3 a_{42} - 2 a_{43} - 30 a_{50} - 15 a_{51} 
+ 15 a_{52} + 10 a_{53} + 10 a_{66} \Big)
\end{align*}
Put the value of  $\mathrm{I}_{c1}$, and $\mathrm{I}_{c2}$ in Eq.~\ref{finalsigmac1}, we get
\begin{equation}
    \sigma_{c1}=-\frac{12 \eta_m\eta_n}{630}\mathrm{I}_{c2}.
\end{equation}
To find the value of $\sigma_{d1}$, we first need to evaluate $\mathrm{I}_{d1}$, $\mathrm{I}_{d2}$, and $\mathrm{I}_{d3}$ from Eq.~\ref{sigmaintc1}, \ref{sigmaintc2} and , \ref{sigmaintc2} respectively. In this case, the dispersion relation is the same for the initial and final band; thus, the value of $\mathrm{I}_{d1}$,  $\mathrm{I}_{d2}$, and $\mathrm{I}_{d3}$ are
\begin{align*}
    \mathrm{I}_{d1}= \frac{\pi^2 N_F^n \tau^2}{ i\hbar^2}
\end{align*}
\begin{align*}
    \mathrm{I}_{d2}= \frac{\pi^2 N_F^n \tau^2}{ i\hbar^2}
\end{align*}
\begin{align*}
    \mathrm{I}_{d3} &= 
\frac{1}{29400 \pi^2} \Big( 25 b_0 + 12 b_{10} - 2 b_{11} - 2 b_{12} - 8 b_{13} 
+ 30 b_{20} + 8 b_{21} + 8 b_{22} + b_{23} + 416 b_{30} - 16 b_{31} \\
\quad &
+ 30 b_{40} - 8 b_{41} - 8 b_{42} + b_{43} 
+ 2 \big( 6 b_{50} + b_{51} + b_{52} - 4 b_{53} \big) + 25 b_{66} \Big)
\end{align*}
Put the value of  $\mathrm{I}_{d1}$, $\mathrm{I}_{d2}$, and $\mathrm{I}_{d3}$ in Eq.~\ref{finalsigmad1}, we get
\begin{equation}
    \sigma_{d1}=-\frac{36 \eta_I \times(\eta_m)^2}{29400}\mathrm{I}_{d3}.
\end{equation}
\subsection{Calculation for lower band for pseudospin-3/2}~\label{Sec_App_B_6}
\textbf{Case-1} Scattering from the lower band of the \( + \) valley, that is, \(|n\rangle = |3/2, 1/2, +\rangle \) to the lower band of the \( - \) valley, that is, \(|n\rangle = |3/2, 1/2, -\rangle \). Since the dispersion is the same for both bands $C_m = \frac{3}{2}$ and $C_m = C_n$, due to this, the DOS are also the same i.e., $\mathrm{N}^{m}_F = \mathrm{N}^{n}_F$. 

\textit{Scattering Time}: To calculate the intravalley and intervalley scattering times, we need to evaluate Eqs.~\ref{intrascattint1} and \ref{intrascattint2}. In this case, the values are given by  
\( \mathrm{I}_{T1} = \pi N_F^m \),  
\( \mathrm{I}_{T2} = \frac{\pi n_0 u_0^2}{4\pi^2} \),  
\( \mathrm{I}_{T3} = \pi N_F^n \), and  
\( \mathrm{I}_{T4} = \frac{\pi n_0 u_I^2}{4\pi^2} \).
Substituting these values into Eqs.~\ref{intrascatt1} and \ref{intrascatt2} gives an intra- and inter- scattering time 
\begin{align}\label{scatttimeL2}
    \frac{1}{\tau_0} &= \frac{2\pi}{\hbar} N_F^m \frac{n_0 u_0^2}{4}, \nonumber \\
    \frac{1}{\tau_I} &= \frac{2\pi}{\hbar} N_F^n \frac{n_0 u_I^2}{4}.
\end{align}
 \textit{Velocity correction}: To calculate a velocity correction, we need the value of $\mathrm{I}_{V1}$ and $\mathrm{I}_{V2}$ from Eq.~\ref{intvelcor} and put those values in Eq.~\ref{expvelcor}. So the value of $\mathrm{I}_{V1}$ and $\mathrm{I}_{V2}$ are 
 \begin{align*}
      \mathrm{I}_{V1}&= \frac{n_0 u_0^2}{4\pi^2}\frac{\pi\cos\theta}{15}\\
     \mathrm{I}_{V2}&= -\frac{n_0 u_I^2}{4\pi^2}\frac{\pi\cos\theta}{15}
 \end{align*}
 and from Eq.~\ref{scatttimeL2}
\begin{align*}
    {\frac{4\hbar}{2\pi\tau_{0}} =  N_F^{m} n_0 u_0^2 }\nonumber\\
    {\frac{4\hbar}{2\pi\tau_{I}} = N_F^{n}  n_0 u_I^2}\nonumber
\end{align*}
after putting all these values in Eq.~\ref{expvelcor}, we get 
\begin{equation}
   {\eta = \frac{15}{14+2\eta_I}}
\end{equation}
\textit{Conductivity correction}: For calculating conductivity correction from Eq.~\ref{conduccorr}, we need to find $\sigma_{a1}$ and $\sigma_{a2}$, which we can find after solving the Bethe-Salpeter equation Eq.~\ref{BSE},
\begin{align*}
    \chi_0(\eta_I) &= \frac{(-1 + \eta_I) (14 + \eta_I) (26 + 9\eta_I) (64 + 21\eta_I)}
{640 (-1 + \eta_I) (52 + \eta_I (86 + \eta_I (21 + \eta_I)))}\\
q^2_0(\eta_I) &= \frac{-64 n (4 + n) (14 + n) (26 + 9n) \eta(n)}
{640 (-1 + n) (52 + n (86 + n (21 + n))) 3 l^2}
\end{align*}
and for calculating $\sigma_{a2}$, we need to evalute the value of integral $\mathrm{I}_{a2}$ from Eq.~\ref{dseconduc}, the value of $\mathrm{I}_{a2} = 1/(20\pi^2)$.Thus from Eq.~\ref{conduccorr}, the conductivity correction is given by 
\begin{align}
    \sigma^{qi}_0(B) &= \frac{112 e^2 \eta[\eta_I]^2}{45 h}\sum_{\mathbf{q}}\frac{\chi_0}{q_0^2 + q^2 }
\end{align}
To calculate the magneto conductivity, we have to replace $\sum_{\mathbf{q}}\frac{\chi_0}{q_0^2 + q^2 }$ in $\sigma^{qi}_0(B)$ by 
\begin{align}
    \Psi_3(B,q_0) &=  \int_0^{1/\ell} \frac{dx}{(2\pi)^2} \bigg[ 
    \psi \left( \frac{\ell_B^2}{\ell^2} \right. \nonumber  \left. + \ell_B^2(q_0^2+ x^2) + \frac{1}{2} \right) 
    - \psi \left( \frac{\ell_B^2}{\ell_\phi^2} \right. \left. + \ell_B^2 (q_0^2+ x^2) + \frac{1}{2} \right)
    \bigg],
\end{align}
where magnetic length is $\ell_{\textit{B}}$ = $\sqrt{{\hbar}/{4e\textit{B}}}$, the magnetic field $B$ is along arbitrary directions. The magnetoconductivity is defined as
\begin{equation}
    \delta\sigma^{qi}(B)=\sigma^{qi}(B)-\sigma^{qi}(0).
\end{equation}

This is the quantum correction in conductivity due to the intravalley Cooperons. Our next task is to find the effect of intervalley cooperons on the magnitude and nature of conductivity.

For finding the effect of intervalley Cooperons on the conductivity correction due to intravalley Cooperons, we need to evaluate $\sigma_{c1}$, $\sigma_{c2}$, and $\sigma_{d1}$ from Eq.~\ref{sigma1}. To find the value of $\sigma_{c1}$, we first need to evaluate $\mathrm{I}_{c1}$,and $\mathrm{I}_{c2}$ from Eq.~\ref{sigmaintc1}, and \ref{sigmaintc2} respectively. In this case, the dispersion relation is the same for the initial and final band; thus, the values of $\mathrm{I}_{c1}$ and $\mathrm{I}_{c2}$ are
\begin{align*}
    \mathrm{I}_{c1}= \frac{4\pi^2 N_F^n\tau^3}{  \hbar^3}
\end{align*}
\begin{align*}
    \mathrm{I}_{c2}&= \frac{1}{630 \pi} \Big( 160 a_0 - 192 a_{11} + 48 a_{12} - 48 a_{13} + 32 a_{14} 
+ 63 a_{21} - 21 a_{22} - 33 a_{23} + 21 a_{24} - 15 a_{25} - 3 a_{26}\\
\quad & 
- 33 a_{27} + 3 a_{28} + 31 a_{29} - 168 a_{31} + 120 a_{32} 
+ 120 a_{33} - 104 a_{34} + 63 a_{41} - 33 a_{42} - 21 a_{43} \\
\quad &
- 15 a_{44} + 3 a_{45} - 33 a_{46} - 3 a_{47} + 31 a_{48} + 21 a_{49} 
- 24 a_{51} - 6 a_{52} + 6 a_{53} + 4 a_{54} + 10 a_{66} \Big)
\end{align*}
Put the value of  $\mathrm{I}_{c1}$, and $\mathrm{I}_{c2}$ in Eq.~\ref{finalsigmac1}, we get
\begin{equation}
    \sigma_{c1}=-\frac{4\eta(\eta_I)^2}{315} \mathrm{I}_{c2}.
\end{equation}
To find the value of $\sigma_{d1}$, we first need to evaluate $\mathrm{I}_{d1}$, $\mathrm{I}_{d2}$, and $\mathrm{I}_{d3}$ from Eq.~\ref{sigmaintc1}, \ref{sigmaintc2} and , \ref{sigmaintc2} respectively. In this case, the dispersion relation is the same for the initial and final band; thus, the value of $\mathrm{I}_{d1}$,  $\mathrm{I}_{d2}$, and $\mathrm{I}_{d3}$ are
\begin{align*}
    \mathrm{I}_{d1}= \frac{2\pi^2 N_F \tau^2}{ i\hbar^2}
\end{align*}
\begin{align*}
    \mathrm{I}_{d2}= \frac{2\pi^2 N_F \tau^2}{ i\hbar^2}
\end{align*}
\begin{align*}
    \mathrm{I}_{d3}&= -\frac{1}{44100 \pi^2} \Big( 6 b_0 + 330 b_{10} + 102 b_{11} + 102 b_{12} 
+ 34 b_{13} + 245 b_{20} + 49 b_{21} - 147 b_{22} + 49 b_{23} + 41 b_{24} \\
\quad & 
+ 9 b_{25} - 147 b_{26} + 9 b_{27} + 141 b_{28} + 980 b_{30} - 588 b_{31} 
- 588 b_{32} + 372 b_{33} + 245 b_{40} - 147 b_{41} - 49 b_{42} + 41 b_{43} \\
\quad & 
- 9 b_{44} - 147 b_{45} - 9 b_{46} + 141 b_{47} - 49 b_{48} + 330 b_{50} 
- 102 b_{51} - 102 b_{52} + 34 b_{53} + 6 b_{66} \Big)
\end{align*}
Put the value of  $\mathrm{I}_{d1}$, $\mathrm{I}_{d2}$, and $\mathrm{I}_{d3}$ in Eq.~\ref{finalsigmad1}, we get
\begin{equation}
    \sigma_{d1}=-\frac{16\eta_I \times\eta(\eta_I)^2}{44100}\mathrm{I}_{d3}.
\end{equation}
\subsection{Effect of intervalley scattering on the pseudospin-\texorpdfstring{$1/2$}{1/2}, and \texorpdfstring{$1$}{1}}\label{spin1and1/2_head}

 \begin{figure*}
    \centering
    \includegraphics{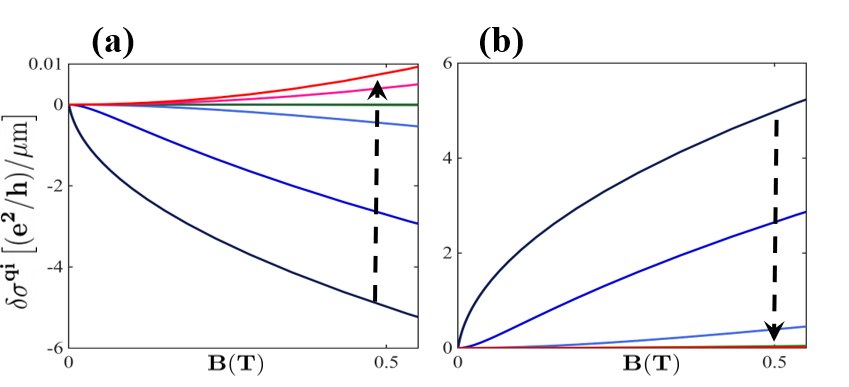}
    \caption{(a) Magnetoconductivity for pseudospin $s=1/2$ fermions with intervalley scattering,  (b) Magnetoconductivity for pseudospin $s=1$ fermions with intervalley scattering. The black curve corresponds to the $\eta_I = 0$ case and arrow in the direction of increasing $\eta_I$.}
    \label{spin1and1/2}
\end{figure*}

In the absence of intervalley scattering ($\eta_I = 0$), the pseudospin-$\tfrac{1}{2}$ Dirac system, which belongs to the symplectic class ($\mathcal{T}^2 = -1$), exhibits the familiar weak-antilocalization (positive magnetoconductivity) at low magnetic field. In contrast, the pseudospin-1 system is in the orthogonal class ($\mathcal{T}^2 = +1$) and shows a weak-localization (negative magnetoconductivity) in the intravalley channel. These behaviors are evident in Fig.~1(a, b), where the black intravalley curve for pseudospin-$\tfrac{1}{2}$ has a sharp positive peak, whereas for pseudospin-1 it has a negative cusp.

When intervalley scattering is turned on ($\eta_I > 0$), the low-field quantum correction is steadily suppressed. For the symplectic pseudospin-$\tfrac{1}{2}$ case, the WAL peak is gradually reduced, and eventually changes sign: we find a WAL $\rightarrow$ WL crossover around $\eta_I \approx 0.71$. By contrast, the orthogonal pseudospin-1 system simply loses its interference correction as $\eta_I \rightarrow 1$, with no sign change. Thus, the pseudospin-1 magnetoconductivity is monotonically suppressed to zero with increasing $\eta_I$, consistent with earlier results~\cite{miao2023weak, lu2015weak}.

%\end{widetext}

%% file: main.bbl
\begin{thebibliography}{46}%
\makeatletter
\providecommand \@ifxundefined [1]{%
 \@ifx{#1\undefined}
}%
\providecommand \@ifnum [1]{%
 \ifnum #1\expandafter \@firstoftwo
 \else \expandafter \@secondoftwo
 \fi
}%
\providecommand \@ifx [1]{%
 \ifx #1\expandafter \@firstoftwo
 \else \expandafter \@secondoftwo
 \fi
}%
\providecommand \natexlab [1]{#1}%
\providecommand \enquote  [1]{``#1''}%
\providecommand \bibnamefont  [1]{#1}%
\providecommand \bibfnamefont [1]{#1}%
\providecommand \citenamefont [1]{#1}%
\providecommand \href@noop [0]{\@secondoftwo}%
\providecommand \href [0]{\begingroup \@sanitize@url \@href}%
\providecommand \@href[1]{\@@startlink{#1}\@@href}%
\providecommand \@@href[1]{\endgroup#1\@@endlink}%
\providecommand \@sanitize@url [0]{\catcode `\\12\catcode `\$12\catcode `\&12\catcode `\#12\catcode `\^12\catcode `\_12\catcode `\%12\relax}%
\providecommand \@@startlink[1]{}%
\providecommand \@@endlink[0]{}%
\providecommand \url  [0]{\begingroup\@sanitize@url \@url }%
\providecommand \@url [1]{\endgroup\@href {#1}{\urlprefix }}%
\providecommand \urlprefix  [0]{URL }%
\providecommand \Eprint [0]{\href }%
\providecommand \doibase [0]{https://doi.org/}%
\providecommand \selectlanguage [0]{\@gobble}%
\providecommand \bibinfo  [0]{\@secondoftwo}%
\providecommand \bibfield  [0]{\@secondoftwo}%
\providecommand \translation [1]{[#1]}%
\providecommand \BibitemOpen [0]{}%
\providecommand \bibitemStop [0]{}%
\providecommand \bibitemNoStop [0]{.\EOS\space}%
\providecommand \EOS [0]{\spacefactor3000\relax}%
\providecommand \BibitemShut  [1]{\csname bibitem#1\endcsname}%
\let\auto@bib@innerbib\@empty
%</preamble>
\bibitem [{\citenamefont {Anderson}(1958)}]{anderson1958absence}%
  \BibitemOpen
  \bibfield  {author} {\bibinfo {author} {\bibfnamefont {P.~W.}\ \bibnamefont {Anderson}},\ }\bibfield  {title} {\bibinfo {title} {Absence of diffusion in certain random lattices},\ }\href@noop {} {\bibfield  {journal} {\bibinfo  {journal} {Physical review}\ }\textbf {\bibinfo {volume} {109}},\ \bibinfo {pages} {1492} (\bibinfo {year} {1958})}\BibitemShut {NoStop}%
\bibitem [{\citenamefont {Abrahams}\ \emph {et~al.}(1979)\citenamefont {Abrahams}, \citenamefont {Anderson}, \citenamefont {Licciardello},\ and\ \citenamefont {Ramakrishnan}}]{abrahams1979scaling}%
  \BibitemOpen
  \bibfield  {author} {\bibinfo {author} {\bibfnamefont {E.}~\bibnamefont {Abrahams}}, \bibinfo {author} {\bibfnamefont {P.~W.}\ \bibnamefont {Anderson}}, \bibinfo {author} {\bibfnamefont {D.~C.}\ \bibnamefont {Licciardello}},\ and\ \bibinfo {author} {\bibfnamefont {T.~V.}\ \bibnamefont {Ramakrishnan}},\ }\bibfield  {title} {\bibinfo {title} {Scaling theory of localization: Absence of quantum diffusion in two dimensions},\ }\href@noop {} {\bibfield  {journal} {\bibinfo  {journal} {Physical Review Letters}\ }\textbf {\bibinfo {volume} {42}},\ \bibinfo {pages} {673} (\bibinfo {year} {1979})}\BibitemShut {NoStop}%
\bibitem [{\citenamefont {Bergmann}(1984)}]{bergmann1984weak}%
  \BibitemOpen
  \bibfield  {author} {\bibinfo {author} {\bibfnamefont {G.}~\bibnamefont {Bergmann}},\ }\bibfield  {title} {\bibinfo {title} {Weak localization in thin films: a time-of-flight experiment with conduction electrons},\ }\href@noop {} {\bibfield  {journal} {\bibinfo  {journal} {Physics Reports}\ }\textbf {\bibinfo {volume} {107}},\ \bibinfo {pages} {1} (\bibinfo {year} {1984})}\BibitemShut {NoStop}%
\bibitem [{\citenamefont {Lee}\ and\ \citenamefont {Ramakrishnan}(1985)}]{lee1985disordered}%
  \BibitemOpen
  \bibfield  {author} {\bibinfo {author} {\bibfnamefont {P.~A.}\ \bibnamefont {Lee}}\ and\ \bibinfo {author} {\bibfnamefont {T.~V.}\ \bibnamefont {Ramakrishnan}},\ }\bibfield  {title} {\bibinfo {title} {Disordered electronic systems},\ }\href@noop {} {\bibfield  {journal} {\bibinfo  {journal} {Reviews of modern physics}\ }\textbf {\bibinfo {volume} {57}},\ \bibinfo {pages} {287} (\bibinfo {year} {1985})}\BibitemShut {NoStop}%
\bibitem [{\citenamefont {Akkermans}\ and\ \citenamefont {Montambaux}(2007)}]{akkermans2007mesoscopic}%
  \BibitemOpen
  \bibfield  {author} {\bibinfo {author} {\bibfnamefont {E.}~\bibnamefont {Akkermans}}\ and\ \bibinfo {author} {\bibfnamefont {G.}~\bibnamefont {Montambaux}},\ }\href@noop {} {\emph {\bibinfo {title} {Mesoscopic physics of electrons and photons}}}\ (\bibinfo  {publisher} {Cambridge university press},\ \bibinfo {year} {2007})\BibitemShut {NoStop}%
\bibitem [{\citenamefont {Altshuler}\ \emph {et~al.}(1980)\citenamefont {Altshuler}, \citenamefont {Khmel'Nitzkii}, \citenamefont {Larkin},\ and\ \citenamefont {Lee}}]{altshuler1980magnetoresistance}%
  \BibitemOpen
  \bibfield  {author} {\bibinfo {author} {\bibfnamefont {B.}~\bibnamefont {Altshuler}}, \bibinfo {author} {\bibfnamefont {D.}~\bibnamefont {Khmel'Nitzkii}}, \bibinfo {author} {\bibfnamefont {A.}~\bibnamefont {Larkin}},\ and\ \bibinfo {author} {\bibfnamefont {P.}~\bibnamefont {Lee}},\ }\bibfield  {title} {\bibinfo {title} {Magnetoresistance and hall effect in a disordered two-dimensional electron gas},\ }\href@noop {} {\bibfield  {journal} {\bibinfo  {journal} {Physical Review B}\ }\textbf {\bibinfo {volume} {22}},\ \bibinfo {pages} {5142} (\bibinfo {year} {1980})}\BibitemShut {NoStop}%
\bibitem [{\citenamefont {Chakravarty}\ and\ \citenamefont {Schmid}(1986)}]{chakravarty1986weak}%
  \BibitemOpen
  \bibfield  {author} {\bibinfo {author} {\bibfnamefont {S.}~\bibnamefont {Chakravarty}}\ and\ \bibinfo {author} {\bibfnamefont {A.}~\bibnamefont {Schmid}},\ }\bibfield  {title} {\bibinfo {title} {Weak localization: The quasiclassical theory of electrons in a random potential},\ }\href@noop {} {\bibfield  {journal} {\bibinfo  {journal} {Physics Reports}\ }\textbf {\bibinfo {volume} {140}},\ \bibinfo {pages} {193} (\bibinfo {year} {1986})}\BibitemShut {NoStop}%
\bibitem [{\citenamefont {Hikami}\ \emph {et~al.}(1980)\citenamefont {Hikami}, \citenamefont {Larkin},\ and\ \citenamefont {Nagaoka}}]{hikami1980spin}%
  \BibitemOpen
  \bibfield  {author} {\bibinfo {author} {\bibfnamefont {S.}~\bibnamefont {Hikami}}, \bibinfo {author} {\bibfnamefont {A.~I.}\ \bibnamefont {Larkin}},\ and\ \bibinfo {author} {\bibfnamefont {Y.}~\bibnamefont {Nagaoka}},\ }\bibfield  {title} {\bibinfo {title} {Spin-orbit interaction and magnetoresistance in the two dimensional random system},\ }\href@noop {} {\bibfield  {journal} {\bibinfo  {journal} {Progress of Theoretical Physics}\ }\textbf {\bibinfo {volume} {63}},\ \bibinfo {pages} {707} (\bibinfo {year} {1980})}\BibitemShut {NoStop}%
\bibitem [{\citenamefont {Lu}\ and\ \citenamefont {Shen}(2015)}]{lu2015weak}%
  \BibitemOpen
  \bibfield  {author} {\bibinfo {author} {\bibfnamefont {H.-Z.}\ \bibnamefont {Lu}}\ and\ \bibinfo {author} {\bibfnamefont {S.-Q.}\ \bibnamefont {Shen}},\ }\bibfield  {title} {\bibinfo {title} {Weak antilocalization and localization in disordered and interacting weyl semimetals},\ }\href@noop {} {\bibfield  {journal} {\bibinfo  {journal} {Physical Review B}\ }\textbf {\bibinfo {volume} {92}},\ \bibinfo {pages} {035203} (\bibinfo {year} {2015})}\BibitemShut {NoStop}%
\bibitem [{\citenamefont {Suzuura}\ and\ \citenamefont {Ando}(2002)}]{suzuura2002crossover}%
  \BibitemOpen
  \bibfield  {author} {\bibinfo {author} {\bibfnamefont {H.}~\bibnamefont {Suzuura}}\ and\ \bibinfo {author} {\bibfnamefont {T.}~\bibnamefont {Ando}},\ }\bibfield  {title} {\bibinfo {title} {Crossover from symplectic to orthogonal class in a two-dimensional honeycomb lattice},\ }\href@noop {} {\bibfield  {journal} {\bibinfo  {journal} {Physical review letters}\ }\textbf {\bibinfo {volume} {89}},\ \bibinfo {pages} {266603} (\bibinfo {year} {2002})}\BibitemShut {NoStop}%
\bibitem [{\citenamefont {Khveshchenko}(2006)}]{khveshchenko2006electron}%
  \BibitemOpen
  \bibfield  {author} {\bibinfo {author} {\bibfnamefont {D.}~\bibnamefont {Khveshchenko}},\ }\bibfield  {title} {\bibinfo {title} {Electron localization properties in graphene},\ }\href@noop {} {\bibfield  {journal} {\bibinfo  {journal} {Physical Review Letters}\ }\textbf {\bibinfo {volume} {97}},\ \bibinfo {pages} {036802} (\bibinfo {year} {2006})}\BibitemShut {NoStop}%
\bibitem [{\citenamefont {McCann}\ \emph {et~al.}(2006)\citenamefont {McCann}, \citenamefont {Kechedzhi}, \citenamefont {Fal’ko}, \citenamefont {Suzuura}, \citenamefont {Ando},\ and\ \citenamefont {Altshuler}}]{mccann2006weak}%
  \BibitemOpen
  \bibfield  {author} {\bibinfo {author} {\bibfnamefont {E.}~\bibnamefont {McCann}}, \bibinfo {author} {\bibfnamefont {K.}~\bibnamefont {Kechedzhi}}, \bibinfo {author} {\bibfnamefont {V.~I.}\ \bibnamefont {Fal’ko}}, \bibinfo {author} {\bibfnamefont {H.}~\bibnamefont {Suzuura}}, \bibinfo {author} {\bibfnamefont {T.}~\bibnamefont {Ando}},\ and\ \bibinfo {author} {\bibfnamefont {B.}~\bibnamefont {Altshuler}},\ }\bibfield  {title} {\bibinfo {title} {Weak-localization magnetoresistance and valley symmetry in graphene},\ }\href@noop {} {\bibfield  {journal} {\bibinfo  {journal} {Physical Review Letters}\ }\textbf {\bibinfo {volume} {97}},\ \bibinfo {pages} {146805} (\bibinfo {year} {2006})}\BibitemShut {NoStop}%
\bibitem [{\citenamefont {Gorbachev}\ \emph {et~al.}(2007)\citenamefont {Gorbachev}, \citenamefont {Tikhonenko}, \citenamefont {Mayorov}, \citenamefont {Horsell},\ and\ \citenamefont {Savchenko}}]{gorbachev2007weak}%
  \BibitemOpen
  \bibfield  {author} {\bibinfo {author} {\bibfnamefont {R.}~\bibnamefont {Gorbachev}}, \bibinfo {author} {\bibfnamefont {F.}~\bibnamefont {Tikhonenko}}, \bibinfo {author} {\bibfnamefont {A.}~\bibnamefont {Mayorov}}, \bibinfo {author} {\bibfnamefont {D.}~\bibnamefont {Horsell}},\ and\ \bibinfo {author} {\bibfnamefont {A.}~\bibnamefont {Savchenko}},\ }\bibfield  {title} {\bibinfo {title} {Weak localization in bilayer graphene},\ }\href@noop {} {\bibfield  {journal} {\bibinfo  {journal} {Physical Review Letters}\ }\textbf {\bibinfo {volume} {98}},\ \bibinfo {pages} {176805} (\bibinfo {year} {2007})}\BibitemShut {NoStop}%
\bibitem [{\citenamefont {Wu}\ \emph {et~al.}(2007)\citenamefont {Wu}, \citenamefont {Li}, \citenamefont {Song}, \citenamefont {Berger},\ and\ \citenamefont {de~Heer}}]{wu2007weak}%
  \BibitemOpen
  \bibfield  {author} {\bibinfo {author} {\bibfnamefont {X.}~\bibnamefont {Wu}}, \bibinfo {author} {\bibfnamefont {X.}~\bibnamefont {Li}}, \bibinfo {author} {\bibfnamefont {Z.}~\bibnamefont {Song}}, \bibinfo {author} {\bibfnamefont {C.}~\bibnamefont {Berger}},\ and\ \bibinfo {author} {\bibfnamefont {W.~A.}\ \bibnamefont {de~Heer}},\ }\bibfield  {title} {\bibinfo {title} {Weak antilocalization in epitaxial graphene: Evidence for chiral electrons},\ }\href@noop {} {\bibfield  {journal} {\bibinfo  {journal} {Physical Review Letters}\ }\textbf {\bibinfo {volume} {98}},\ \bibinfo {pages} {136801} (\bibinfo {year} {2007})}\BibitemShut {NoStop}%
\bibitem [{\citenamefont {Tikhonenko}\ \emph {et~al.}(2008)\citenamefont {Tikhonenko}, \citenamefont {Horsell}, \citenamefont {Gorbachev},\ and\ \citenamefont {Savchenko}}]{tikhonenko2008weak}%
  \BibitemOpen
  \bibfield  {author} {\bibinfo {author} {\bibfnamefont {F.}~\bibnamefont {Tikhonenko}}, \bibinfo {author} {\bibfnamefont {D.}~\bibnamefont {Horsell}}, \bibinfo {author} {\bibfnamefont {R.}~\bibnamefont {Gorbachev}},\ and\ \bibinfo {author} {\bibfnamefont {A.}~\bibnamefont {Savchenko}},\ }\bibfield  {title} {\bibinfo {title} {Weak localization in graphene flakes},\ }\href@noop {} {\bibfield  {journal} {\bibinfo  {journal} {Physical Review Letters}\ }\textbf {\bibinfo {volume} {100}},\ \bibinfo {pages} {056802} (\bibinfo {year} {2008})}\BibitemShut {NoStop}%
\bibitem [{\citenamefont {Tkachov}\ and\ \citenamefont {Hankiewicz}(2011)}]{tkachov2011weak}%
  \BibitemOpen
  \bibfield  {author} {\bibinfo {author} {\bibfnamefont {G.}~\bibnamefont {Tkachov}}\ and\ \bibinfo {author} {\bibfnamefont {E.}~\bibnamefont {Hankiewicz}},\ }\bibfield  {title} {\bibinfo {title} {Weak antilocalization in hgte quantum wells and topological surface states: Massive versus massless dirac fermions},\ }\href@noop {} {\bibfield  {journal} {\bibinfo  {journal} {Physical Review B}\ }\textbf {\bibinfo {volume} {84}},\ \bibinfo {pages} {035444} (\bibinfo {year} {2011})}\BibitemShut {NoStop}%
\bibitem [{\citenamefont {Lu}\ \emph {et~al.}(2011)\citenamefont {Lu}, \citenamefont {Shi},\ and\ \citenamefont {Shen}}]{lu2011competition}%
  \BibitemOpen
  \bibfield  {author} {\bibinfo {author} {\bibfnamefont {H.-Z.}\ \bibnamefont {Lu}}, \bibinfo {author} {\bibfnamefont {J.}~\bibnamefont {Shi}},\ and\ \bibinfo {author} {\bibfnamefont {S.-Q.}\ \bibnamefont {Shen}},\ }\bibfield  {title} {\bibinfo {title} {Competition between weak localization and antilocalization in topological surface states},\ }\href@noop {} {\bibfield  {journal} {\bibinfo  {journal} {Physical Review Letters}\ }\textbf {\bibinfo {volume} {107}},\ \bibinfo {pages} {076801} (\bibinfo {year} {2011})}\BibitemShut {NoStop}%
\bibitem [{\citenamefont {Lu}\ \emph {et~al.}(2013)\citenamefont {Lu}, \citenamefont {Yao}, \citenamefont {Xiao},\ and\ \citenamefont {Shen}}]{lu2013intervalley}%
  \BibitemOpen
  \bibfield  {author} {\bibinfo {author} {\bibfnamefont {H.-Z.}\ \bibnamefont {Lu}}, \bibinfo {author} {\bibfnamefont {W.}~\bibnamefont {Yao}}, \bibinfo {author} {\bibfnamefont {D.}~\bibnamefont {Xiao}},\ and\ \bibinfo {author} {\bibfnamefont {S.-Q.}\ \bibnamefont {Shen}},\ }\bibfield  {title} {\bibinfo {title} {Intervalley scattering and localization behaviors of spin-valley coupled dirac fermions},\ }\href@noop {} {\bibfield  {journal} {\bibinfo  {journal} {Physical Review Letters}\ }\textbf {\bibinfo {volume} {110}},\ \bibinfo {pages} {016806} (\bibinfo {year} {2013})}\BibitemShut {NoStop}%
\bibitem [{\citenamefont {Lu}\ and\ \citenamefont {Shen}(2014)}]{lu2014finite}%
  \BibitemOpen
  \bibfield  {author} {\bibinfo {author} {\bibfnamefont {H.-Z.}\ \bibnamefont {Lu}}\ and\ \bibinfo {author} {\bibfnamefont {S.-Q.}\ \bibnamefont {Shen}},\ }\bibfield  {title} {\bibinfo {title} {Finite-temperature conductivity and magnetoconductivity of topological insulators},\ }\href@noop {} {\bibfield  {journal} {\bibinfo  {journal} {Physical Review Letters}\ }\textbf {\bibinfo {volume} {112}},\ \bibinfo {pages} {146601} (\bibinfo {year} {2014})}\BibitemShut {NoStop}%
\bibitem [{\citenamefont {Fu}\ \emph {et~al.}(2019)\citenamefont {Fu}, \citenamefont {Wang},\ and\ \citenamefont {Shen}}]{fu2019quantum}%
  \BibitemOpen
  \bibfield  {author} {\bibinfo {author} {\bibfnamefont {B.}~\bibnamefont {Fu}}, \bibinfo {author} {\bibfnamefont {H.-W.}\ \bibnamefont {Wang}},\ and\ \bibinfo {author} {\bibfnamefont {S.-Q.}\ \bibnamefont {Shen}},\ }\bibfield  {title} {\bibinfo {title} {Quantum interference theory of magnetoresistance in dirac materials},\ }\href@noop {} {\bibfield  {journal} {\bibinfo  {journal} {Physical Review Letters}\ }\textbf {\bibinfo {volume} {122}},\ \bibinfo {pages} {246601} (\bibinfo {year} {2019})}\BibitemShut {NoStop}%
\bibitem [{\citenamefont {Castro~Neto}\ \emph {et~al.}(2009)\citenamefont {Castro~Neto}, \citenamefont {Guinea}, \citenamefont {Peres}, \citenamefont {Novoselov},\ and\ \citenamefont {Geim}}]{castro2009electronic}%
  \BibitemOpen
  \bibfield  {author} {\bibinfo {author} {\bibfnamefont {A.~H.}\ \bibnamefont {Castro~Neto}}, \bibinfo {author} {\bibfnamefont {F.}~\bibnamefont {Guinea}}, \bibinfo {author} {\bibfnamefont {N.~M.}\ \bibnamefont {Peres}}, \bibinfo {author} {\bibfnamefont {K.~S.}\ \bibnamefont {Novoselov}},\ and\ \bibinfo {author} {\bibfnamefont {A.~K.}\ \bibnamefont {Geim}},\ }\bibfield  {title} {\bibinfo {title} {The electronic properties of graphene},\ }\href@noop {} {\bibfield  {journal} {\bibinfo  {journal} {Reviews of modern physics}\ }\textbf {\bibinfo {volume} {81}},\ \bibinfo {pages} {109} (\bibinfo {year} {2009})}\BibitemShut {NoStop}%
\bibitem [{\citenamefont {Armitage}\ \emph {et~al.}(2018)\citenamefont {Armitage}, \citenamefont {Mele},\ and\ \citenamefont {Vishwanath}}]{armitage2018weyl}%
  \BibitemOpen
  \bibfield  {author} {\bibinfo {author} {\bibfnamefont {N.~P.}\ \bibnamefont {Armitage}}, \bibinfo {author} {\bibfnamefont {E.~J.}\ \bibnamefont {Mele}},\ and\ \bibinfo {author} {\bibfnamefont {A.}~\bibnamefont {Vishwanath}},\ }\bibfield  {title} {\bibinfo {title} {Weyl and dirac semimetals in three-dimensional solids},\ }\href@noop {} {\bibfield  {journal} {\bibinfo  {journal} {Reviews of Modern Physics}\ }\textbf {\bibinfo {volume} {90}},\ \bibinfo {pages} {015001} (\bibinfo {year} {2018})}\BibitemShut {NoStop}%
\bibitem [{\citenamefont {Vafek}\ and\ \citenamefont {Vishwanath}(2014)}]{vafek2014dirac}%
  \BibitemOpen
  \bibfield  {author} {\bibinfo {author} {\bibfnamefont {O.}~\bibnamefont {Vafek}}\ and\ \bibinfo {author} {\bibfnamefont {A.}~\bibnamefont {Vishwanath}},\ }\bibfield  {title} {\bibinfo {title} {Dirac fermions in solids: From high-t c cuprates and graphene to topological insulators and weyl semimetals},\ }\href@noop {} {\bibfield  {journal} {\bibinfo  {journal} {Annu. Rev. Condens. Matter Phys.}\ }\textbf {\bibinfo {volume} {5}},\ \bibinfo {pages} {83} (\bibinfo {year} {2014})}\BibitemShut {NoStop}%
\bibitem [{\citenamefont {Rosen}\ \emph {et~al.}(2019)\citenamefont {Rosen}, \citenamefont {Yudhistira}, \citenamefont {Sharma}, \citenamefont {Salehi}, \citenamefont {Kastner}, \citenamefont {Oh}, \citenamefont {Adam},\ and\ \citenamefont {Goldhaber-Gordon}}]{rosen2019absence}%
  \BibitemOpen
  \bibfield  {author} {\bibinfo {author} {\bibfnamefont {I.~T.}\ \bibnamefont {Rosen}}, \bibinfo {author} {\bibfnamefont {I.}~\bibnamefont {Yudhistira}}, \bibinfo {author} {\bibfnamefont {G.}~\bibnamefont {Sharma}}, \bibinfo {author} {\bibfnamefont {M.}~\bibnamefont {Salehi}}, \bibinfo {author} {\bibfnamefont {M.}~\bibnamefont {Kastner}}, \bibinfo {author} {\bibfnamefont {S.}~\bibnamefont {Oh}}, \bibinfo {author} {\bibfnamefont {S.}~\bibnamefont {Adam}},\ and\ \bibinfo {author} {\bibfnamefont {D.}~\bibnamefont {Goldhaber-Gordon}},\ }\bibfield  {title} {\bibinfo {title} {Absence of strong localization at low conductivity in the topological surface state of low-disorder sb 2 te 3},\ }\href@noop {} {\bibfield  {journal} {\bibinfo  {journal} {Physical Review B}\ }\textbf {\bibinfo {volume} {99}},\ \bibinfo {pages} {201101} (\bibinfo {year} {2019})}\BibitemShut {NoStop}%
\bibitem [{\citenamefont {Bradlyn}\ \emph {et~al.}(2016)\citenamefont {Bradlyn}, \citenamefont {Cano}, \citenamefont {Wang}, \citenamefont {Vergniory}, \citenamefont {Felser}, \citenamefont {Cava},\ and\ \citenamefont {Bernevig}}]{bradlyn2016beyond}%
  \BibitemOpen
  \bibfield  {author} {\bibinfo {author} {\bibfnamefont {B.}~\bibnamefont {Bradlyn}}, \bibinfo {author} {\bibfnamefont {J.}~\bibnamefont {Cano}}, \bibinfo {author} {\bibfnamefont {Z.}~\bibnamefont {Wang}}, \bibinfo {author} {\bibfnamefont {M.}~\bibnamefont {Vergniory}}, \bibinfo {author} {\bibfnamefont {C.}~\bibnamefont {Felser}}, \bibinfo {author} {\bibfnamefont {R.~J.}\ \bibnamefont {Cava}},\ and\ \bibinfo {author} {\bibfnamefont {B.~A.}\ \bibnamefont {Bernevig}},\ }\bibfield  {title} {\bibinfo {title} {Beyond dirac and weyl fermions: Unconventional quasiparticles in conventional crystals},\ }\href@noop {} {\bibfield  {journal} {\bibinfo  {journal} {Science}\ }\textbf {\bibinfo {volume} {353}},\ \bibinfo {pages} {aaf5037} (\bibinfo {year} {2016})}\BibitemShut {NoStop}%
\bibitem [{\citenamefont {Bradley}(1972)}]{bradley1972c}%
  \BibitemOpen
  \bibfield  {author} {\bibinfo {author} {\bibfnamefont {C.}~\bibnamefont {Bradley}},\ }\href@noop {} {\bibinfo {title} {C racknell, ap: The mathematical theory of symmetry in solids, clarendon}} (\bibinfo {year} {1972})\BibitemShut {NoStop}%
\bibitem [{\citenamefont {Wieder}\ \emph {et~al.}(2016)\citenamefont {Wieder}, \citenamefont {Kim}, \citenamefont {Rappe},\ and\ \citenamefont {Kane}}]{wieder2016double}%
  \BibitemOpen
  \bibfield  {author} {\bibinfo {author} {\bibfnamefont {B.~J.}\ \bibnamefont {Wieder}}, \bibinfo {author} {\bibfnamefont {Y.}~\bibnamefont {Kim}}, \bibinfo {author} {\bibfnamefont {A.}~\bibnamefont {Rappe}},\ and\ \bibinfo {author} {\bibfnamefont {C.}~\bibnamefont {Kane}},\ }\bibfield  {title} {\bibinfo {title} {Double dirac semimetals in three dimensions},\ }\href@noop {} {\bibfield  {journal} {\bibinfo  {journal} {Physical review letters}\ }\textbf {\bibinfo {volume} {116}},\ \bibinfo {pages} {186402} (\bibinfo {year} {2016})}\BibitemShut {NoStop}%
\bibitem [{\citenamefont {Ezawa}(2016)}]{ezawa2016pseudospin}%
  \BibitemOpen
  \bibfield  {author} {\bibinfo {author} {\bibfnamefont {M.}~\bibnamefont {Ezawa}},\ }\bibfield  {title} {\bibinfo {title} {Pseudospin-3 2 fermions, type-ii weyl semimetals, and critical weyl semimetals in tricolor cubic lattices},\ }\href@noop {} {\bibfield  {journal} {\bibinfo  {journal} {Physical Review B}\ }\textbf {\bibinfo {volume} {94}},\ \bibinfo {pages} {195205} (\bibinfo {year} {2016})}\BibitemShut {NoStop}%
\bibitem [{\citenamefont {Chang}\ \emph {et~al.}(2018)\citenamefont {Chang}, \citenamefont {Wieder}, \citenamefont {Schindler}, \citenamefont {Sanchez}, \citenamefont {Belopolski}, \citenamefont {Huang}, \citenamefont {Singh}, \citenamefont {Wu}, \citenamefont {Chang}, \citenamefont {Neupert} \emph {et~al.}}]{chang2018topological}%
  \BibitemOpen
  \bibfield  {author} {\bibinfo {author} {\bibfnamefont {G.}~\bibnamefont {Chang}}, \bibinfo {author} {\bibfnamefont {B.~J.}\ \bibnamefont {Wieder}}, \bibinfo {author} {\bibfnamefont {F.}~\bibnamefont {Schindler}}, \bibinfo {author} {\bibfnamefont {D.~S.}\ \bibnamefont {Sanchez}}, \bibinfo {author} {\bibfnamefont {I.}~\bibnamefont {Belopolski}}, \bibinfo {author} {\bibfnamefont {S.-M.}\ \bibnamefont {Huang}}, \bibinfo {author} {\bibfnamefont {B.}~\bibnamefont {Singh}}, \bibinfo {author} {\bibfnamefont {D.}~\bibnamefont {Wu}}, \bibinfo {author} {\bibfnamefont {T.-R.}\ \bibnamefont {Chang}}, \bibinfo {author} {\bibfnamefont {T.}~\bibnamefont {Neupert}}, \emph {et~al.},\ }\bibfield  {title} {\bibinfo {title} {Topological quantum properties of chiral crystals},\ }\href@noop {} {\bibfield  {journal} {\bibinfo  {journal} {Nature materials}\ }\textbf {\bibinfo {volume} {17}},\ \bibinfo {pages} {978} (\bibinfo {year} {2018})}\BibitemShut {NoStop}%
\bibitem [{\citenamefont {Ahmad}\ and\ \citenamefont {Sharma}(2025)}]{ahmad2025longitudinal}%
  \BibitemOpen
  \bibfield  {author} {\bibinfo {author} {\bibfnamefont {A.}~\bibnamefont {Ahmad}}\ and\ \bibinfo {author} {\bibfnamefont {G.}~\bibnamefont {Sharma}},\ }\bibfield  {title} {\bibinfo {title} {Longitudinal magnetoconductance of higher-pseudospin fermions},\ }\href@noop {} {\bibfield  {journal} {\bibinfo  {journal} {Physical Review B}\ }\textbf {\bibinfo {volume} {112}},\ \bibinfo {pages} {045135} (\bibinfo {year} {2025})}\BibitemShut {NoStop}%
\bibitem [{\citenamefont {Takane}\ \emph {et~al.}(2019)\citenamefont {Takane}, \citenamefont {Wang}, \citenamefont {Souma}, \citenamefont {Nakayama}, \citenamefont {Nakamura}, \citenamefont {Oinuma}, \citenamefont {Nakata}, \citenamefont {Iwasawa}, \citenamefont {Cacho}, \citenamefont {Kim} \emph {et~al.}}]{takane2019observation}%
  \BibitemOpen
  \bibfield  {author} {\bibinfo {author} {\bibfnamefont {D.}~\bibnamefont {Takane}}, \bibinfo {author} {\bibfnamefont {Z.}~\bibnamefont {Wang}}, \bibinfo {author} {\bibfnamefont {S.}~\bibnamefont {Souma}}, \bibinfo {author} {\bibfnamefont {K.}~\bibnamefont {Nakayama}}, \bibinfo {author} {\bibfnamefont {T.}~\bibnamefont {Nakamura}}, \bibinfo {author} {\bibfnamefont {H.}~\bibnamefont {Oinuma}}, \bibinfo {author} {\bibfnamefont {Y.}~\bibnamefont {Nakata}}, \bibinfo {author} {\bibfnamefont {H.}~\bibnamefont {Iwasawa}}, \bibinfo {author} {\bibfnamefont {C.}~\bibnamefont {Cacho}}, \bibinfo {author} {\bibfnamefont {T.}~\bibnamefont {Kim}}, \emph {et~al.},\ }\bibfield  {title} {\bibinfo {title} {Observation of chiral fermions with a large topological charge and associated fermi-arc surface states in cosi},\ }\href@noop {} {\bibfield  {journal} {\bibinfo  {journal} {Physical review letters}\ }\textbf {\bibinfo {volume} {122}},\ \bibinfo {pages} {076402} (\bibinfo {year} {2019})}\BibitemShut {NoStop}%
\bibitem [{\citenamefont {Sanchez}\ \emph {et~al.}(2019)\citenamefont {Sanchez}, \citenamefont {Belopolski}, \citenamefont {Cochran}, \citenamefont {Xu}, \citenamefont {Yin}, \citenamefont {Chang}, \citenamefont {Xie}, \citenamefont {Manna}, \citenamefont {S{\"u}{\ss}}, \citenamefont {Huang} \emph {et~al.}}]{sanchez2019topological}%
  \BibitemOpen
  \bibfield  {author} {\bibinfo {author} {\bibfnamefont {D.~S.}\ \bibnamefont {Sanchez}}, \bibinfo {author} {\bibfnamefont {I.}~\bibnamefont {Belopolski}}, \bibinfo {author} {\bibfnamefont {T.~A.}\ \bibnamefont {Cochran}}, \bibinfo {author} {\bibfnamefont {X.}~\bibnamefont {Xu}}, \bibinfo {author} {\bibfnamefont {J.-X.}\ \bibnamefont {Yin}}, \bibinfo {author} {\bibfnamefont {G.}~\bibnamefont {Chang}}, \bibinfo {author} {\bibfnamefont {W.}~\bibnamefont {Xie}}, \bibinfo {author} {\bibfnamefont {K.}~\bibnamefont {Manna}}, \bibinfo {author} {\bibfnamefont {V.}~\bibnamefont {S{\"u}{\ss}}}, \bibinfo {author} {\bibfnamefont {C.-Y.}\ \bibnamefont {Huang}}, \emph {et~al.},\ }\bibfield  {title} {\bibinfo {title} {Topological chiral crystals with helicoid-arc quantum states},\ }\href@noop {} {\bibfield  {journal} {\bibinfo  {journal} {Nature}\ }\textbf {\bibinfo {volume} {567}},\ \bibinfo {pages} {500} (\bibinfo {year} {2019})}\BibitemShut {NoStop}%
\bibitem [{\citenamefont {Schr{\"o}ter}\ \emph {et~al.}(2019)\citenamefont {Schr{\"o}ter}, \citenamefont {Pei}, \citenamefont {Vergniory}, \citenamefont {Sun}, \citenamefont {Manna}, \citenamefont {De~Juan}, \citenamefont {Krieger}, \citenamefont {S{\"u}ss}, \citenamefont {Schmidt}, \citenamefont {Dudin} \emph {et~al.}}]{schroter2019chiral}%
  \BibitemOpen
  \bibfield  {author} {\bibinfo {author} {\bibfnamefont {N.~B.}\ \bibnamefont {Schr{\"o}ter}}, \bibinfo {author} {\bibfnamefont {D.}~\bibnamefont {Pei}}, \bibinfo {author} {\bibfnamefont {M.~G.}\ \bibnamefont {Vergniory}}, \bibinfo {author} {\bibfnamefont {Y.}~\bibnamefont {Sun}}, \bibinfo {author} {\bibfnamefont {K.}~\bibnamefont {Manna}}, \bibinfo {author} {\bibfnamefont {F.}~\bibnamefont {De~Juan}}, \bibinfo {author} {\bibfnamefont {J.~A.}\ \bibnamefont {Krieger}}, \bibinfo {author} {\bibfnamefont {V.}~\bibnamefont {S{\"u}ss}}, \bibinfo {author} {\bibfnamefont {M.}~\bibnamefont {Schmidt}}, \bibinfo {author} {\bibfnamefont {P.}~\bibnamefont {Dudin}}, \emph {et~al.},\ }\bibfield  {title} {\bibinfo {title} {Chiral topological semimetal with multifold band crossings and long fermi arcs},\ }\href@noop {} {\bibfield  {journal} {\bibinfo  {journal} {Nature Physics}\ }\textbf {\bibinfo {volume} {15}},\ \bibinfo {pages} {759} (\bibinfo {year} {2019})}\BibitemShut {NoStop}%
\bibitem [{\citenamefont {Zhu}\ \emph {et~al.}(2017)\citenamefont {Zhu}, \citenamefont {Zhang}, \citenamefont {Yan}, \citenamefont {Xing},\ and\ \citenamefont {Zhu}}]{zhu2017emergent}%
  \BibitemOpen
  \bibfield  {author} {\bibinfo {author} {\bibfnamefont {Y.-Q.}\ \bibnamefont {Zhu}}, \bibinfo {author} {\bibfnamefont {D.-W.}\ \bibnamefont {Zhang}}, \bibinfo {author} {\bibfnamefont {H.}~\bibnamefont {Yan}}, \bibinfo {author} {\bibfnamefont {D.-Y.}\ \bibnamefont {Xing}},\ and\ \bibinfo {author} {\bibfnamefont {S.-L.}\ \bibnamefont {Zhu}},\ }\bibfield  {title} {\bibinfo {title} {Emergent pseudospin-1 maxwell fermions with a threefold degeneracy in optical lattices},\ }\href@noop {} {\bibfield  {journal} {\bibinfo  {journal} {Physical Review A}\ }\textbf {\bibinfo {volume} {96}},\ \bibinfo {pages} {033634} (\bibinfo {year} {2017})}\BibitemShut {NoStop}%
\bibitem [{\citenamefont {Singh}\ and\ \citenamefont {Sharma}(2023)}]{singh2023quantum}%
  \BibitemOpen
  \bibfield  {author} {\bibinfo {author} {\bibfnamefont {A.}~\bibnamefont {Singh}}\ and\ \bibinfo {author} {\bibfnamefont {G.}~\bibnamefont {Sharma}},\ }\bibfield  {title} {\bibinfo {title} {Quantum interference of pseudospin-1 fermions},\ }\href@noop {} {\bibfield  {journal} {\bibinfo  {journal} {Physical Review B}\ }\textbf {\bibinfo {volume} {108}},\ \bibinfo {pages} {195426} (\bibinfo {year} {2023})}\BibitemShut {NoStop}%
\bibitem [{\citenamefont {Miao}\ \emph {et~al.}(2023)\citenamefont {Miao}, \citenamefont {Tu},\ and\ \citenamefont {Zhou}}]{miao2023weak}%
  \BibitemOpen
  \bibfield  {author} {\bibinfo {author} {\bibfnamefont {S.}~\bibnamefont {Miao}}, \bibinfo {author} {\bibfnamefont {D.}~\bibnamefont {Tu}},\ and\ \bibinfo {author} {\bibfnamefont {J.}~\bibnamefont {Zhou}},\ }\bibfield  {title} {\bibinfo {title} {Weak localization in disordered spin-1 chiral fermions},\ }\href@noop {} {\bibfield  {journal} {\bibinfo  {journal} {Chinese Physics B}\ }\textbf {\bibinfo {volume} {32}},\ \bibinfo {pages} {017502} (\bibinfo {year} {2023})}\BibitemShut {NoStop}%
\bibitem [{\citenamefont {Sakurai}\ and\ \citenamefont {Napolitano}(2020)}]{sakurai2020modern}%
  \BibitemOpen
  \bibfield  {author} {\bibinfo {author} {\bibfnamefont {J.~J.}\ \bibnamefont {Sakurai}}\ and\ \bibinfo {author} {\bibfnamefont {J.}~\bibnamefont {Napolitano}},\ }\href@noop {} {\emph {\bibinfo {title} {Modern quantum mechanics}}}\ (\bibinfo  {publisher} {Cambridge university press},\ \bibinfo {year} {2020})\BibitemShut {NoStop}%
\bibitem [{\citenamefont {Varshalovich}\ \emph {et~al.}(1988)\citenamefont {Varshalovich}, \citenamefont {Moskalev},\ and\ \citenamefont {Khersonskii}}]{varshalovich1988quantum}%
  \BibitemOpen
  \bibfield  {author} {\bibinfo {author} {\bibfnamefont {D.~A.}\ \bibnamefont {Varshalovich}}, \bibinfo {author} {\bibfnamefont {A.~N.}\ \bibnamefont {Moskalev}},\ and\ \bibinfo {author} {\bibfnamefont {V.~K.}\ \bibnamefont {Khersonskii}},\ }\href@noop {} {\emph {\bibinfo {title} {Quantum theory of angular momentum}}}\ (\bibinfo  {publisher} {World Scientific},\ \bibinfo {year} {1988})\BibitemShut {NoStop}%
\bibitem [{\citenamefont {Edmonds}(1996)}]{edmonds1996angular}%
  \BibitemOpen
  \bibfield  {author} {\bibinfo {author} {\bibfnamefont {A.~R.}\ \bibnamefont {Edmonds}},\ }\href@noop {} {\emph {\bibinfo {title} {Angular momentum in quantum mechanics}}},\ Vol.~\bibinfo {volume} {4}\ (\bibinfo  {publisher} {Princeton university press},\ \bibinfo {year} {1996})\BibitemShut {NoStop}%
\bibitem [{\citenamefont {Bruus}\ and\ \citenamefont {Flensberg}(2004)}]{bruus2004many}%
  \BibitemOpen
  \bibfield  {author} {\bibinfo {author} {\bibfnamefont {H.}~\bibnamefont {Bruus}}\ and\ \bibinfo {author} {\bibfnamefont {K.}~\bibnamefont {Flensberg}},\ }\href@noop {} {\emph {\bibinfo {title} {Many-body quantum theory in condensed matter physics: an introduction}}}\ (\bibinfo  {publisher} {Oxford university press},\ \bibinfo {year} {2004})\BibitemShut {NoStop}%
\bibitem [{\citenamefont {Velick{\`y}}(1969)}]{velicky1969theory}%
  \BibitemOpen
  \bibfield  {author} {\bibinfo {author} {\bibfnamefont {B.}~\bibnamefont {Velick{\`y}}},\ }\bibfield  {title} {\bibinfo {title} {Theory of electronic transport in disordered binary alloys: coherent-potential approximation},\ }\href@noop {} {\bibfield  {journal} {\bibinfo  {journal} {Physical Review}\ }\textbf {\bibinfo {volume} {184}},\ \bibinfo {pages} {614} (\bibinfo {year} {1969})}\BibitemShut {NoStop}%
\bibitem [{\citenamefont {Rammer}\ and\ \citenamefont {Smith}(1986)}]{rammer1986quantum}%
  \BibitemOpen
  \bibfield  {author} {\bibinfo {author} {\bibfnamefont {J.}~\bibnamefont {Rammer}}\ and\ \bibinfo {author} {\bibfnamefont {H.}~\bibnamefont {Smith}},\ }\bibfield  {title} {\bibinfo {title} {Quantum field-theoretical methods in transport theory of metals},\ }\href@noop {} {\bibfield  {journal} {\bibinfo  {journal} {Reviews of modern physics}\ }\textbf {\bibinfo {volume} {58}},\ \bibinfo {pages} {323} (\bibinfo {year} {1986})}\BibitemShut {NoStop}%
\bibitem [{\citenamefont {Rammer}(1991)}]{rammer1991quantum}%
  \BibitemOpen
  \bibfield  {author} {\bibinfo {author} {\bibfnamefont {J.}~\bibnamefont {Rammer}},\ }\bibfield  {title} {\bibinfo {title} {Quantum transport theory of electrons in solids: A single-particle approach},\ }\href@noop {} {\bibfield  {journal} {\bibinfo  {journal} {Reviews of Modern Physics}\ }\textbf {\bibinfo {volume} {63}},\ \bibinfo {pages} {781} (\bibinfo {year} {1991})}\BibitemShut {NoStop}%
\bibitem [{\citenamefont {Mahan}(2013)}]{mahan2013many}%
  \BibitemOpen
  \bibfield  {author} {\bibinfo {author} {\bibfnamefont {G.~D.}\ \bibnamefont {Mahan}},\ }\href@noop {} {\emph {\bibinfo {title} {Many-particle physics}}}\ (\bibinfo  {publisher} {Springer Science \& Business Media},\ \bibinfo {year} {2013})\BibitemShut {NoStop}%
\bibitem [{\citenamefont {Nielsen}\ and\ \citenamefont {Ninomiya}(1981)}]{nielsen1981no}%
  \BibitemOpen
  \bibfield  {author} {\bibinfo {author} {\bibfnamefont {H.~B.}\ \bibnamefont {Nielsen}}\ and\ \bibinfo {author} {\bibfnamefont {M.}~\bibnamefont {Ninomiya}},\ }\href@noop {} {\emph {\bibinfo {title} {No-go theorum for regularizing chiral fermions}}},\ \bibinfo {type} {Tech. Rep.}\ (\bibinfo  {institution} {Science Research Council},\ \bibinfo {year} {1981})\BibitemShut {NoStop}%
\bibitem [{\citenamefont {Xu}\ \emph {et~al.}(2020)\citenamefont {Xu}, \citenamefont {Fang}, \citenamefont {S{\'a}nchez-Mart{\'\i}nez}, \citenamefont {Venderbos}, \citenamefont {Ni}, \citenamefont {Qiu}, \citenamefont {Manna}, \citenamefont {Wang}, \citenamefont {Paglione}, \citenamefont {Bernhard} \emph {et~al.}}]{xu2020optical}%
  \BibitemOpen
  \bibfield  {author} {\bibinfo {author} {\bibfnamefont {B.}~\bibnamefont {Xu}}, \bibinfo {author} {\bibfnamefont {Z.}~\bibnamefont {Fang}}, \bibinfo {author} {\bibfnamefont {M.-{\'A}.}\ \bibnamefont {S{\'a}nchez-Mart{\'\i}nez}}, \bibinfo {author} {\bibfnamefont {J.~W.}\ \bibnamefont {Venderbos}}, \bibinfo {author} {\bibfnamefont {Z.}~\bibnamefont {Ni}}, \bibinfo {author} {\bibfnamefont {T.}~\bibnamefont {Qiu}}, \bibinfo {author} {\bibfnamefont {K.}~\bibnamefont {Manna}}, \bibinfo {author} {\bibfnamefont {K.}~\bibnamefont {Wang}}, \bibinfo {author} {\bibfnamefont {J.}~\bibnamefont {Paglione}}, \bibinfo {author} {\bibfnamefont {C.}~\bibnamefont {Bernhard}}, \emph {et~al.},\ }\bibfield  {title} {\bibinfo {title} {Optical signatures of multifold fermions in the chiral topological semimetal cosi},\ }\href@noop {} {\bibfield  {journal} {\bibinfo  {journal} {Proceedings of the National Academy of Sciences}\ }\textbf {\bibinfo {volume} {117}},\ \bibinfo {pages} {27104} (\bibinfo {year} {2020})}\BibitemShut {NoStop}%
\end{thebibliography}%
